\RequirePackage{lineno}
\documentclass[aps,twocolumn,superscriptaddress,preprintnumbers,amsmath,amssymb]{revtex4}
\pdfoutput=1 % if your are submitting a pdflatex (i.e. if you have % images in pdf, png or jpg format)

\usepackage{verbatim}
\usepackage{overpic}
\usepackage{graphicx}
\usepackage{bm}    
\usepackage{amsmath,amssymb}
\usepackage{enumerate}
\usepackage{lineno}
\usepackage[section]{placeins}
\usepackage{color}
\usepackage[figuresright]{rotating}
\usepackage{setspace}
\usepackage{booktabs} 
\usepackage{threeparttable}
\usepackage[colorlinks,linkcolor=blue,anchorcolor=green,citecolor=blue]{hyperref}

\usepackage{notoccite}
\graphicspath{{figures/}}
\usepackage{float}
\usepackage{epsfig}
\usepackage{epstopdf}
\usepackage{indentfirst}
\usepackage{array}
\usepackage{dcolumn}
\usepackage{multirow}
\usepackage{pifont} 
\usepackage{comment}
\usepackage{threeparttable} 
\usepackage{orcidlink}

\newcommand{\BR}{{\cal B}}
\newcommand{\chibJ}{{\chi_{b1,b2}}}

\begin{document}

\vspace*{-0\baselineskip}

\title{\quad\\[1.0cm] \boldmath 
First measurement of the masses of the $\Upsilon_1(1D)$ and $\Upsilon_3(1D)$ states and the energy dependence of the cross sections for $e^+e^-\to\Upsilon_J(1D)\eta$ and $e^+e^-\to\Upsilon_J(1D)\pi^+\pi^-$}

%\author{Authors ...}
%\collaboration{The Belle II Collaboration}
\author{M.~Abumusabh\,\orcidlink{0009-0004-1031-5425}} % 26883
  \author{I.~Adachi\,\orcidlink{0000-0003-2287-0173}} % 2590
  \author{K.~Adamczyk\,\orcidlink{0000-0001-6208-0876}} % 2239
  \author{A.~Aggarwal\,\orcidlink{0000-0002-5623-3896}} % 24463
% \author{L.~Aggarwal\,\orcidlink{0000-0002-0909-7537}} % 10083
  \author{H.~Ahmed\,\orcidlink{0000-0003-3976-7498}} % 11323
% \author{J.~K.~Ahn\,\orcidlink{0000-0002-5795-2243}} % 7423
  \author{Y.~Ahn\,\orcidlink{0000-0001-6820-0576}} % 14363
% \author{H.~Aihara\,\orcidlink{0000-0002-1907-5964}} % 2223
  \author{M.~Akdag\,\orcidlink{0009-0004-3728-1077}} % 27563
  \author{N.~Akopov\,\orcidlink{0000-0002-4425-2096}} % 9443
  \author{S.~Alghamdi\,\orcidlink{0000-0001-7609-112X}} % 27804
  \author{M.~Alhakami\,\orcidlink{0000-0002-2234-8628}} % 28103
% \author{A.~Aloisio\,\orcidlink{0000-0002-3883-6693}} % 2194
% \author{S.~Al~Said\,\orcidlink{0000-0002-4895-3869}} % 6823
% \author{A.~Alsharari\,\orcidlink{0000-0002-6993-1597}} % 27803
% ...
% 13403
% \author{S.~Zou\,\orcidlink{0000-0003-3377-7222}} % 19363
\collaboration{The Belle and Belle II Collaborations}

\begin{abstract}
We study the processes $e^+e^-\to\Upsilon_J(1D)\eta$ and $e^+e^-\to\Upsilon_J(1D)\pi^+\pi^-$ at center-of-mass 
energies $\sqrt{s}$=(10.73  -- 11.02) GeV using a $142.5\,\mathrm{fb}^{-1}$ data sample, including 122~fb$^{-1}$ near the $\Upsilon$(10860) peak ($\sqrt{s}$  = 10.866 GeV),  collected with the Belle detector
at the KEKB asymmetric-energy $e^+e^-$ collider.
From the peak sample, the products of Born cross section times branching fraction are obtained for $\sigma_{\rm Born}(e^+e^-\to\Upsilon_J(1D)\eta)$ or $\sigma_{\rm Born}(e^+e^-\to\Upsilon_J(1D)\pi^+\pi^-)$ and $\BR(\Upsilon_J(1D)\to\chi_{b1}\gamma)$ or $\BR(\Upsilon_J(1D)\to\chi_{b2}\gamma)$ for each $\Upsilon_J(1D)$ state.
The corresponding branching fractions for $\Upsilon(10860)$ decays are also obtained.
The significances of the $\Upsilon_1(1D)$, $\Upsilon_2(1D)$, and $\Upsilon_3(1D)$ signals are 4.8$\sigma$, ${>}10\sigma$, and 3.0$\sigma$, respectively, including systematic uncertainties.
The mass for $\Upsilon_2(1D)$ is measured to be $(10167.0\pm 1.0\pm 0.2)$ MeV/$c^2$, where the first and second uncertainties are statistical and systematic.
The mass splittings $\Delta m_{12}=m(\Upsilon_2(1D))-m(\Upsilon_1(1D))$ and $\Delta m_{23}=m(\Upsilon_3(1D))-m(\Upsilon_2(1D))$ are $(11.8\pm1.5\pm0.4)$ MeV/$c^2$ and $(7.6\pm2.4\pm0.6)$ MeV/$c^2$, respectively.~We determine the energy dependence of the cross sections for $e^+e^-\to\Upsilon_J(1D)\eta$ and $e^+e^-\to\Upsilon_J(1D)\pi^+\pi^-$ for the $\Upsilon_1(1D)$, $\Upsilon_2(1D)$, and $\Upsilon_3(1D)$ states, combined. 
\end{abstract}

\maketitle

%\linenumbers
%%%%%%%%%%%%%%%%%%%%%%%%%end
As a prominent member of the heavy quarkonium family, bottomonium is an ideal probe for the nonperturbative behavior of quantum chromodynamics~\cite{1534,2981,2603.09315,0412158}.
To date, only one $D$-wave bottomonium state, $\Upsilon_2(1D)$, has been observed, by CLEO and BaBar~\cite{70032001,111102}.
The relativized quark model, nonrelativistic constituent quark model, screened potential model, and the modified relativized quark model with a ﻿screening effect have predicted the mass splittings of the $\Upsilon_J(1D)$~\cite{074027,054034,074002,915}.
However, the states $\Upsilon_1(1D)$ and $\Upsilon_3(1D)$ have not yet been observed.

Within the hadronic loop mechanism, the branching fractions for $\Upsilon(10860)\to\Upsilon_J(1D)\eta$, $\Upsilon(11020)\to\Upsilon_J(1D)\eta$, and $\Upsilon(10860)\to\Upsilon_J(1D)\pi^+\pi^-$ range in magnitude from $10^{-4}$ to $10^{-3}$~\cite{094039,094018,114037}.
In a 121.4 fb$^{-1}$ sample at $\sqrt{s}$ = 10.866 GeV, Belle studied $e^+e^-\to\Upsilon_J(1D)\eta$ and $e^+e^-\to\Upsilon_J(1D)\pi^+\pi^-$ in the recoil mass spectra of $\eta$ and $\pi^+\pi^-$~\cite{633,032001}, where the $\Upsilon_1(1D)$, $\Upsilon_2(1D)$, and $\Upsilon_3(1D)$ states were treated as a single component or with fixed mass splittings. 
In this work, we separate $\Upsilon_1(1D)$, $\Upsilon_2(1D)$, and $\Upsilon_3(1D)$ states through exclusive reconstruction, applying a kinematic constraint, and performing a three-dimensional (3D) fit.
To date, there are no reported experimental results for $e^+e^-\to\Upsilon_J(1D)\eta$ and $e^+e^-\to\Upsilon_J(1D)\pi^+\pi^-$ in the energy region around the $\Upsilon(11020)$ resonance.

In this Letter, we measure the cross sections for $e^+e^-\to\Upsilon_J(1D)\eta$ and $e^+e^-\to\Upsilon_J(1D)\pi^+\pi^-$ using Belle data at $\sqrt{s}$ from 10.73 to 11.02 GeV. 
We report first evidence for the $\Upsilon_1(1D)$ and $\Upsilon_3(1D)$ states, as well as measurements of the $\Upsilon_J(1D)$ masses.
For the energy point at $\sqrt{s}$ = 10.866 GeV, we determine the product of the Born cross section for 
$e^+e^-\to\Upsilon_J(1D)\eta$ ($e^+e^-\to\Upsilon_J(1D)\pi^+\pi^-$) and the branching fraction of $\Upsilon_J(1D)\to\chi_{b1,b2}\gamma$.
The corresponding branching fractions for the $\Upsilon(10860)$ decays are also found.

The data reported here were recorded by the Belle detector~\cite{Belle1,Belle2} at the KEKB asymmetric-energy $e^+e^-$ collider~\cite{KEKB1,KEKB2}.
We use $\approx$1 fb$^{-1}$ taken at each of 18 center-of-mass (c.m.) energies in the range 10.73 GeV -- 11.02 GeV (energy scan data) and a 122 fb$^{-1}$ sample collected at the $\Upsilon(10860)$, $\sqrt{s}$=10865.8 MeV ($\Upsilon(10860)$ on-resonance data).
Thus, there are 19 energy points in total at which we measure cross sections for $e^+e^-\to\Upsilon_J(1D)\eta$ and $\Upsilon_J(1D)\pi^+\pi^-$. 
A detailed description of the Belle detector can be found in Refs.~\cite{Belle1,Belle2}.

We generate signal Monte Carlo (MC) events with c.m.\ energies from 10.73 to 11.02~GeV using the {\sc evtgen} generator~\cite{152} to determine the reconstruction efficiencies and signal shapes. 
Initial-state radiation (ISR) at next-to-leading order precision in quantum electrodynamics is simulated with {\sc phokhara}~\cite{71}.
For $e^+e^-\to\Upsilon_J(1D)\eta$, the polar angle of the $\eta$ in the $e^+e^-$ c.m.\ system ($\theta_\eta$) is distributed according to $(1+{\rm cos}^2\theta_\eta)$, $(1-{\frac{1}{7}}{\rm cos}^2\theta_\eta)$, and $(1+{\rm cos}^2\theta_\eta)$ for $\Upsilon_1(1D)\eta$, $\Upsilon_2(1D)\eta$, and $\Upsilon_3(1D)\eta$ production, respectively.
For $e^+e^-\to\Upsilon_J(1D)\pi^+\pi^-$, the polar angle of the $\pi^+\pi^-$ system in the $e^+e^-$ c.m.\ system ($\theta_{\pi^+\pi^-}$) is distributed uniformly.
Generic MC samples of $e^+e^-\to q\bar q$  ($q=u,d,s,c$) with the {\sc kkmc}~\cite{260} and {\sc pythia} packages and $\Upsilon(10860)\to B_{(s)}^{(*)}\bar B_{(s)}^{(*)}$ produced with 4 times the luminosity of the data are used to estimate backgrounds.
The simulated events are processed with a detector simulation based on {\sc geant3}~\cite{geant3}.

We reconstruct the decay chains~\cite{074027,054034,074002,915}
\begin{gather*}
e^+e^-\to\Upsilon_J(1D)\eta,~e^+e^-\to\Upsilon_J(1D)\pi^+\pi^-, \\
\eta\to\gamma\gamma,~\eta\to\pi^+\pi^-\pi^0, \\
\Upsilon_1(1D)\to\chi_{b1}\gamma,~\Upsilon_2(1D)\to\chi_{b1,b2}\gamma,~\Upsilon_3(1D)\to\chi_{b2}\gamma, \\
\chi_{b1,b2}\to\Upsilon(1S)\gamma,~{\rm and}~\Upsilon(1S)\to\ell^+\ell^-~(\ell=\mu~{\rm or}~e).
\end{gather*}The $\chi_{b0}$ is not included due to a small branching fraction for $\chi_{b0}\to\Upsilon(1S)\gamma$~\cite{112010,054003,PDG}.
Event selection and criteria for tracks and particle identification follow the methods used in Refs.~\cite{2510.25461,220}.

We require the energies of photons from $\pi^0$ and $\eta$ candidates to be greater than 50 MeV and 100 MeV, respectively.
The signal regions for $\pi^0\to\gamma\gamma$, $\eta\to\gamma\gamma$, and $\eta\to\pi^+\pi^-\pi^0$ are $110<M(\gamma\gamma)<150$ MeV/$c^2$ ($\pm$2.5$\sigma$), $0.51<M(\gamma\gamma)<0.58$ GeV/$c^2$ ($\pm$2.5$\sigma$), and $0.535<M(\pi^+\pi^-\pi^0)<0.560$ GeV/$c^2$ ($\pm$2.5$\sigma$), respectively, where the symbol $M$ denotes invariant mass and the $\sigma$ is the mass resolution.

The $\Upsilon(1S)$ signal regions are 9.15 $<$ $M(e^+e^-)$ $<$ 9.66 GeV/$c^2$ and 9.26 $<$ $M(\mu^+\mu^-)$ $<$ 9.66 GeV/$c^2$ (approximately $\pm$2.5$\sigma$). 

Photon pairs remaining in the event are used to reconstruct $\chibJ$ and $\Upsilon_J(1D)$ candidates.
The photon with the higher energy ($\gamma_H$) in the $e^+e^-$ c.m.\ frame is used to reconstruct $\chibJ\to\Upsilon({\rm 1S})\gamma$, and the other ($\gamma_L$) to reconstruct  $\Upsilon_J(1D)\to \chibJ\gamma$.
To calibrate the photon energy resolution function, three control channels $D^{*0} \to D^0\gamma$, $\pi^0 \to \gamma\gamma$, and $\eta \to \gamma\gamma$ are used~\cite{232002}.

We require the numbers of charged tracks to be two, four, and four in the channels $e^+e^-\to\Upsilon_J(1D)\eta(\to\gamma\gamma)$, $e^+e^-\to\Upsilon_J(1D)\eta(\to\pi^+\pi^-\pi^0)$, and $e^+e^-\to\Upsilon_J(1D)\pi^+\pi^-$, respectively.
For $e^+e^-\to\Upsilon_J(1D)\pi^+\pi^-$, background from $\gamma\to e^+e^-$ conversions is removed by requiring ${\rm cos}\theta_{\pi^+\pi^-} < 0.95$, where $\theta_{\pi^+\pi^-}$ is the angle between $\pi^+$ and $\pi^-$ in the laboratory frame.

To reduce backgrounds with additional particles in the final state, a 5C kinematic fit using {\sc kfit}~\cite{kfit} is performed, where the four-momentum of the final state system is constrained to match the initial $e^+e^-$ c.m.\ system, and the invariant mass of the lepton pair is constrained to the $\Upsilon(1S)$ nominal mass~\cite{PDG}.
The $\chi^2_{\rm 5C}/{n_{\rm dof}}$ value of the 5C fit is required to be less than 20 which retains 98\% of the signal.
In events with multiple candidates, only the candidate with the smallest value of $\chi^2_{\rm 5C}$ is retained.
The fractions of selected events with multiple candidates are found in signal MC to be 8\%, 2\%, and 1\% for $e^+e^-\to\Upsilon_J(1D)\eta(\to\gamma\gamma)$, $e^+e^-\to\Upsilon_J(1D)\eta(\to\pi^+\pi^-\pi^0)$, and $e^+e^-\to\Upsilon_J(1D)\pi^+\pi^-$, respectively. These values are consistent with the multiple candidate rates observed in the data. The fractions of correctly reconstructed candidates in $e^+e^-\to\Upsilon_J(1D)\eta(\to\gamma\gamma)$, $e^+e^-\to\Upsilon_J(1D)\eta(\to\pi^+\pi^-\pi^0)$, and $e^+e^-\to\Upsilon_J(1D)\pi^+\pi^-$ are found to be 94\%, 97\%, and 98\%, respectively.

For the data at $\sqrt{s}$ = 10.866 GeV, we simultaneously fit to the $e^+e^-\to\Upsilon_J(1D)\eta$ and $e^+e^-\to\Upsilon_J(1D)\pi^+\pi^-$ channels, where the masses of the $\Upsilon_J(1D)$ are constrained to be the same.~For $e^+e^-\to\Upsilon_J(1D)\eta$ or $e^+e^-\to\Upsilon_J(1D)\pi^+\pi^-$, we perform a 3D unbinned extended maximum-likelihood fit to the $M^{\prime}(\ell^+\ell^-\gamma_H)=M(\ell^+\ell^-\gamma_H) - M(\ell^+\ell^-) + m(\Upsilon(1S))$ ($\chi_{b1,b2}$ mass), $\Delta M = M(\ell^+\ell^-\gamma_{H}\gamma_{L}) - M(\ell^+\ell^-\gamma_H)$ (mass difference between $\chi_{b1,b2}$ and $\Upsilon_{J}(1D)$), and $M^{\prime}_{\rm rec}(\eta)$ = $M_{\rm rec}(\eta)+M(\gamma\gamma,\,\pi^+\pi^-\pi^0)-m(\eta)$ or $M_{\rm rec}(\pi^+\pi^-)$ ($\Upsilon_{J}(1D)$ mass) distributions before the kinematic fit to extract signal yields.
Hereinafter, the variables $M_{\rm rec}$ and $m$ denote the recoil mass and the nominal mass~\cite{PDG}.
The magnitude of the maximum correlation coefficient among the above three observables is 1\% from simulation.
The fitting function is a sum of the following components: 3D signal peaks and corresponding combinatorial backgrounds for $\Upsilon_1(1D)\to\chi_{b1}\gamma$, $\Upsilon_2(1D)\to\chi_{b1}\gamma$, $\Upsilon_2(1D)\to\chi_{b2}\gamma$, and $\Upsilon_3(1D)\to\chi_{b2}\gamma$ in $M^{\prime}(\ell^+\ell^-\gamma_H)$, $\Delta M$, $M^{\prime}_{\rm rec}(\eta)~{\rm or}~M_{\rm rec}(\pi^+\pi^-)$, respectively.
The 3D signal and background probability density functions (PDFs) are each the product of the three 1D PDFs.

The signal shapes in $M^{\prime}(\ell^+\ell^-\gamma_H)$ and $\Delta M$  are each described by a crystal ball (CB) function~\cite{CB}.
The signal shapes in $M^{\prime}_{\rm rec}(\eta)$ and $M_{\rm rec}(\pi^+\pi^-)$  are described by double Gaussian functions.
Except for the masses of the $\Upsilon_J(1D)$ states, all parameters are fixed according to signal MC simulations.
A first-order polynomial function with free parameters is used to describe the combinatorial background.
The justification for using this polynomial is verified through the control channels $e^+e^-\to\Upsilon(2S)(\to\chi_{b1,b2}\gamma)\eta$ and $e^+e^-\to\Upsilon(2S)(\to\chi_{b1,b2}\gamma)\pi^+\pi^-$.
The yield ratio $N^{\rm sig}(\eta\to\gamma\gamma)/N^{\rm sig}(\eta\to\pi^+\pi^-\pi^0)$ is fixed at 4.2 in the 3D fit according to the products of branching fractions and efficiencies. The ratio of the yields $N^{\rm sig}(\Upsilon_2(1D)\to\chi_{b1}\gamma)/N^{\rm sig}(\Upsilon_2(1D)\to\chi_{b2}\gamma)$ in $e^+e^-\to\Upsilon_J(1D)\eta$ and $e^+e^-\to\Upsilon_J(1D)\pi^+\pi^-$ is floated as a common parameter.

\begin{figure*}[htbp]
\centering
\includegraphics[width=4.9cm]{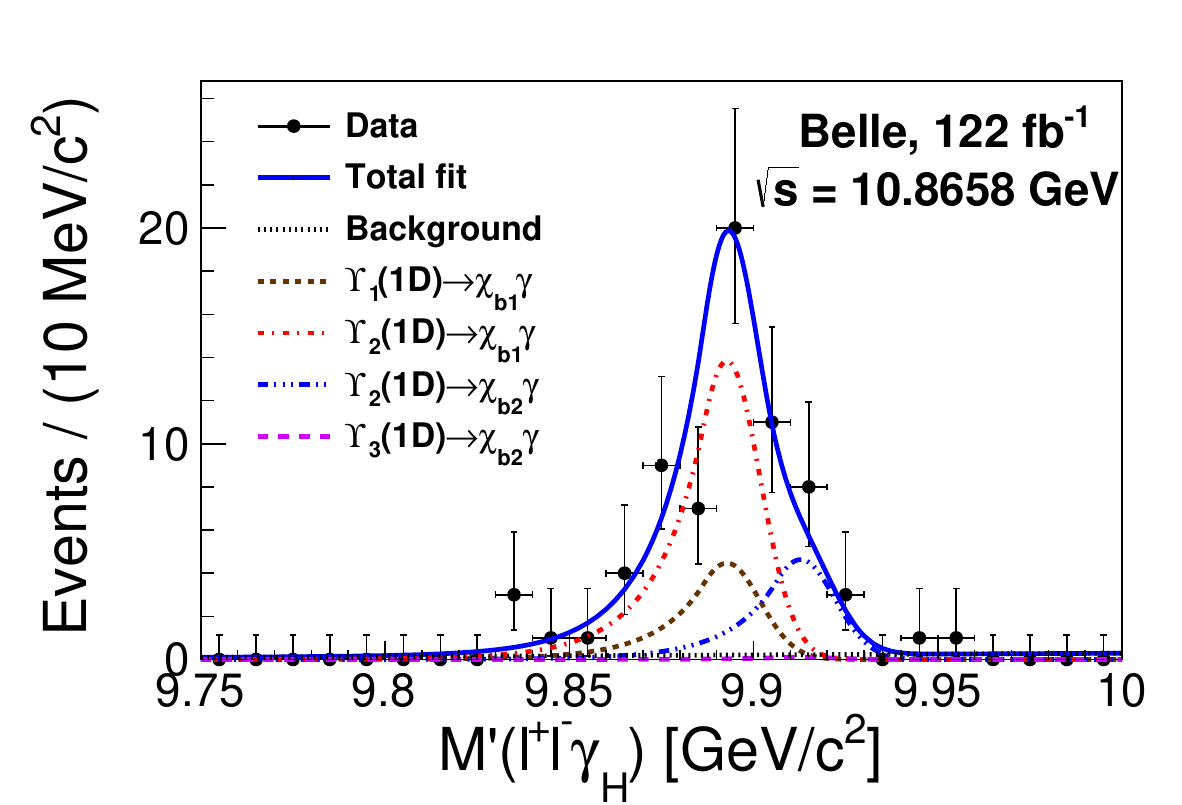}
\put(-23, 55){\small \bf (a)}
\includegraphics[width=4.9cm]{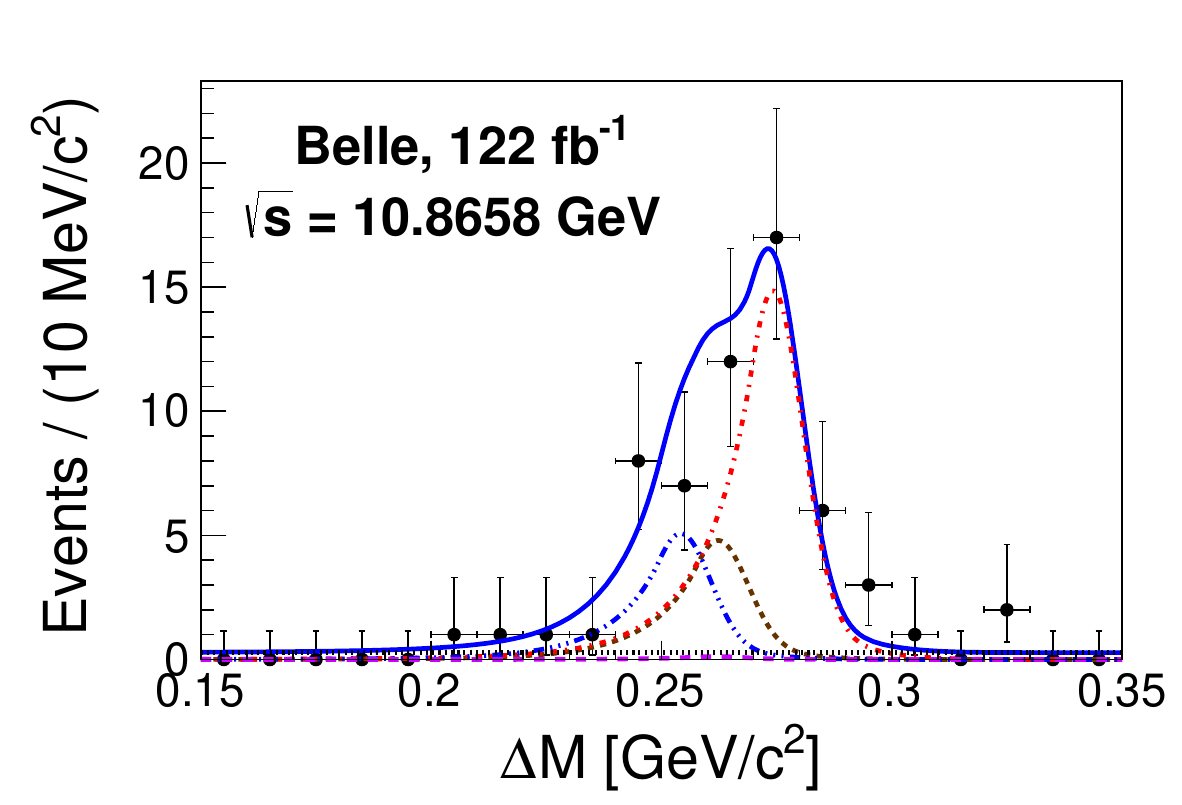}
\put(-23, 75){\small \bf (b)}
\includegraphics[width=4.9cm]{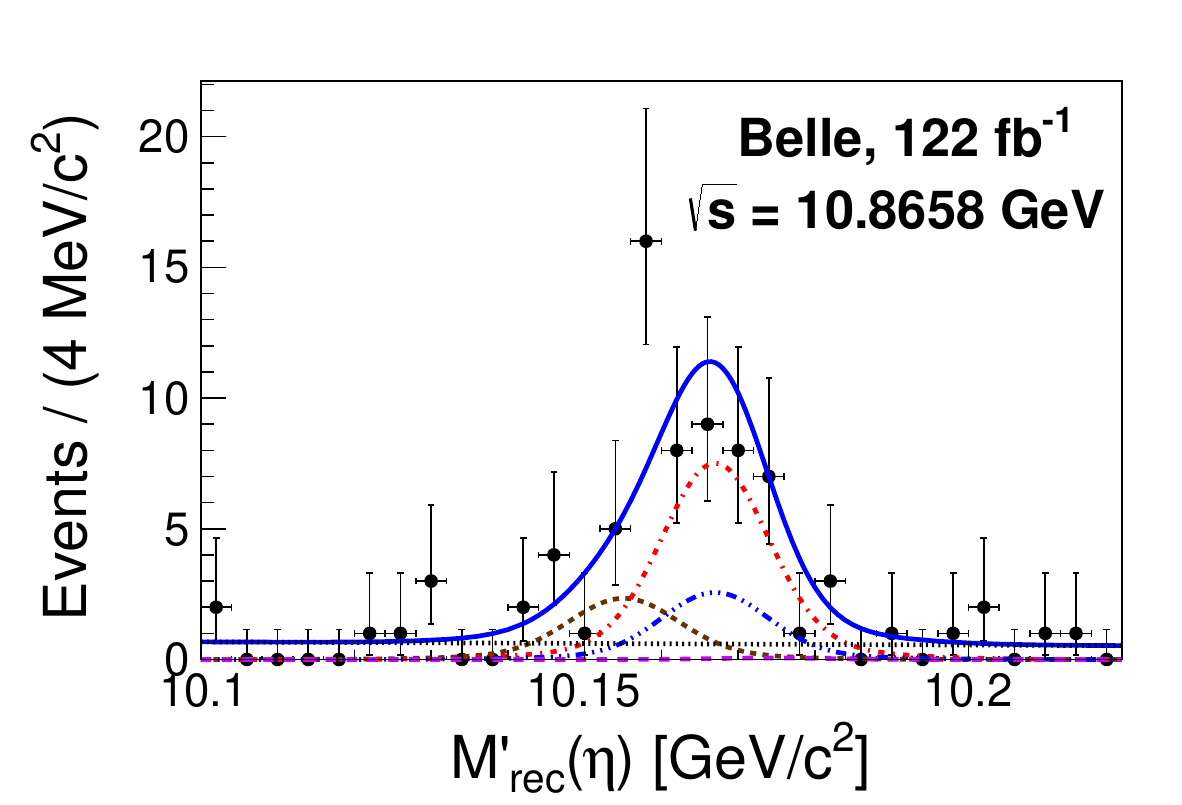}
\put(-23, 55){\small \bf (c)}

\vspace{0.2cm}
\includegraphics[width=4.9cm]{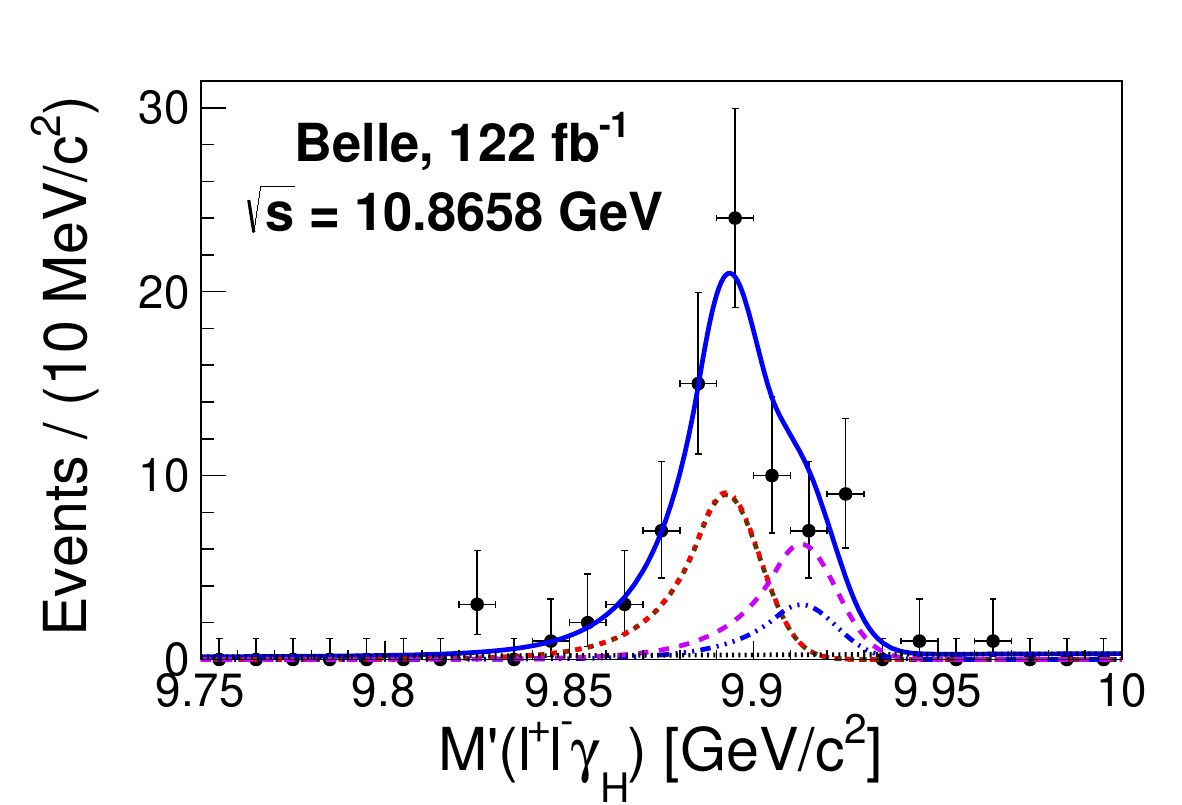}
\put(-23, 75){\small \bf (d)}
\includegraphics[width=4.9cm]{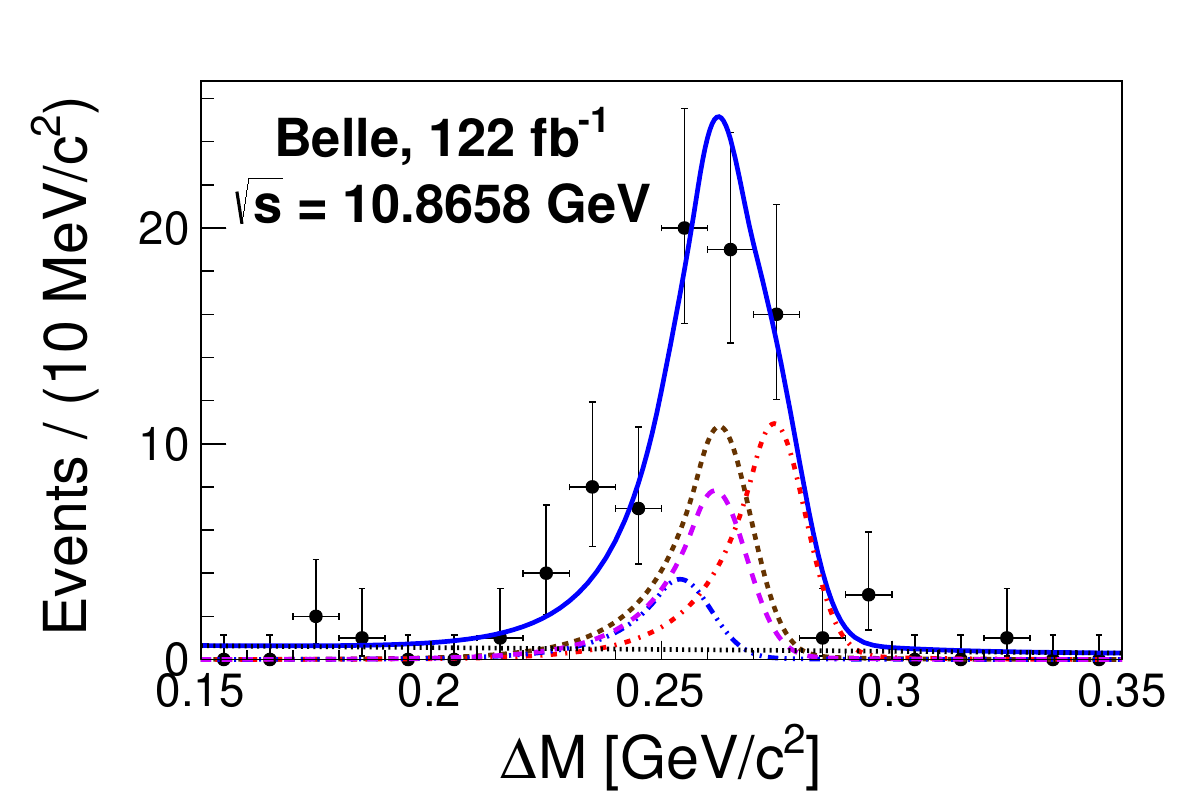}
\put(-23, 75){\small \bf (e)}
\includegraphics[width=4.9cm]{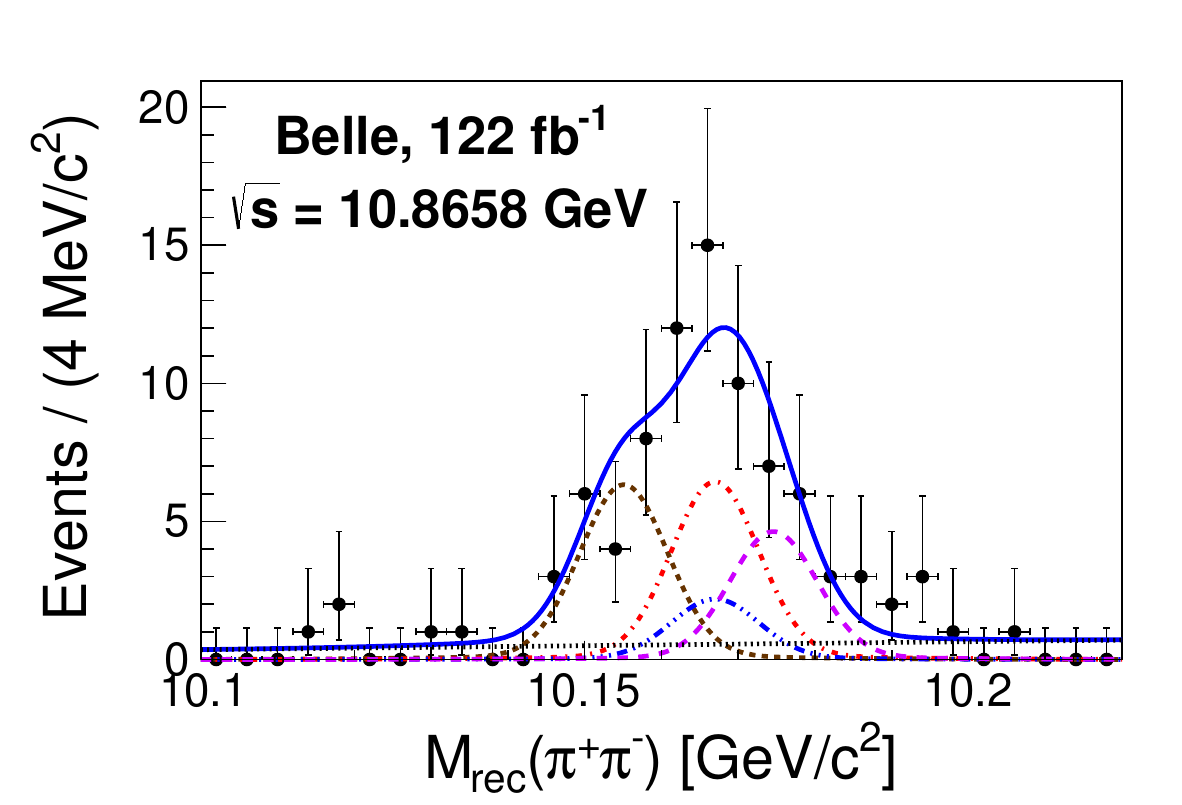}
\put(-23, 75){\small \bf (f)}
\caption{Distributions of (a) $M^{\prime}(\ell^+\ell^-\gamma_H)$, (b) $\Delta M$, and (c) $M^{\prime}_{\rm rec}(\eta)$ in $e^+e^-\to\Upsilon_J(1D)\eta$; and (d) $M^{\prime}(\ell^+\ell^-\gamma_H)$, (e) $\Delta M$, and (f) $M_{\rm rec}(\pi^+\pi^-)$ in $e^+e^-\to\Upsilon_J(1D)\pi^+\pi^-$ in data at $\sqrt{s}$ = 10.8658 GeV with 3D fit results overlaid.}
\label{3Dfit}
\end{figure*}

Fit projections for the signal region are shown in Fig.~\ref{3Dfit}.
The signal yields are listed in Table~\ref{table1}.
The statistical significances of the signals are
calculated from the difference of the
logarithmic likelihoods~\cite{Wilks}, $-2\ln(\mathcal{L}_0/\mathcal{L}_{\rm max})$,
where $\mathcal{L}_0$ and $\mathcal{L}_{\rm max}$ are the maximized likelihoods without
and with a signal component, respectively, taking into
account the difference in the number of degrees of freedom
($\Delta {\rm ndf}$ = 4 for the $\Upsilon_2(1D)$ and 3 for the $\Upsilon_{1,3}(1D)$).
The statistical significances are 5.1$\sigma$, $>$10$\sigma$, and $3.1\sigma$ for $\Upsilon_1(1D)$, $\Upsilon_2(1D)$, and $\Upsilon_3(1D)$, respectively, and the significances including systematic uncertainties are 4.8$\sigma$, $>$10$\sigma$, and $3.0\sigma$.
We varied the fit configurations over all additive systematic variations (discussed below), and conservatively took the lowest values as the significances with systematics included.

The mass of $\Upsilon_2(1D)$ is found to be $(10167.0\pm 1.0\pm 0.2)$ MeV/$c^2$. The first and second uncertainties are statistical and systematic, throughout this paper.
The mass difference from the world average mass of $(10163.7\pm1.4)$ MeV/$c^2$~\cite{PDG} is $1.8\sigma$.~The mass splittings $\Delta m_{12}=m(\Upsilon_2(1D))-m(\Upsilon_1(1D))$ and $\Delta m_{23}=m(\Upsilon_3(1D))-m(\Upsilon_2(1D))$ are $(11.8\pm1.5\pm0.4)$ MeV/$c^2$ and $(7.6\pm2.4\pm0.6)$ MeV/$c^2$, respectively.
The branching fraction ratio $\BR(\Upsilon_2(1D)\to\chi_{b2}\gamma)/\BR(\Upsilon_2(1D)\to\chi_{b1}\gamma)$ is $0.66\pm0.27\pm0.15$, which is consistent with the prediction of  $\approx$0.3 in Refs.~\cite{074027,054034,074002,915}.~The distribution of $M(\pi^+\pi^-)$ in $e^+e^-\to\Upsilon_J(1D)\pi^+\pi^-$ is shown in the Supplemental Material~\cite{SM}.

For each $\Upsilon_J(1D)$ state and $\Upsilon_J(1D)$ radiative decay, we calculate the product of the
Born cross section for $e^+e^-\to\Upsilon_J(1D)\eta$ and the branching fraction for $\Upsilon_J(1D)\to\chibJ\gamma$, using

\vspace{-0.6cm}
\begin{equation}\label{eq.1}
\begin{split}
\sigma^{(1)}_{\rm Born}=\frac{N^{\rm sig}\,|1-\Pi|^2}{{\cal L}\,\sum^{2}_{i=1}\varepsilon_i\BR^{\rm int}_i\,(1+\delta_{\rm ISR})},
\end{split}
\end{equation}where the index $i$ runs over channels $\eta\to\gamma\gamma$ and $\eta\to\pi^+\pi^-\pi^0$;
${\cal L}$ is the integrated luminosity of the data sample;
$\varepsilon_i$ is the reconstruction efficiency;
$\BR^{\rm int}_i$ is the product of the branching fractions of all intermediate states.
The $(1+\delta_{\rm ISR})$ is the radiative correction factor~\cite{466,260,2605} and the numerical values are listed in the Supplemental Material~\cite{SM}.
In calculating the radiative correction factor, the energy dependency of the Born cross section is described according to the measurement in this work (discussed below).
The $|1-\Pi|^2$ = 0.93 at each energy point is the vacuum polarization factor~\cite{083001,585}.
We calculate $\sigma^{(2)}_{\rm Born}$ for $e^+e^-\to\Upsilon_{J}(1D)\pi^+\pi^-$ using a similar equation.
The values of $\sigma^{(1)}_{\rm Born}$ and $\sigma^{(2)}_{\rm Born}$ are listed in Table~\ref{table1}.

Assuming the observed signals are all from $\Upsilon(10860)$ decays, we determine the product of branching fractions of 
$\Upsilon(10860)\to\Upsilon_J(1D)\eta$ and $\Upsilon_J(1D)\to\chibJ\gamma$ with the following formula

\vspace{-0.6cm}
\begin{equation}
\begin{split}
\BR^{(1)}_{\Upsilon(10860)}=\frac{N^{\rm sig}}{N_{\Upsilon(10860)}\,\sum^{2}_{i=1}\varepsilon_i\BR^{\rm int}_i}.
\end{split}
\end{equation}
The total number of $\Upsilon(10860)$ events is calculated as $N_{\Upsilon(10860)}={\cal L}\,\sigma(e^+e^-\to b\bar b)$ = $(4.2\pm0.2)\times10^7$, where 
$\sigma(e^+e^-\to b\bar b) = 0.34\pm0.02~{\rm nb}$~\cite{031101}.
We calculate the $\BR^{(2)}_{\Upsilon(10860)}$ for $e^+e^-\to\Upsilon_{J}(1D)\pi^+\pi^-$ using a similar equation.~The values of $\BR^{(1)}_{\Upsilon(10860)}$ and $\BR^{(2)}_{\Upsilon(10860)}$ are listed in Table~\ref{table1}.

\begin{table*}[htbp]
\renewcommand\arraystretch{1.2}
\setlength{\tabcolsep}{2pt}
\centering
\footnotesize
\caption{Results for $e^+e^-\to\Upsilon_J(1D)\eta$ and $e^+e^-\to\Upsilon_J(1D)\pi^+\pi^-$ at $\sqrt{s}$ = 10865.8 MeV. 
All symbols are explained in the text.
}\label{table1}
\vspace{0.2cm}
\begin{tabular}{cccccc}
\hline\hline
Channel&$\varepsilon_{\eta\to\gamma\gamma,\eta\to\pi^+\pi^-\pi^0}$&$N^{\rm sig}$&~~~$\BR^{(1)}_{\Upsilon(10860)}$($\times10^{-3}$)~~~&~~~$\sigma^{(1)}_{\rm Born}$(pb)~~~&\\\hline
$e^+e^-\to\Upsilon_1(1D)\eta,\,\Upsilon_1(1D)\to\chi_{b1}\gamma$&0.102,\,0.041&$12.2\pm6.3$  &$0.34\pm0.18\pm0.06$&$0.17\pm0.09\pm0.03$\\
$e^+e^-\to\Upsilon_2(1D)\eta,\,\Upsilon_2(1D)\to\chi_{b1}\gamma$&0.103,\,0.042&$37.8\pm8.0$  &$1.05\pm0.22\pm0.12$&$0.52\pm0.11\pm0.06$\\
$e^+e^-\to\Upsilon_2(1D)\eta,\,\Upsilon_2(1D)\to\chi_{b2}\gamma$&0.103,\,0.042&$12.9\pm6.0$  &$0.70\pm0.33\pm0.18$&$0.34\pm0.17\pm0.09$\\
$e^+e^-\to\Upsilon_3(1D)\eta,\,\Upsilon_3(1D)\to\chi_{b2}\gamma$&0.104,\,0.043&$0.3\pm2.8$    &$0.02\pm0.15\pm0.02$&$0.01\pm0.07\pm0.01$\\
\hline
Channel&$\varepsilon_{\pi^+\pi^-}$&$N^{\rm sig}$&$\BR^{(2)}_{\Upsilon(10860)}$($\times10^{-3}$)&$\sigma^{(2)}_{\rm Born}$(pb)\\\hline
$e^+e^-\to\Upsilon_1(1D)\pi^+\pi^-,\,\Upsilon_1(1D)\to\chi_{b1}\gamma$&0.177&$24.7\pm6.5$       &$0.19\pm0.05\pm0.03$&$0.09\pm0.02\pm0.01$\\
$e^+e^-\to\Upsilon_2(1D)\pi^+\pi^-,\,\Upsilon_2(1D)\to\chi_{b1}\gamma$&0.180&$25.0\pm6.7$       &$0.19\pm0.06\pm0.03$&$0.09\pm0.03\pm0.01$\\
$e^+e^-\to\Upsilon_2(1D)\pi^+\pi^-,\,\Upsilon_2(1D)\to\chi_{b2}\gamma$&0.181&$8.5\pm4.2$         &$0.13\pm0.06\pm0.04$&$0.06\pm0.03\pm0.02$\\
$e^+e^-\to\Upsilon_3(1D)\pi^+\pi^-,\,\Upsilon_3(1D)\to\chi_{b2}\gamma$&0.182&$17.9\pm6.7$       &$0.27\pm0.09\pm0.03$&$0.13\pm0.05\pm0.02$\\
\hline\hline
\end{tabular}
\end{table*}

For the other energy points, the $M^{\prime}_{\rm rec}(\eta)$ and $M_{\rm rec}(\pi^+\pi^-)$ distributions after applying the signal region requirements 
of $9.83<M^{\prime}(\ell^+\ell^-\gamma_H)<9.94$ GeV/$c^2$ and $0.20<\Delta M<0.29$ GeV/$c^2$ (about 98\% of signal candidates are retained) are shown in the Supplemental Material~\cite{SM}.
These distributions are very sparse and, instead of fitting, we use event counting.
We combine the $\Upsilon_1(1D)$, $\Upsilon_2(1D)$, and $\Upsilon_3(1D)$ states due to significant cross-feed among them.

We count the number of events, $N^{\rm obs}$, in the $\Upsilon_J(1D)$ signal region of (10.13 -- 10.19) GeV/$c^2$. The signal yield is defined as $N^{\rm sig}={\rm max}(0,N^{\rm obs}-N^{\rm bg})$. To determine the number of background events, $N^{\rm bg}$, we perform a 3D fit to the data sample that combines all energies, including 10.866 GeV. 
At all energy points except $\sqrt{s}$ = 10.866 GeV, the total numbers of $N^{\rm obs}$ and $N^{\rm bg}$ in $e^+e^-\to\Upsilon_J(1D)\eta$ ($e^+e^-\to\Upsilon_J(1D)\pi^+\pi^-$) are 3.0 (8.0) and 1.7 (1.4), respectively.
The background is assumed to be of non-resonant origin, and for each energy, the $N^{\rm bg}$ value is assigned based on the corresponding luminosity and the background yield in the combined sample.
The statistical uncertainties on $N^{\rm sig}$ are assigned based on 68.3\% confidence intervals provided by the Poissonian
limit estimator (POLE) program~\cite{012002};
these are equivalent to the unified approach in Ref.~\cite{3873}.

The product of the Born cross section for $e^+e^-\to\Upsilon_{J}(1D)\eta$, the branching fraction for $\Upsilon_J(1D)\to\chibJ\gamma$, and the branching fraction for $\chibJ\to\Upsilon(1S)\gamma$ ($\sigma^{(3)}_{\rm Born}$) is calculated using Eq.~(\ref{eq.1}), where $\BR^{\rm int}_i$ = $\BR(\eta\to\gamma\gamma)\,\BR(\Upsilon(1S)\to\ell^+\ell^-)$ or $\BR(\eta\to\pi^+\pi^-\pi^0)\,\BR(\pi^0\to\gamma\gamma)\,\BR(\Upsilon(1S)\to\ell^+\ell^-)$.
We calculate the corresponding $\sigma^{(4)}_{\rm Born}$ for $e^+e^-\to\Upsilon_{J}(1D)\pi^+\pi^-$ using a similar equation.
At $\sqrt{s}=10.866$ GeV, we use the same counting method to determine $\sigma^{(3)}_{\rm Born}$ and $\sigma^{(4)}_{\rm Born}$.
The upper limits are calculated directly using POLE, with the same configuration as in Ref.~\cite{2510.25461}. 
The signal yields, Born cross sections, and upper limits on the Born cross sections are listed in the Supplemental Material~\cite{SM}.

The systematic uncertainties in the measurements of cross sections for $e^+e^-\to\Upsilon_J(1D)\eta$ and $e^+e^-\to\Upsilon_J(1D)\pi^+\pi^-$ 
include contributions from the photon energy calibration, fit model, reconstruction efficiency, radiative correction factor,
angular distributions, beam-energy calibration, trigger simulation, integrated luminosity, and branching fractions of intermediate states. The additive systematic uncertainties are those from the photon energy calibration and fit model.
The other sources of systematic uncertainties are multiplicative.

We study the additive systematic uncertainties from the photon energy calibration and fitting procedure as follows.
The resulting uncertainties of the peak positions are 1.1 MeV/$c^2$ in both the $M^{\prime}(\ell^+\ell^-\gamma_H)$ and $\Delta M$ distributions; the uncertainties in the resolutions are 0.9 and 1.1 MeV/$c^2$ in the $M^{\prime}(\ell^+\ell^-\gamma_H)$ and $\Delta M$ distributions~\cite{232002}.
For the results obtained with the fitting method, we change the signal peak positions and resolutions by $\pm1\sigma$.
We increase the order of the polynomial describing the background by one and change the fit interval. 
We vary the ratio $N^{\rm sig}(\eta\to\gamma\gamma)/N^{\rm sig}(\eta\to\pi^+\pi^-\pi^0)$ from 4.2 to 4.0 and 4.4 to account for the uncertainties of the efficiencies and the $\eta$ decay branching fractions, and take the difference in the yield as a systematic uncertainty.
For each fitted quantity, we varied all the above configurations simultaneously and conservatively consider the largest deviation from the nominal result as the additive systematic uncertainty for the corresponding energy point. 

The contributions to the multiplicative systematic uncertainty are given in Table~\ref{sys1}. Detection efficiency uncertainties include momentum-dependent tracking uncertainties (1.0\% per pion and 0.4\% per lepton, derived from $D^{*+}\to D^0(\to K_S^0\pi^+\pi^-)\pi^+$), pion identiﬁcation (1.1\% per pion, derived from $D^{*+}\to D^0(\to K^-\pi^+)\pi^{+}$), lepton identification (1.6\% per electron and 1.2\% per muon, derived from $\gamma\gamma \to \ell^+\ell^-$ ($\ell$ = e, $\mu$)), photon reconstruction (2.0\% per photon, derived from $e^+e^- \to \gamma e^+e^-$), and $\pi^0$ reconstruction (2.3\% per $\pi^0$, derived from $\tau^- \to \pi^-\pi^0\nu_\tau$). 

When the uniform angular distributions are changed to $1\pm{\rm cos}^2\theta_{\eta}$ and $1\pm{\rm cos}^2\theta_{\pi^+\pi^-}$, the maximal deviations in the efficiency are 1.7\% and 4.8\% for $e^+e^-\to\Upsilon_J(1D)\eta$ and $e^+e^-\to\Upsilon_J(1D)\pi^+\pi^-$~\cite{109354}, respectively. 
We assume a flat distribution of $\theta_{\pi^+}$ in the $\pi^+\pi^-$ system in $e^+e^-\to\Upsilon_J(1D)\pi^+\pi^-$. 
We change the uniform distribution to $1\pm{\rm cos}^2\theta_{\pi^+}$, and take the maximal deviation of 4.5\% in the efficiency as the systematic uncertainty due to angular distribution. The trigger uncertainty is neglected because the trigger efficiencies for $e^+e^-\to\Upsilon_J(1D)\eta$ and $e^+e^-\to\Upsilon_J(1D)\pi^+\pi^-$ are 99.3\% and 100.0\%.

The uncertainties in the center of mass energies are about 1 MeV.
We change the collision energies in the 5C kinematic fit by $\pm1\sigma$ and find the deviations in the signal yields are less than 1.0\%, which is negligible.
In calculating the radiative correction factor, the measured energy dependence of the cross sections is used.~We change all of the parameters in the fit function by $\pm1\sigma$ according to the fitted results from the distributions of $\sigma(e^+e^-\to\Upsilon_J(1D)\eta)$ and $\sigma(e^+e^-\to\Upsilon_J(1D)\pi^+\pi^-)$ as a function of $e^+e^-$ c.m.\ energy, and take the maximum difference on the radiative correction factor as the resulting uncertainty. 
Belle measures luminosity at 1.4\% precision using wide angle Bhabha events.
The branching fractions of all intermediate decays are taken from Ref.~\cite{PDG}.
We add all
the multiplicative systematic uncertainties in quadrature to obtain the
final multiplicative systematic uncertainty, as listed in Table~\ref{sys1}. 

\begin{table}
\centering
\small
\caption{The multiplicative systematic uncertainties (\%) in the measurements of cross
sections for $e^+e^-\to\Upsilon_J(1D)\eta$ and $e^+e^-\to\Upsilon_J(1D)\pi^+\pi^-$.}\label{sys1}
\begin{tabular}{lc c c}
\hline\hline
Source & $\Upsilon_J(1D)\eta$ & $\Upsilon_J(1D)\pi^+\pi^-$ \\\hline
Efficiency & 8.6 & 6.2 \\
Angular distributions & 1.7 & 6.6 \\
Radiative correction factor & 2.0 & 6.3 \\
Luminosity & 1.4 & 1.4 \\
Branching fractions & 6.7 & 6.6 \\\hline
Total & 11.3 & 12.9 \\
\hline\hline
\end{tabular}
\end{table}

We perform all possible combinations of the above variations of the fit, and conservatively consider the largest deviation of fit results as the systematic
uncertainty in the measurement of $\Upsilon_J(1D)$ masses.
The uncertainty due to the fit interval is dominant.

\begin{figure}[htbp]
\centering
\includegraphics[width=4.4cm]{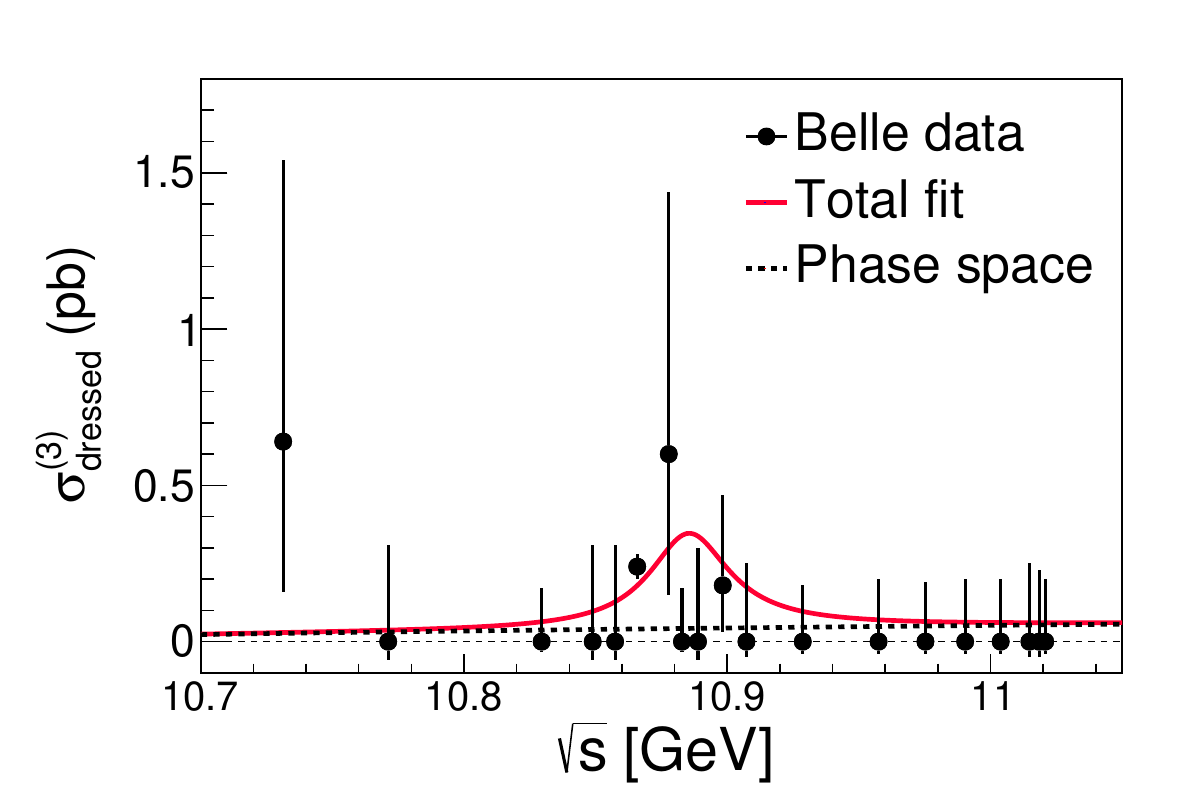}
\put(-88, 65){\small \bf (a)}
\includegraphics[width=4.4cm]{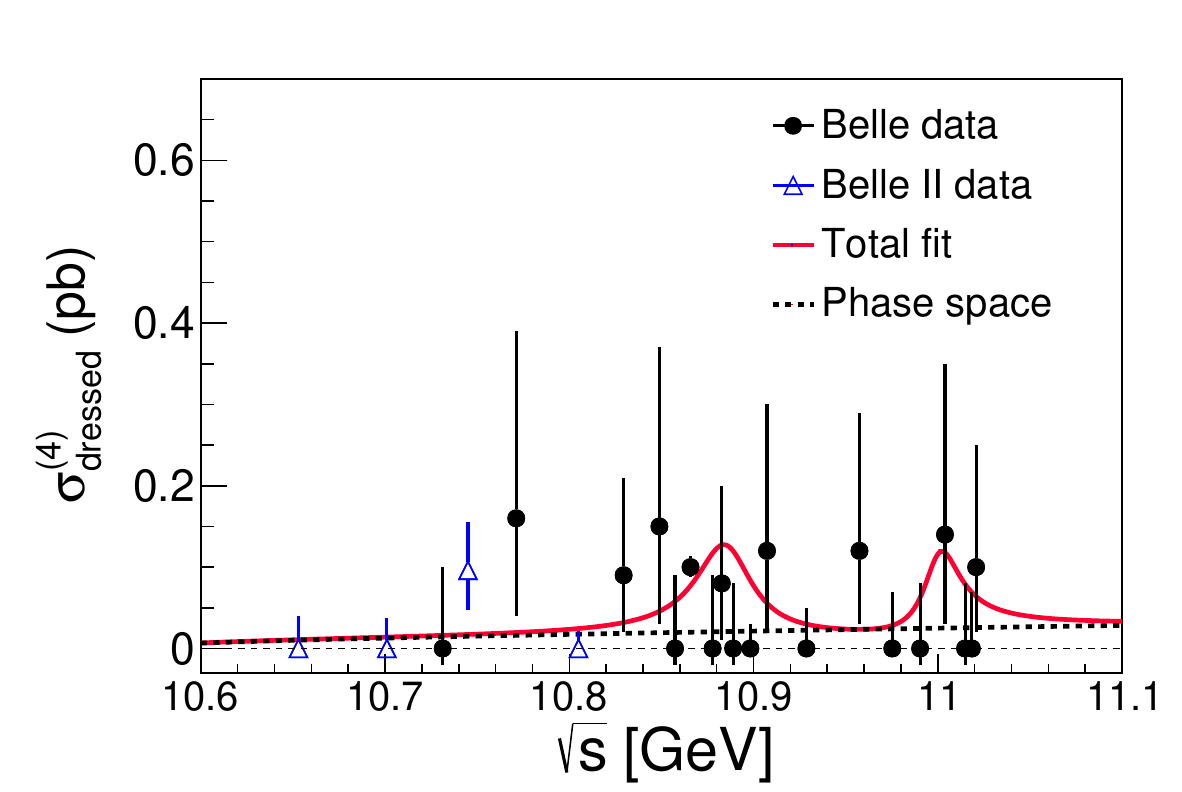}
\put(-98, 65){\small \bf (b)}
\caption{Energy dependences of (a) $\sigma^{(3)}_{\rm dressed}$ and (b) $\sigma^{(4)}_{\rm dressed}$ for $e^+e^-\to\Upsilon_J(1D)\eta$ and $e^+e^-\to\Upsilon_J(1D)\pi^+\pi^-$.
Filled circles show the measurements at Belle, and triangles show the measurements at Belle II~\cite{2602.19807}.
Error bars represent the total uncorrelated uncertainties (statistical and additive uncertainties). Curves show the fit results to the Belle and Belle II data points.}
\label{dependence}
\end{figure}

The dressed cross sections ($\sigma_{\rm dressed}$) for $e^+e^-\to\Upsilon_J(1D)\eta$ or $e^+e^-\to\Upsilon_J(1D)\pi^+\pi^-$ as a function of c.m.\ energy are shown in Fig.~\ref{dependence}. The $\sigma_{\rm dressed}$ can be obtained by removing the vacuum polarization factor.
The triangles show the dressed cross sections for $e^+e^-\to\Upsilon_J(1D)\pi^+\pi^-$ at Belle II, which are obtained according to the recoil mass distributions of $\pi^+\pi^-$~\cite{2602.19807} (see the Supplemental Material~\cite{SM} for details).
The profile likelihoods for each energy point are given in the Supplemental Material~\cite{SM}.
The fit function to the Belle and Belle II data points is the sum of the phase-space and a coherent sum of the $\Upsilon(10860)$ and $\Upsilon(11020)$ Breit-Wigner amplitudes:

\vspace{-0.6cm}
\begin{equation} \label{eq:BW}
{\Bigl |}\sum _{i=1}^2\frac{\sqrt{12\pi\Gamma_{ee,i}\BR_{i}\Gamma_i}}{s-M_i^2-iM_i\Gamma_i}\sqrt{\frac{\Phi(\sqrt{s})}{\Phi(M_i)}}e^{i\phi}{\Bigr |}^2+c_{\rm NR}\Phi(\sqrt{s}),
\end{equation}where the
index $i$ runs over the $\Upsilon(10860)$ and $\Upsilon(11020)$ states; $M_i$, $\Gamma_i$, and $\Gamma_{ee,i}$ are the mass, total width, and electronic width of the $\Upsilon$ states; $\Phi$ is the phase-space factor; $\phi$ is the complex phase; and $c_{\rm NR}$ is a constant.
Only one solution is found from the fit since the interference is weak due to the large mass difference between the $\Upsilon(10860)$ and $\Upsilon(11020)$ and the insignificant signal of the $\Upsilon(11020)$. Their masses and widths are fixed to the world-average values~\cite{PDG}. 
The parameters $\Gamma_{ee,i}\BR_{i}$ and $\phi$ are floated.
The fitted results are
shown in Fig.~\ref{dependence} and summarized in Table~\ref{tab:dependence}.
Since the signal significances of fit outputs are less than $3\sigma$, the upper limits at 90\% C.L. are also given in Table~\ref{tab:dependence}.
The value of $\phi$ is $(-0.5\pm0.6)$~rad for $e^+e^-\to\Upsilon_J(1D)\pi^+\pi^-$.
Systematic uncertainties in the cross section energy dependence fit results include the
beam-energy calibration, the fit model, and the multiplicative systematic uncertainties of cross section measurements~\cite{2510.25461}.
For $e^+e^-\to\Upsilon_J(1D)\pi^+\pi^-$, the Belle and Belle II-specific correlated multiplicative uncertainties are taken into account separately~\cite{2510.25461}.
The uncertainties from the masses and widths of the $\Upsilon(10860)$ and $\Upsilon(11020)$ are dominant.

\begin{table}[htbp]
\centering
\scriptsize
\caption{
Fitted values for $\Gamma_{ee}\BR(\Upsilon(10860,11020)\to\Upsilon_J(1D)\eta)\BR_f$ and $\Gamma_{ee}\BR(\Upsilon(10860,11020)\to\Upsilon_J(1D)\pi^+\pi^-)\BR_f$ in eV, where $\BR_f=\BR(\Upsilon_J(1D)\to\chibJ\gamma)\BR(\chibJ\to\Upsilon(1S)\gamma)$. The upper limits at 90\% C.L. are shown in square brackets.
}\label{tab:dependence}
\vspace{0.2cm}
\begin{tabular}{cc}
\hline\hline
$\Gamma_{ee}\BR(\Upsilon(10860)\to\Upsilon_J(1D)\eta)\BR_f$ & $(0.091\pm0.039\pm0.016)$ [$<0.14$]\\
$\Gamma_{ee}\BR(\Upsilon(11020)\to\Upsilon_J(1D)\eta)\BR_f$ & $(0.000\pm0.018\pm0.005)$ [$<0.04$] \\
$\Gamma_{ee}\BR(\Upsilon(10860)\to\Upsilon_J(1D)\pi^+\pi^-)\BR_f$ & $(0.034\pm0.013\pm0.006)$ [$<0.06$] \\
$\Gamma_{ee}\BR(\Upsilon(11020)\to\Upsilon_J(1D)\pi^+\pi^-)\BR_f$ & $(0.020\pm0.016\pm0.005)$ [$<0.05$] \\
\hline\hline
\end{tabular}
\end{table}

\begin{figure}[htbp]
\centering
\hspace*{-0.7cm}
\includegraphics[width=7cm]{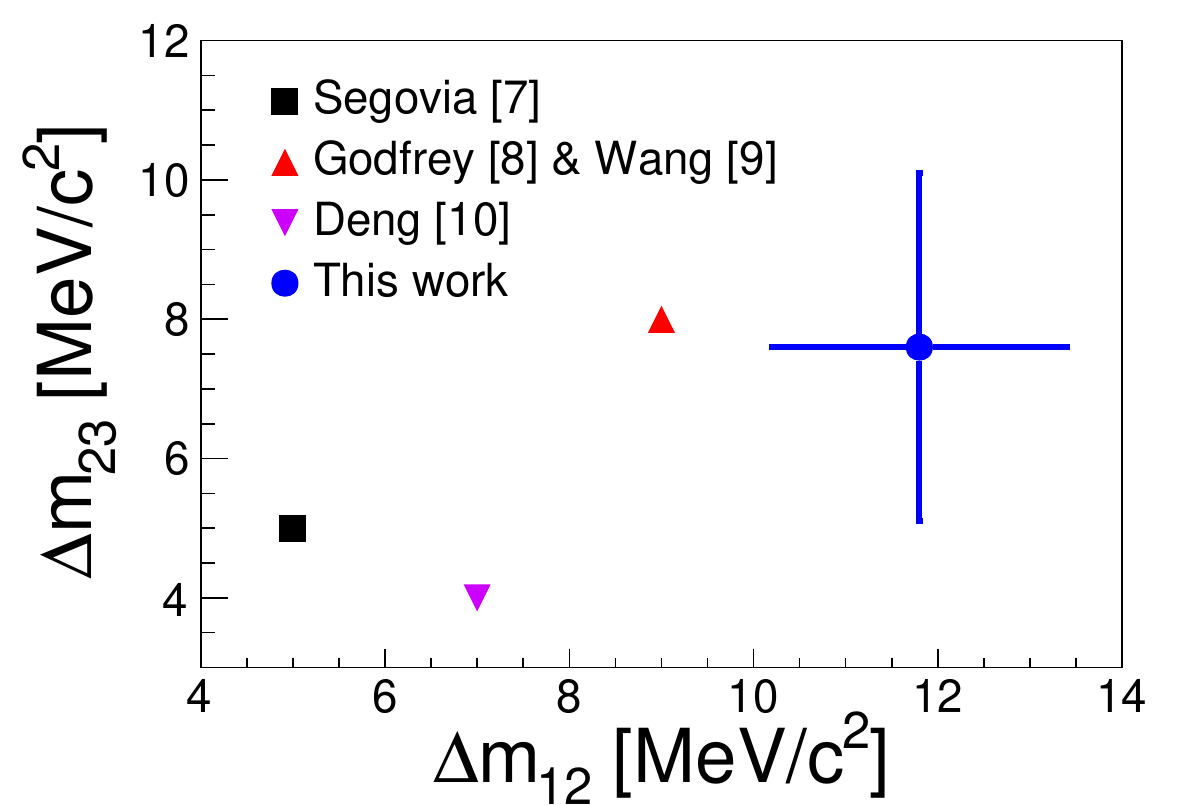}
\caption{The comparison of mass splittings $\Delta m_{12}$ and $\Delta m_{23}$ with different theoretical predictions.
The predictions are derived from nonrelativistic constituent quark model in Segovia~\cite{074027}, relativized quark model in Godfrey~\cite{054034}, the modified relativized quark model with a ﻿screening effect in Wang~\cite{915}, and screened potential model in Deng~\cite{074002}. The correlation coefficient between $\Delta m_{12}$ and $\Delta m_{23}$ is $-0.06$.
}
\label{mass}
\end{figure}

In summary, we measure the cross sections for $e^+e^-\to\Upsilon_J(1D)\eta$ and $e^+e^-\to\Upsilon_J(1D)\pi^+\pi^-$ using Belle 
data collected at c.m.\ energies between 10.73 and 11.02 GeV.
We report first evidence for the $\Upsilon_1(1D)$ and $\Upsilon_3(1D)$ states.
In the $\eta$ channel, $\Upsilon_3(1D)$ production is strongly suppressed and $\Upsilon_1(1D)$ production is slightly suppressed relative to $\Upsilon_2(1D)$, which is consistent with the prediction in Ref.~\cite{094039}. In the $\pi^+\pi^-$ channel, however, we find comparable production rates for the $\Upsilon_1(1D)$, $\Upsilon_2(1D)$, and $\Upsilon_3(1D)$ states.
The mass for $\Upsilon_2(1D)$ is $(10167.0\pm 1.0\pm 0.2)$ MeV/$c^2$.
We compare the mass splittings $\Delta m_{12}$ and $\Delta m_{23}$ with different theoretical predictions, as shown in Fig.~\ref{mass}.

%----------- Long version, for most papers ----------- 
This work, based on data collected using the Belle detector, which was
operated until June 2010, was supported by 
the Ministry of Education, Culture, Sports, Science, and
Technology (MEXT) of Japan, the Japan Society for the 
Promotion of Science (JSPS), and the Tau-Lepton Physics 
Research Center of Nagoya University; 
the Australian Research Council including grants
DP210101900, % Urquijo
DP210102831, % Sevior
DE220100462, % Hsu
LE210100098, % Infrastructure
LE230100085; % Infrastructure
Austrian Federal Ministry of Education, Science and Research (FWF) and
FWF Austrian Science Fund No.~P~31361-N36;
National Key R\&D Program of China under Contract No.~2022YFA1601903,
2024YFA1610503, and 2024YFA1610504;
National Natural Science Foundation of China and research grants
No.~12475076,
No.~11575017,
No.~11761141009, 
No.~11705209, 
No.~11975076, 
No.~12135005, 
No.~12150004, 
No.~12161141008, 
and
No.~12175041, 
and Shandong Provincial Natural Science Foundation Project ZR2022JQ02;
Fundamental Research Funds of China for the Central Universities under contract Nos. 2242025RCB0014 and RF1028623046;
the Czech Science Foundation Grant No. 22-18469S;
Horizon 2020 ERC Advanced Grant No.~884719 and ERC Starting Grant No.~947006 ``InterLeptons'' (European Union);
the Carl Zeiss Foundation, the Deutsche Forschungsgemeinschaft, the
Excellence Cluster Universe, and the VolkswagenStiftung;
the Department of Atomic Energy (Project Identification No. RTI 4002), the Department of Science and Technology of India,
and the UPES (India) SEED finding programs Nos. UPES/R\&D-SEED-INFRA/17052023/01 and UPES/R\&D-SOE/20062022/06; 
the Istituto Nazionale di Fisica Nucleare of Italy; 
National Research Foundation (NRF) of Korea Grants
No.~2021R1-F1A-1064008,
No.~2022R1-A2C-1003993,
No.~RS-2018-NR031074,
No.~RS-2021-NR060129,
No.~RS-2024-00354342
No.~RS-2025-02219521,
No.~RS-2026-25471491,
No.~RS-2026-25480677,
and
No.~RS-2026-25486791,
Radiation Science Research Institute,
Foreign Large-Size Research Facility Application Supporting project,
the Global Science Experimental Data Hub Center, the Korea Institute of Science and
Technology Information (K26L1M2C3) and KREONET/GLORIAD;
the Polish Ministry of Science and Higher Education and 
the National Science Center;
the Ministry of Science and Higher Education of the Russian Federation
and the HSE University Basic Research Program, Moscow; % from 15.04.2021
University of Tabuk research grants
S-1440-0321, S-0256-1438, and S-0280-1439 (Saudi Arabia);
the Slovenian Research Agency Grant Nos. J1-50010 and P1-0135;
Ikerbasque, Basque Foundation for Science, and the State Agency for Research
of the Spanish Ministry of Science and Innovation through Grant No. PID2022-136510NB-C33 (Spain);
the Swiss National Science Foundation; 
the Ministry of Education and the National Science and Technology Council of Taiwan;
and the United States Department of Energy and the National Science Foundation.
These acknowledgements are not to be interpreted as an endorsement of any
statement made by any of our institutes, funding agencies, governments, or
their representatives.
We thank the KEKB group for the excellent operation of the
accelerator; the KEK cryogenics group for the efficient
operation of the solenoid; and the KEK computer group and the Pacific Northwest National
Laboratory (PNNL) Environmental Molecular Sciences Laboratory (EMSL)
computing group for strong computing support; and the National
Institute of Informatics, and Science Information NETwork 6 (SINET6) for
valuable network support.

Numerical data corresponding to the results presented are available as HEPData.
The full Belle II data are not publicly available.
The collaboration will consider requests for access to the data that support this article.

\onecolumngrid
\newpage

\huge	
%\centering
{\bf \centering Supplemental Material}
\vspace{1.0cm}

\normalsize

\begin{itemize}

\item For other energy points (except at $\sqrt{s}$ = 10.866 GeV), the $M^{\prime}_{\rm rec}(\eta)$ and $M_{\rm rec}(\pi^+\pi^-)$ distributions for $e^+e^-\to\Upsilon_J(1D)\eta$  and $e^+e^-\to\Upsilon_J(1D)\pi^+\pi^-$ in data at Belle are shown in Figs.~\ref{scaneta} and~\ref{scanpipi}.

\begin{figure}[htbp]
\centering
\includegraphics[width=2.9cm]{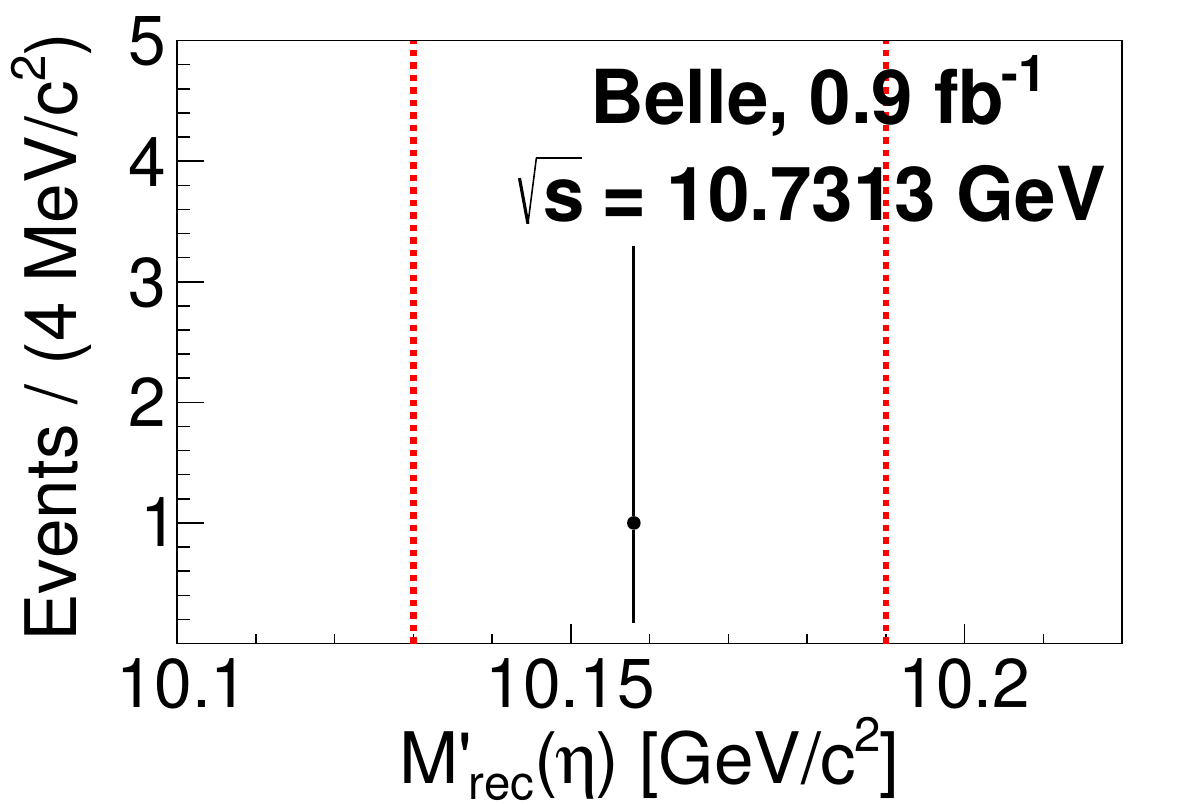}
\put(-85, 60){\large \bf Preliminary}
\includegraphics[width=2.9cm]{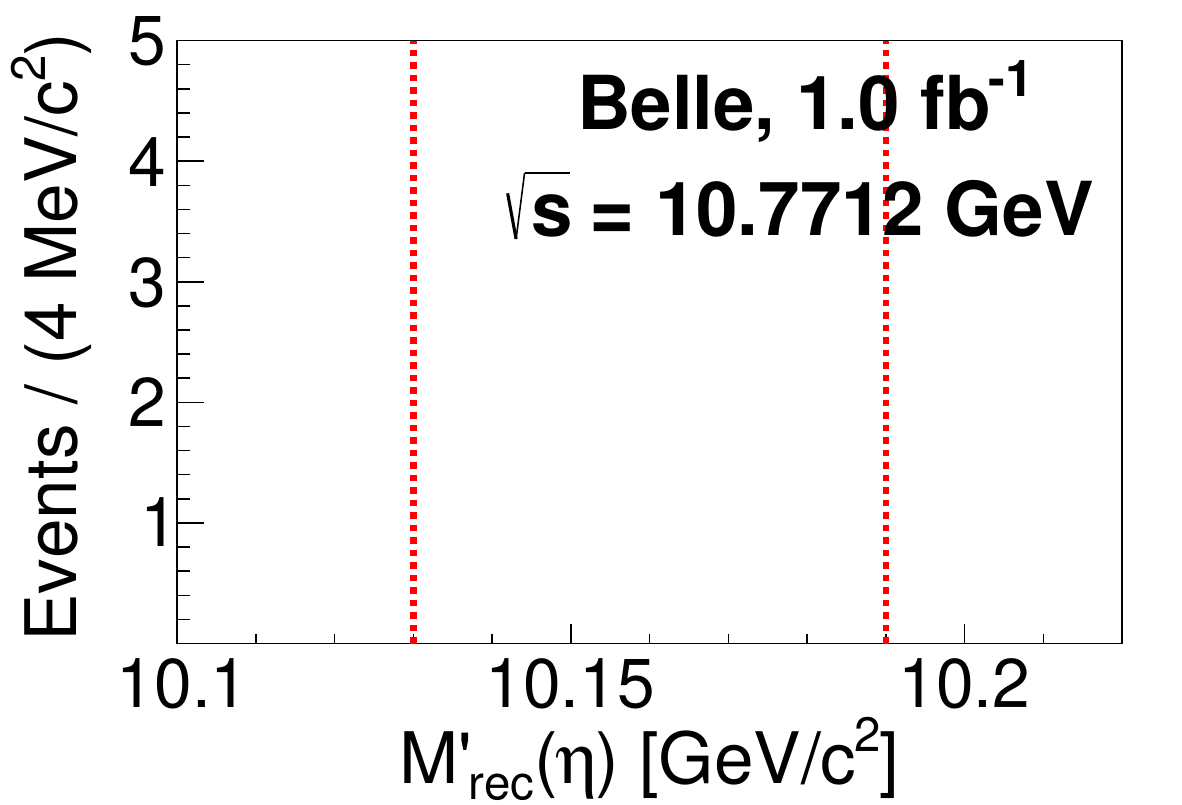}
\includegraphics[width=2.9cm]{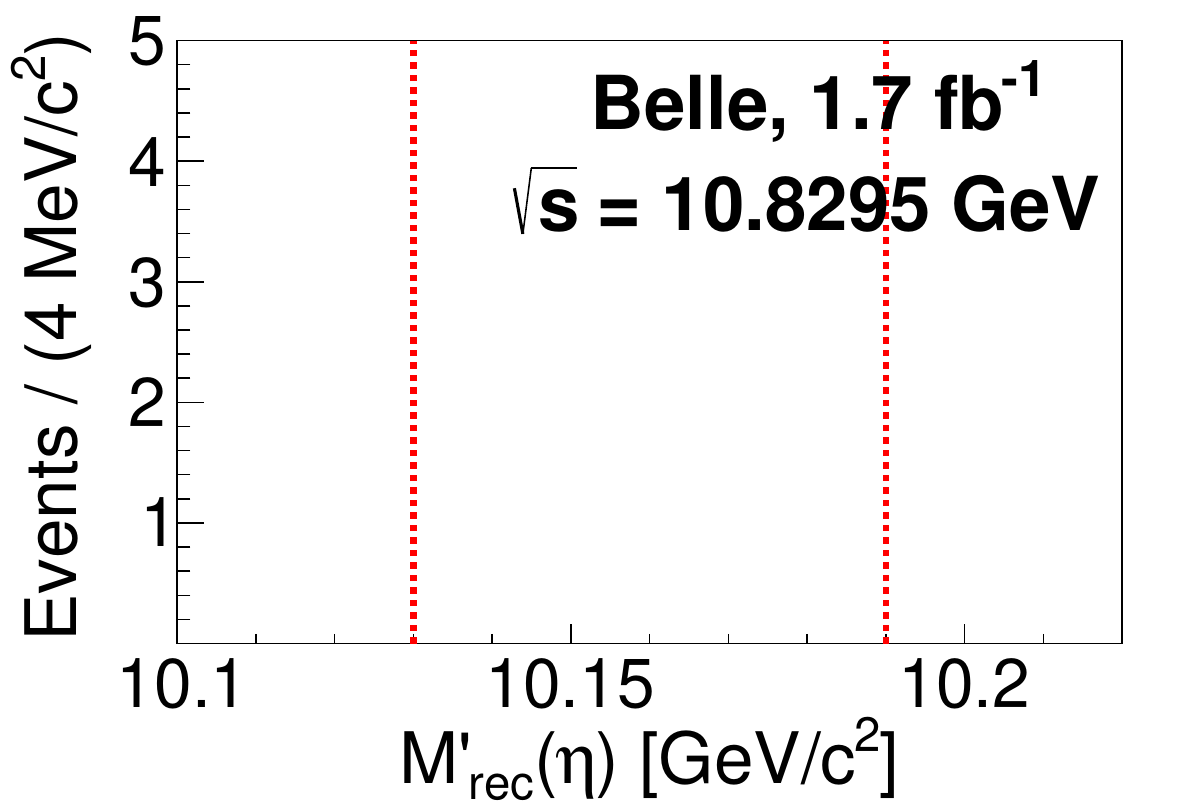}
\includegraphics[width=2.9cm]{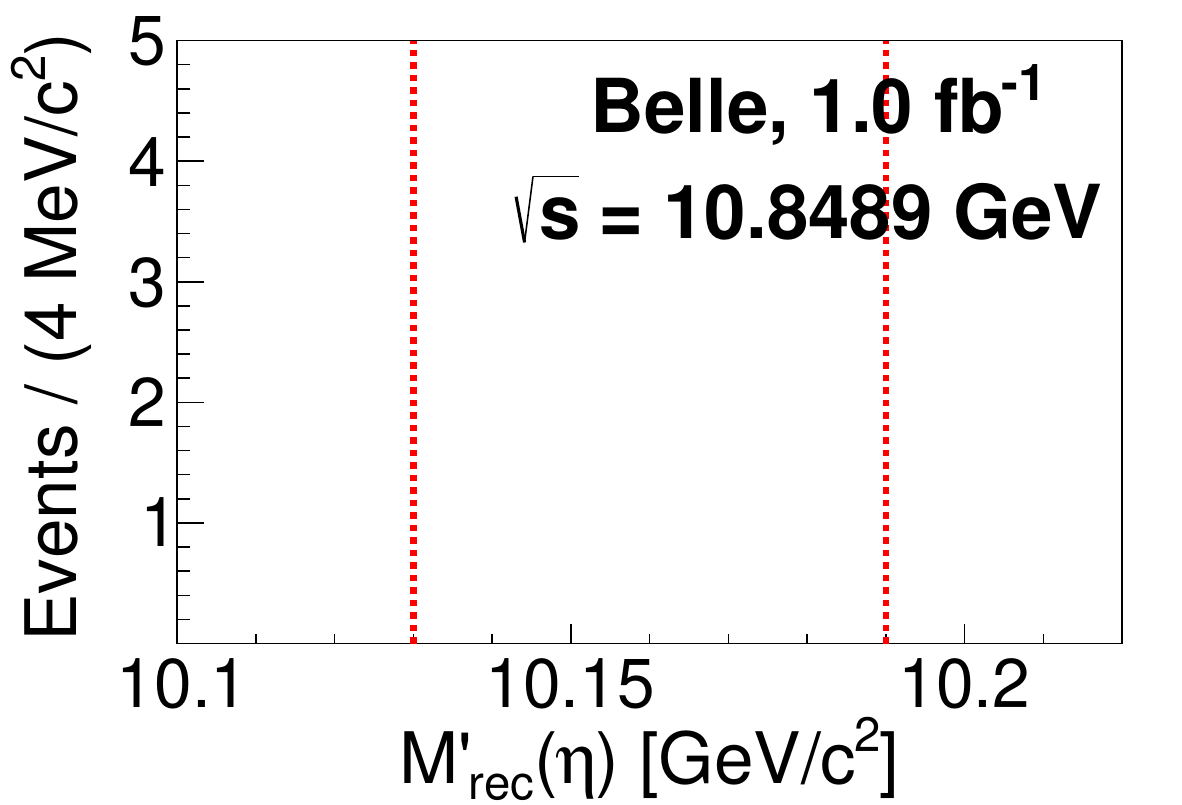}
\includegraphics[width=2.9cm]{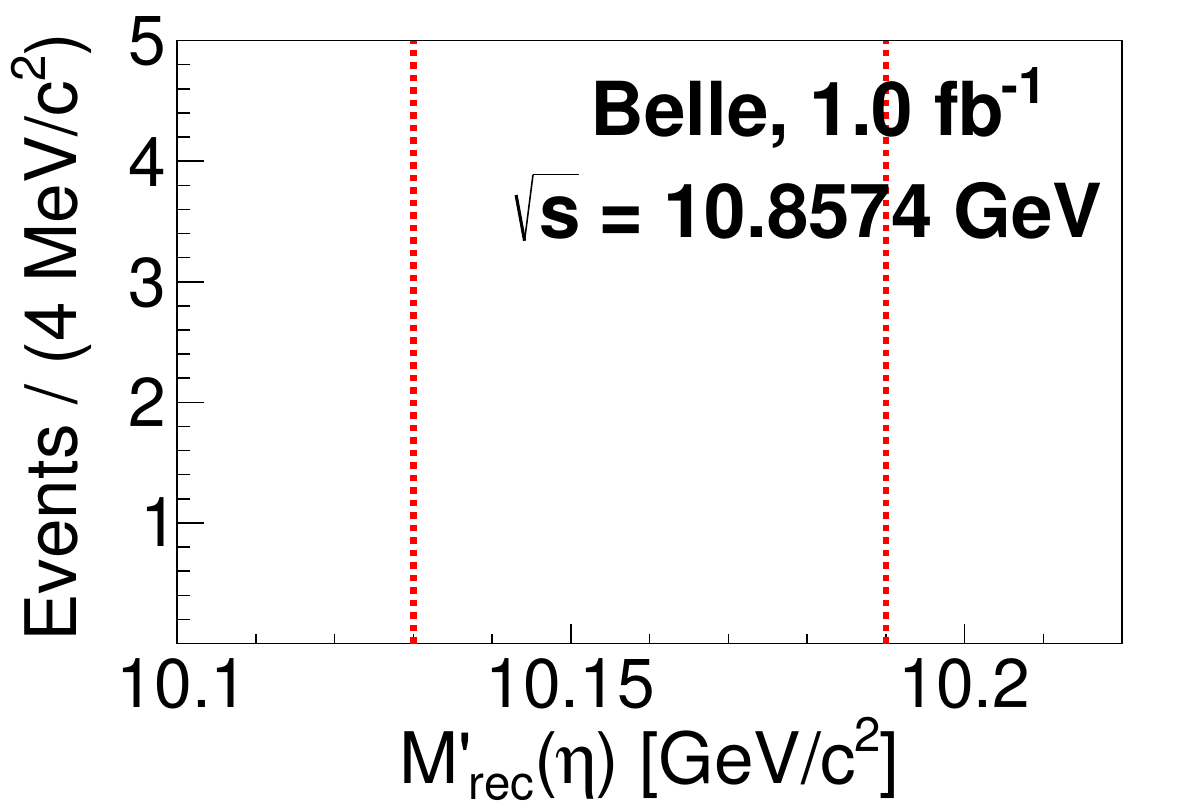}
\includegraphics[width=2.9cm]{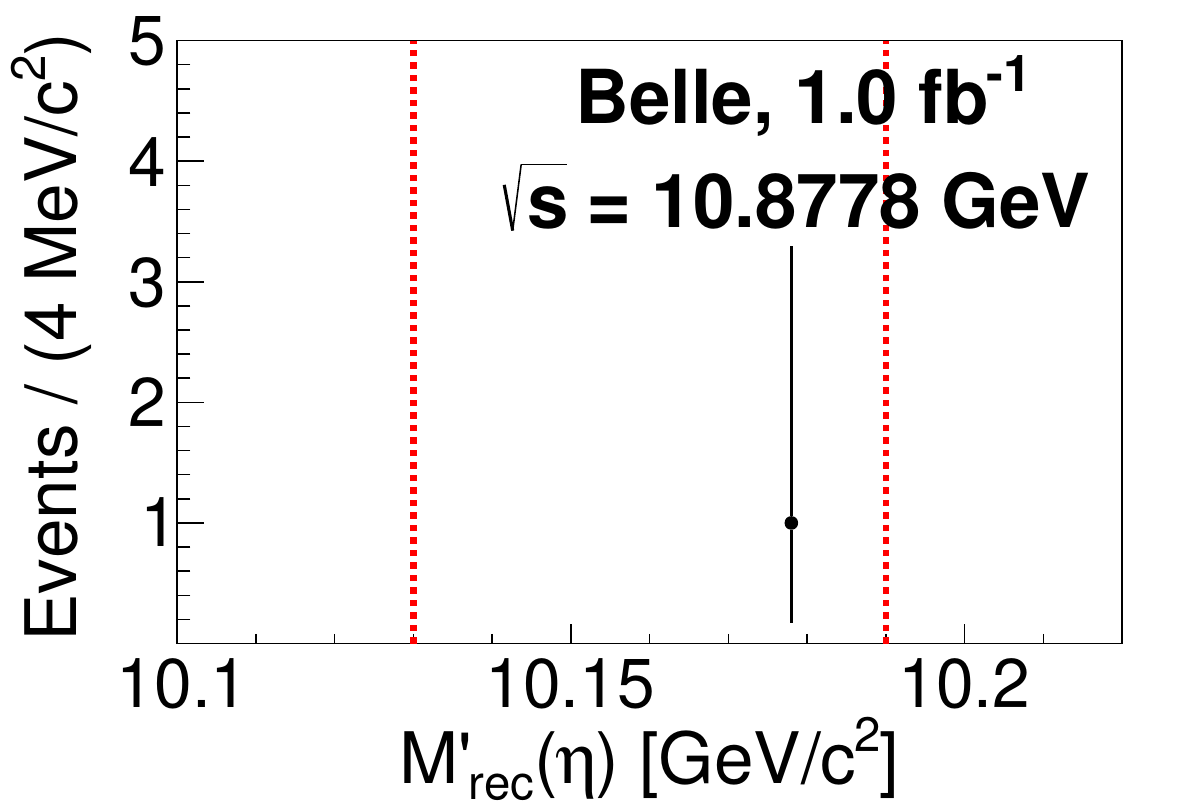}

\includegraphics[width=2.9cm]{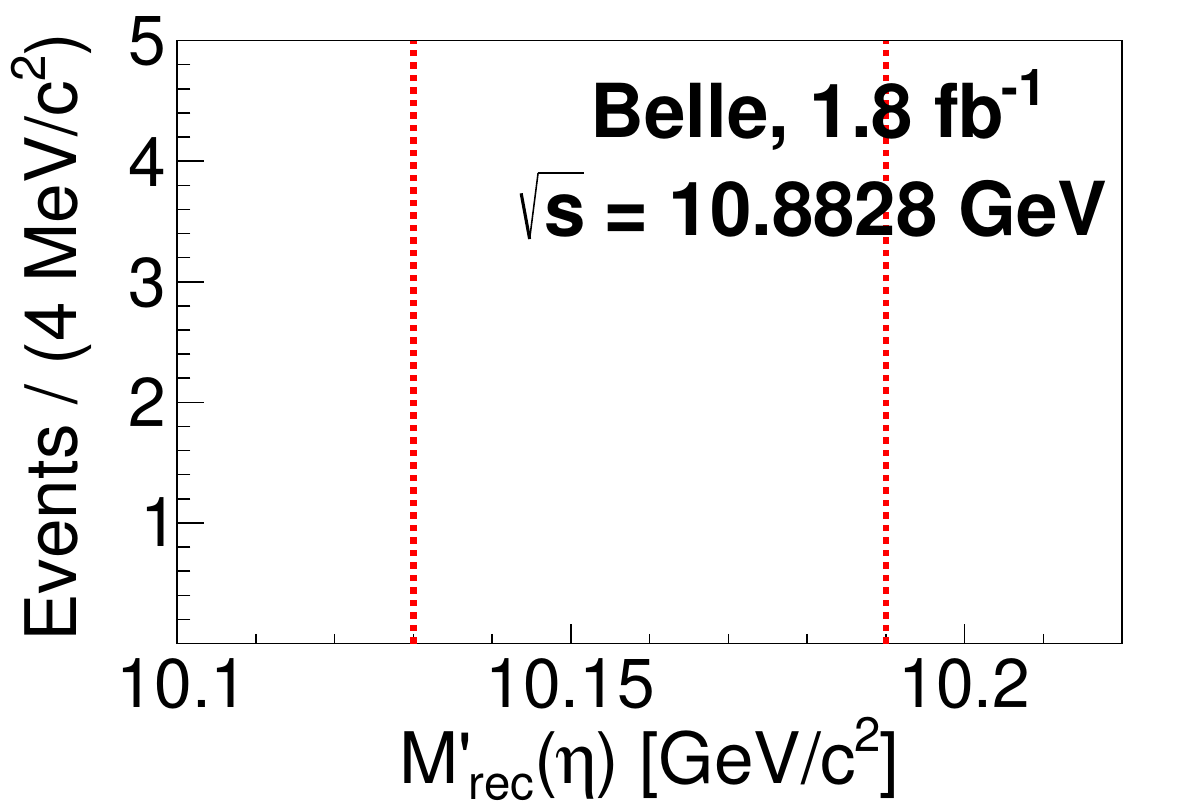}
\includegraphics[width=2.9cm]{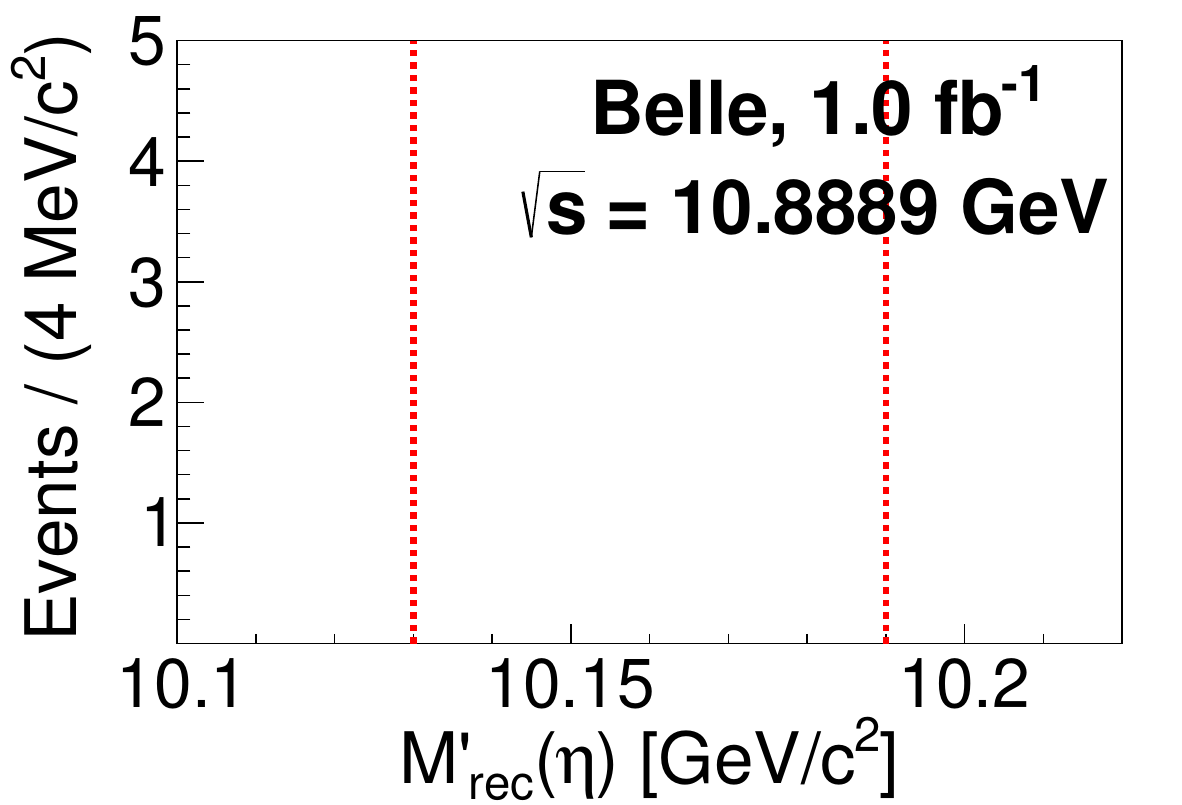}
\includegraphics[width=2.9cm]{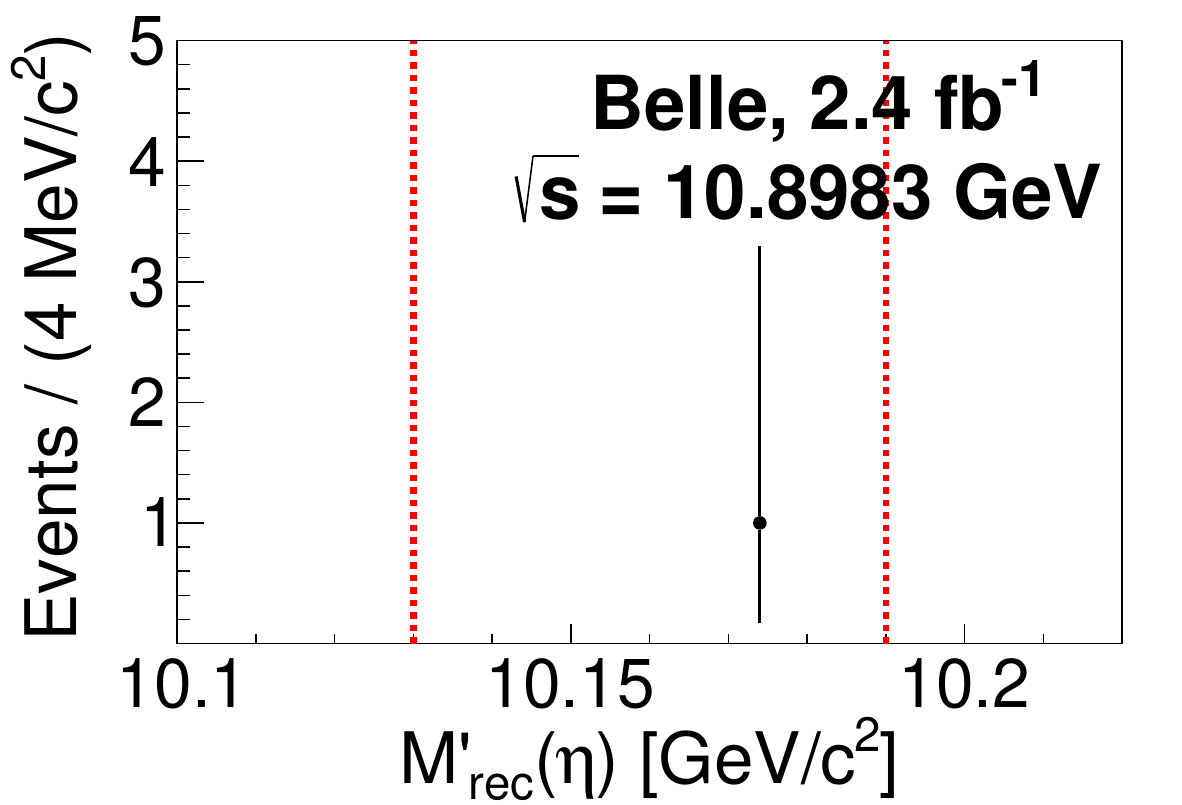}
\includegraphics[width=2.9cm]{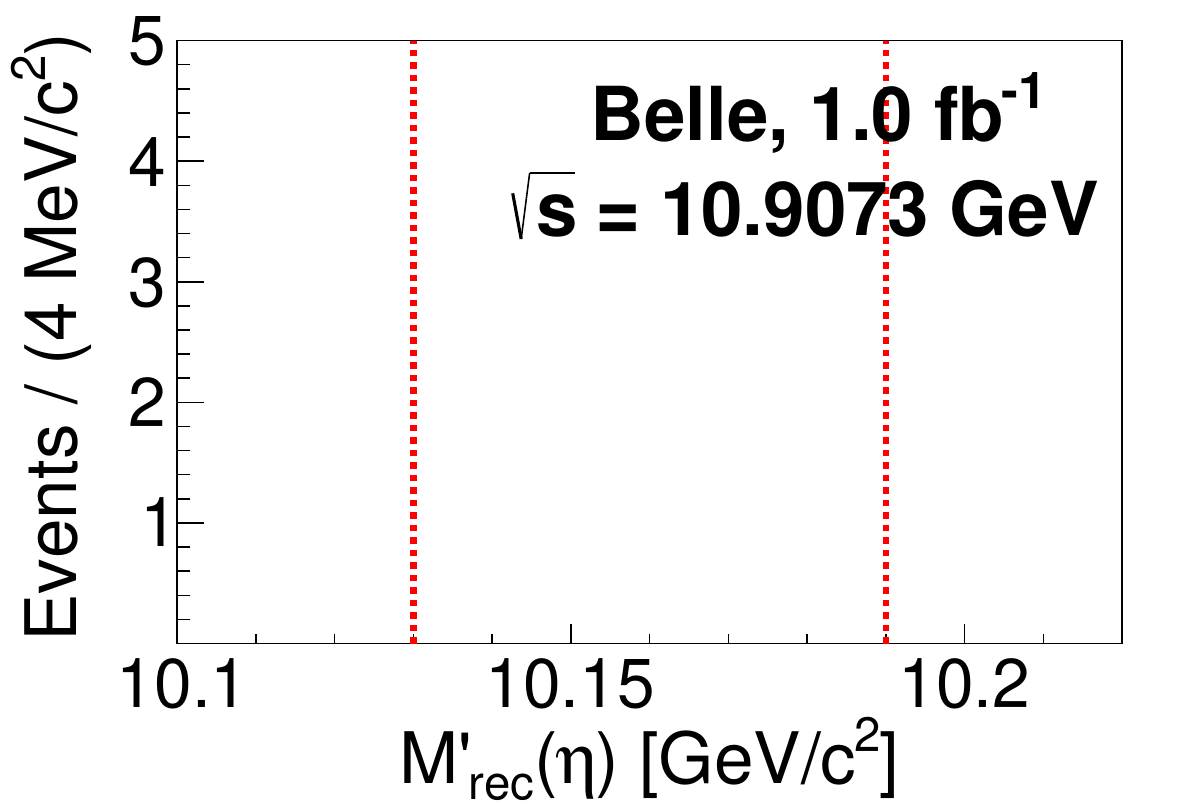}
\includegraphics[width=2.9cm]{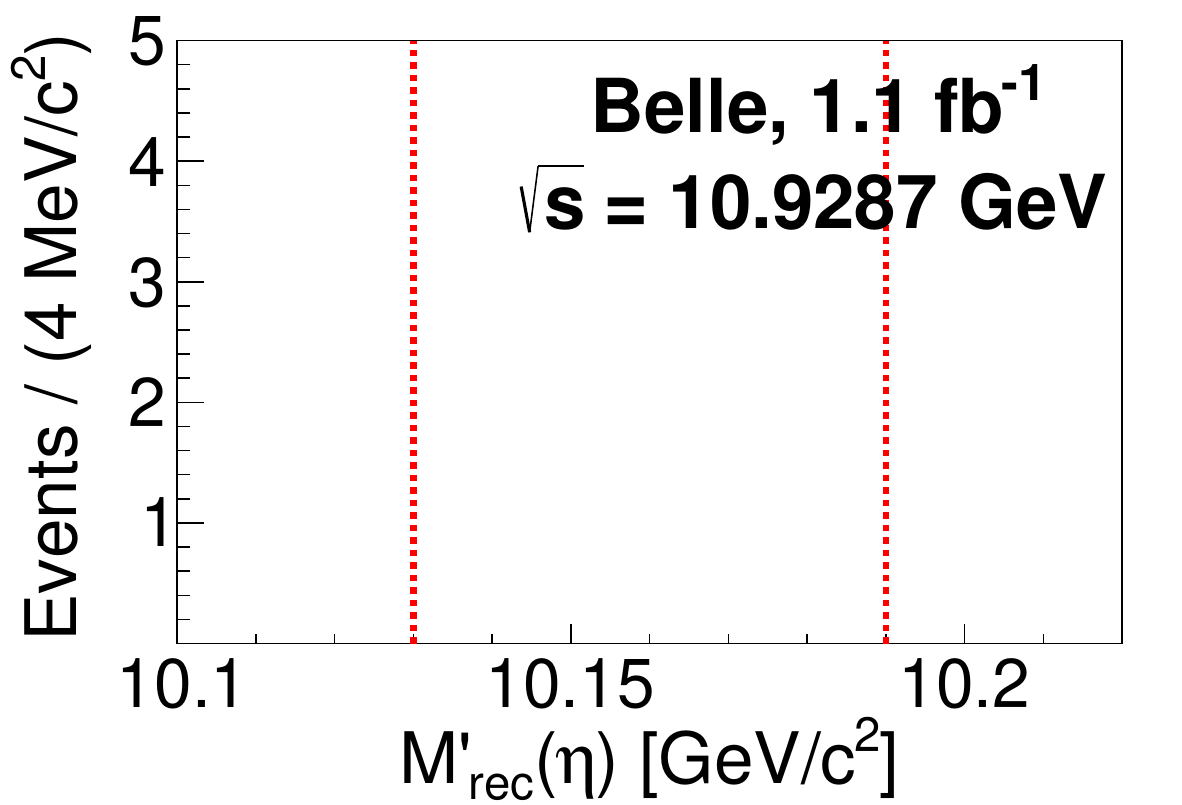}
\includegraphics[width=2.9cm]{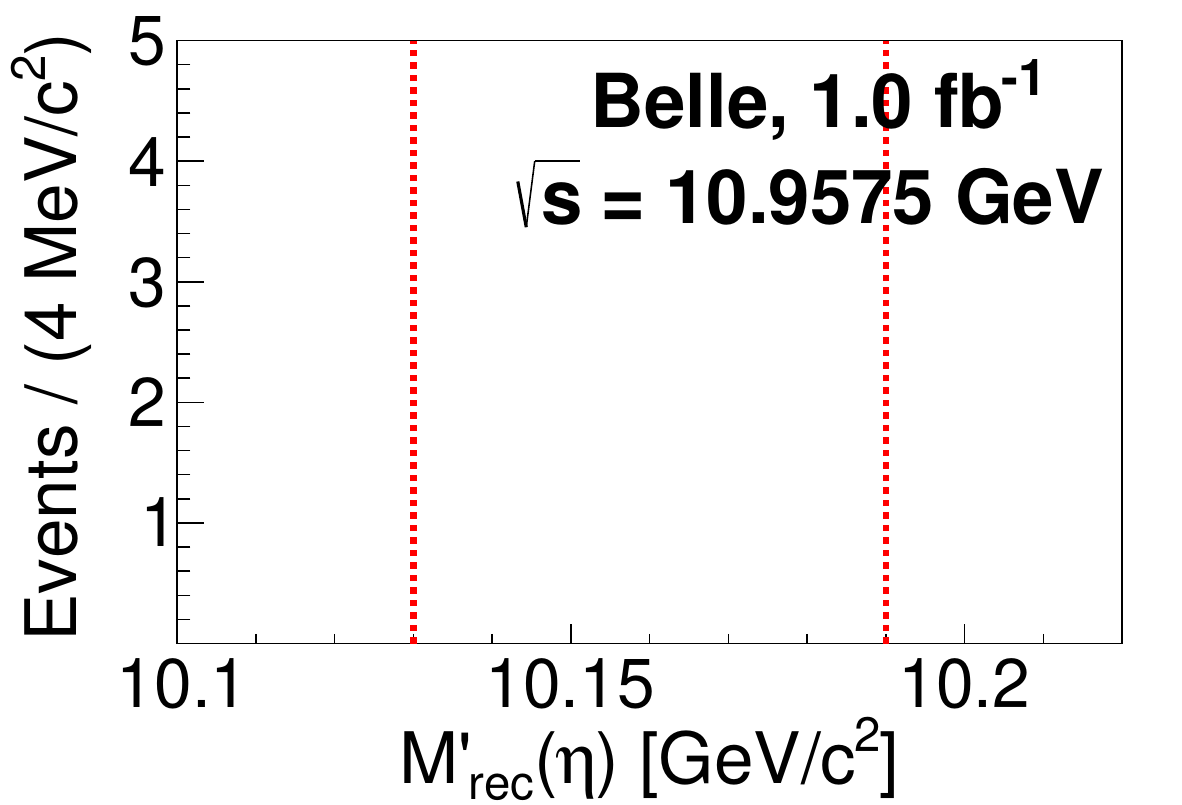}

\includegraphics[width=2.9cm]{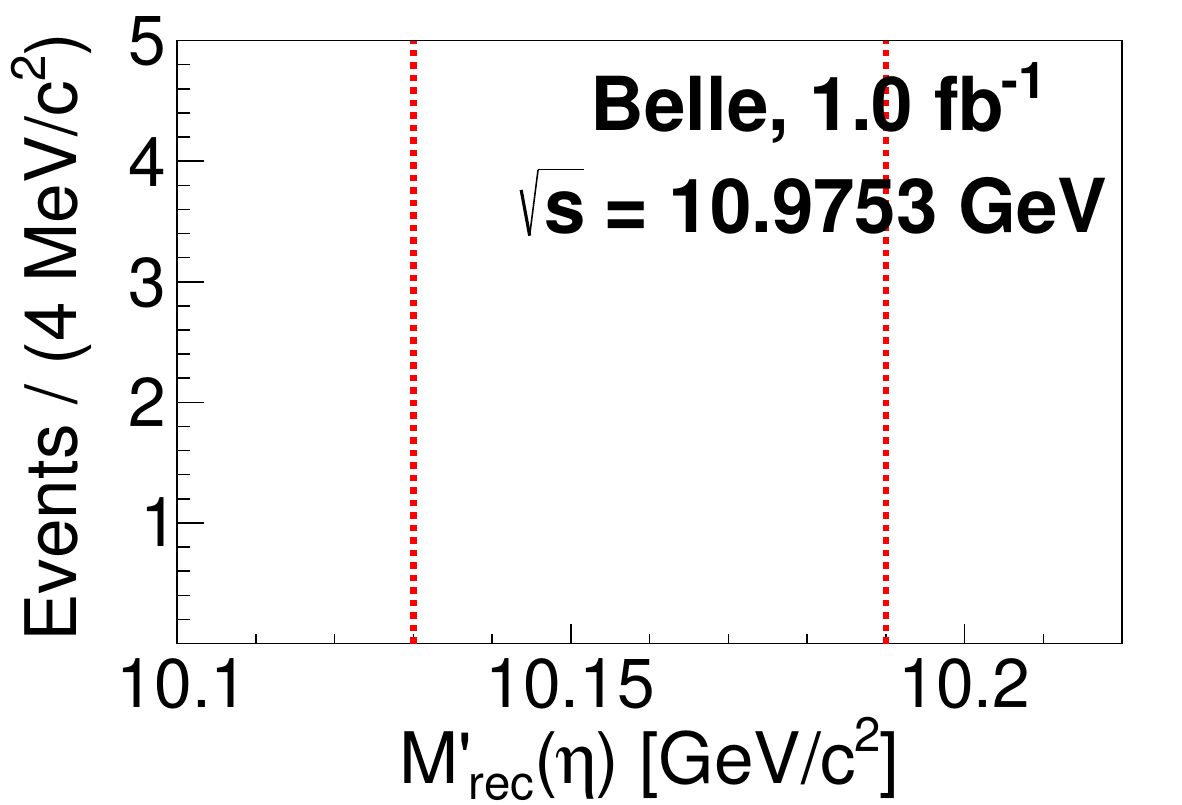}
\includegraphics[width=2.9cm]{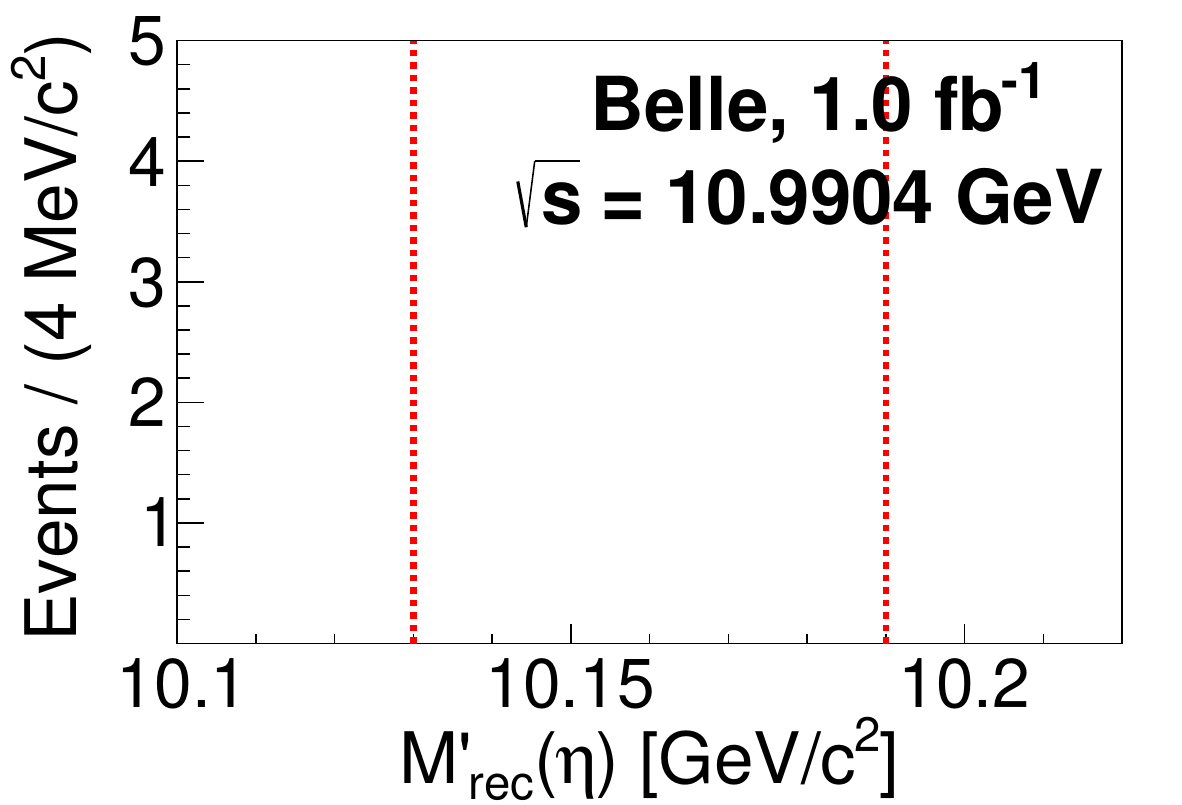}
\includegraphics[width=2.9cm]{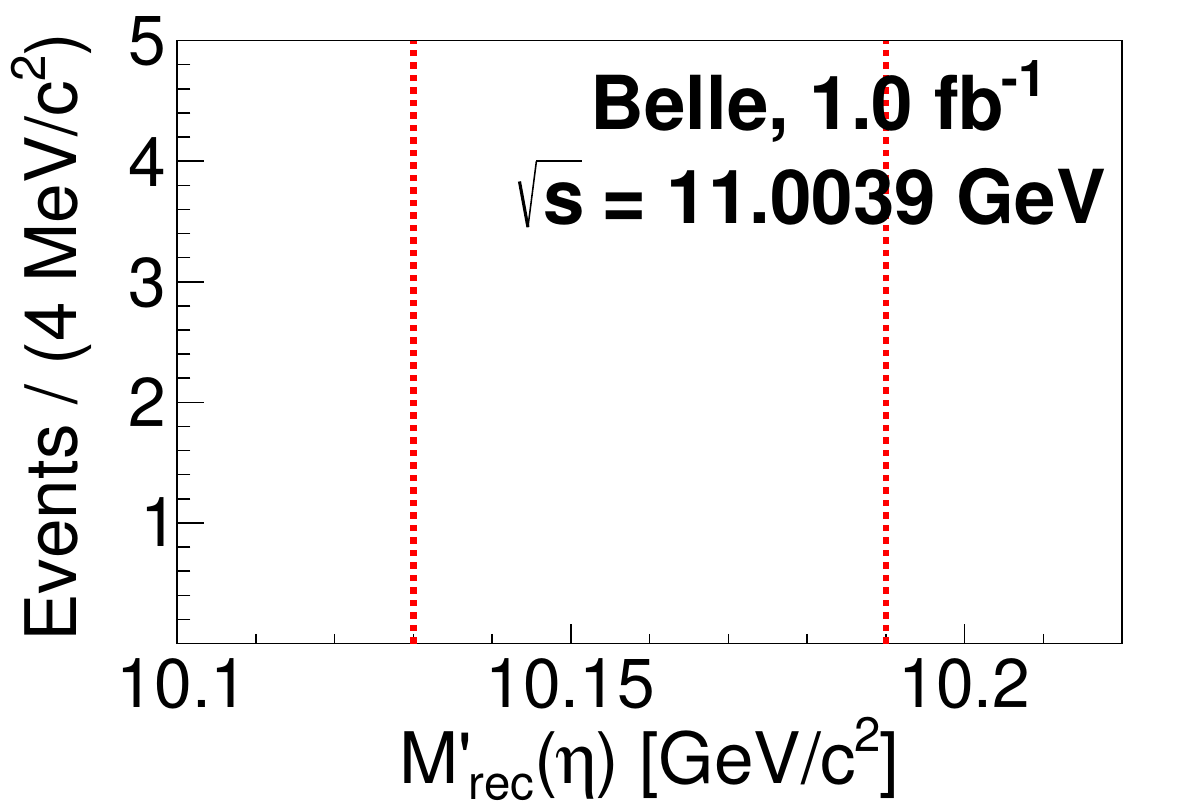}
\includegraphics[width=2.9cm]{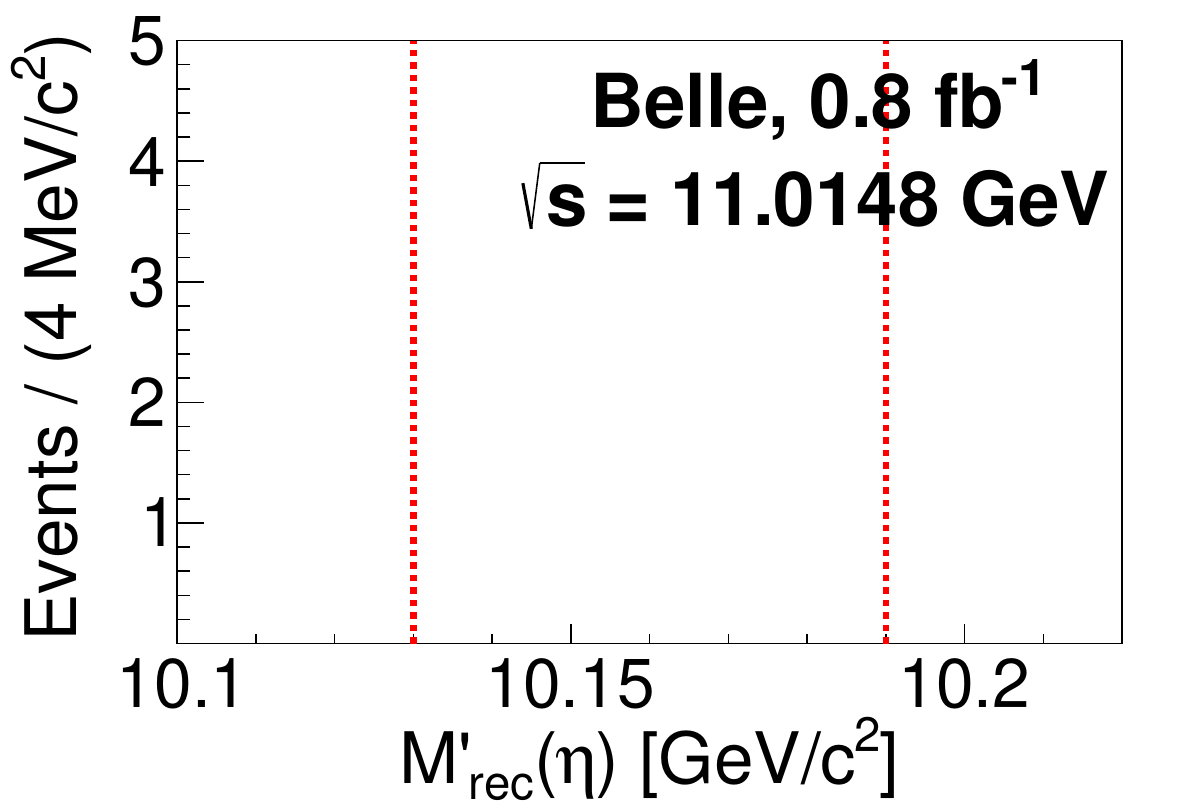}
\includegraphics[width=2.9cm]{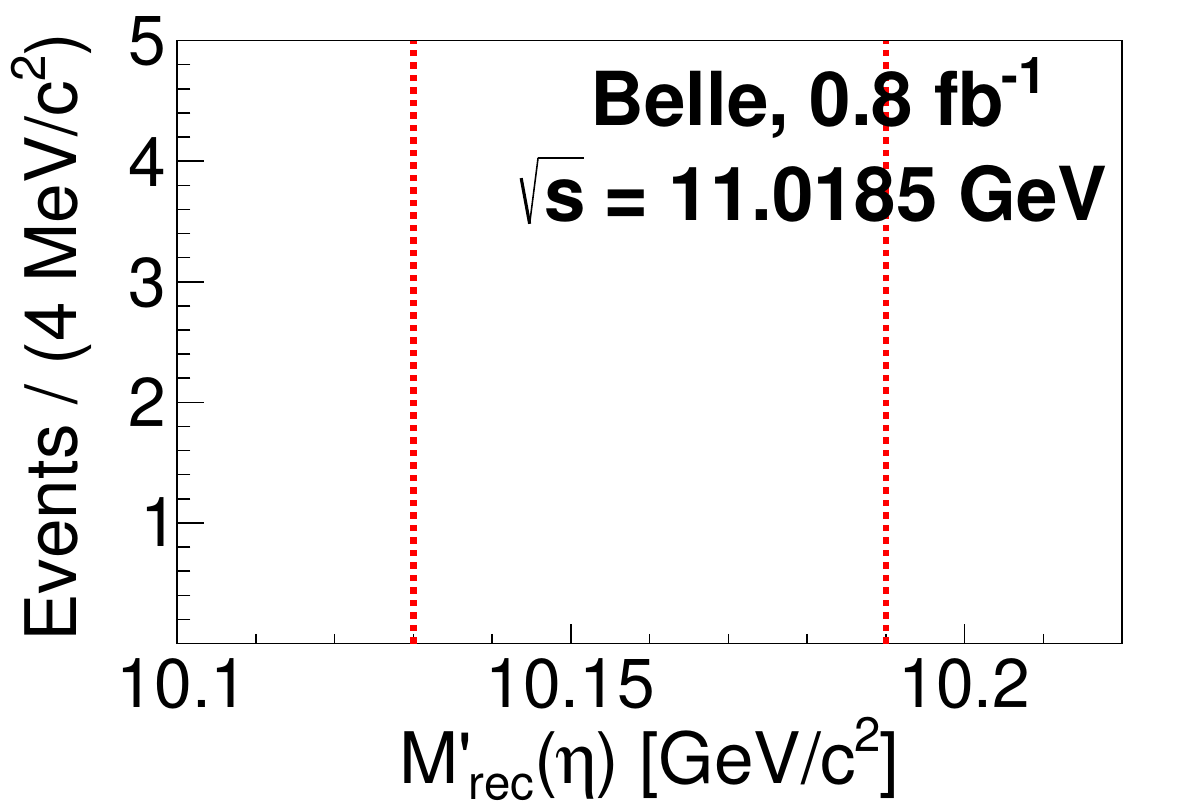}
\includegraphics[width=2.9cm]{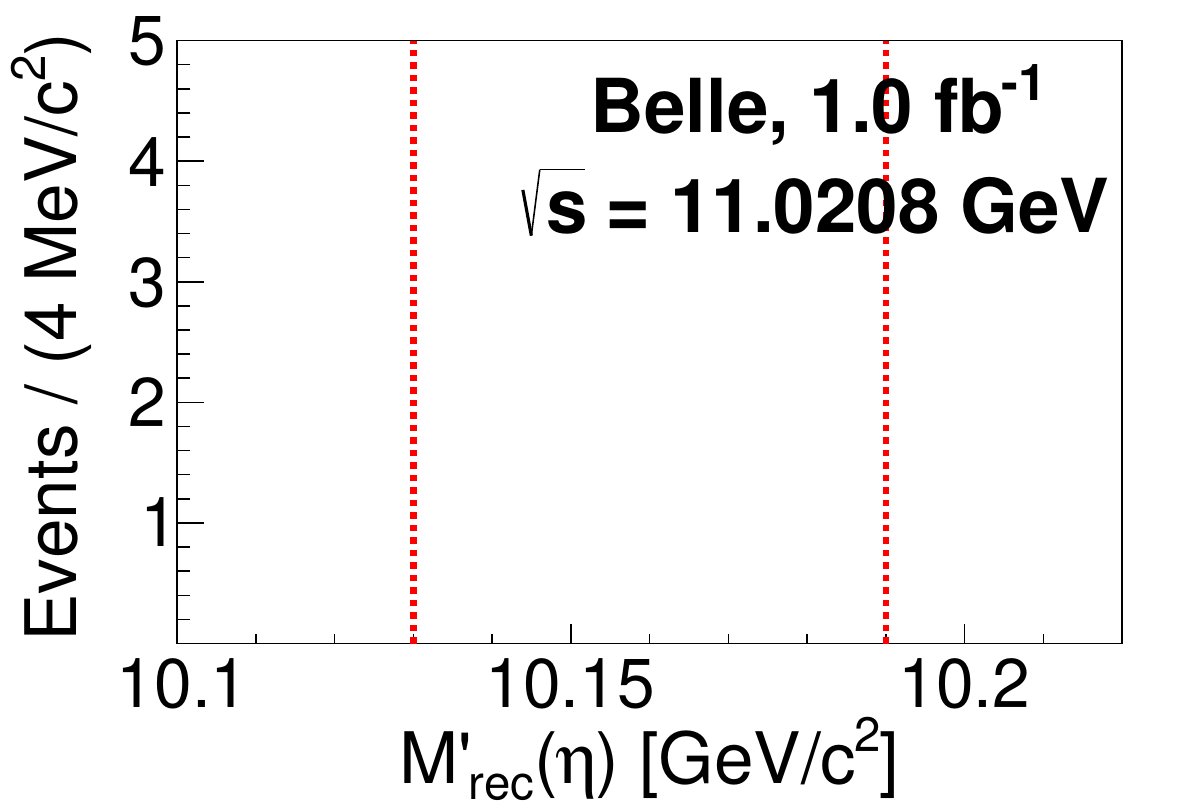}
\caption{The $M^{\prime}_{\rm rec}(\eta)$ distributions in data at each energy point at Belle. The vertical dashed lines show the $\Upsilon_J(1D)$ signal region.}\label{scaneta}
\end{figure}

\begin{figure}[htbp]
\centering
\includegraphics[width=2.9cm]{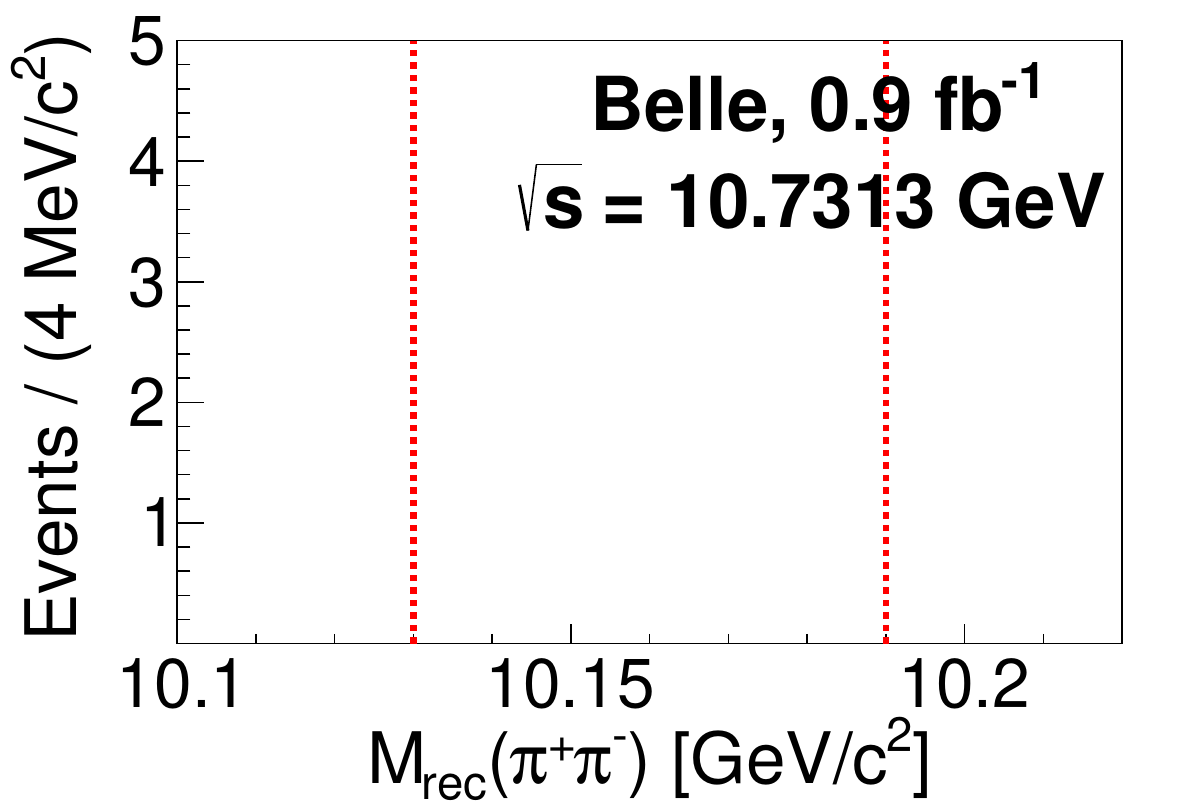}
\put(-85, 60){\large \bf Preliminary}
\includegraphics[width=2.9cm]{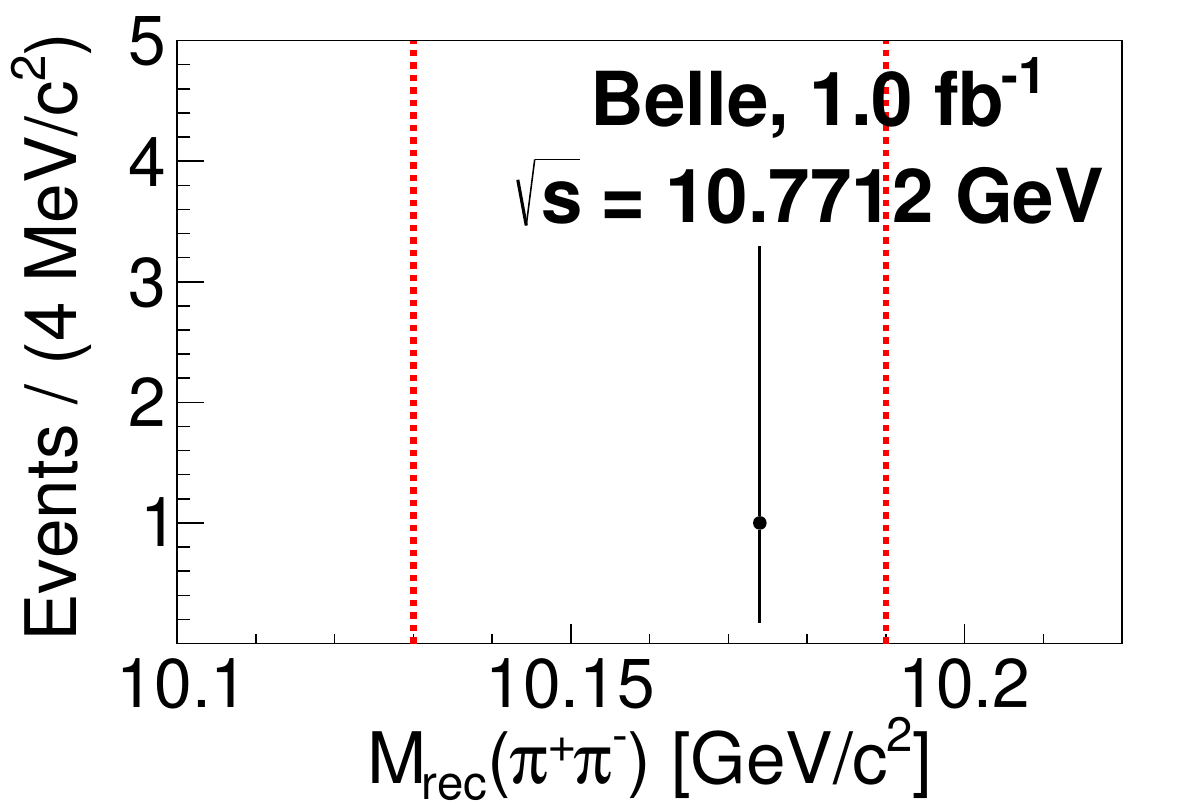}
\includegraphics[width=2.9cm]{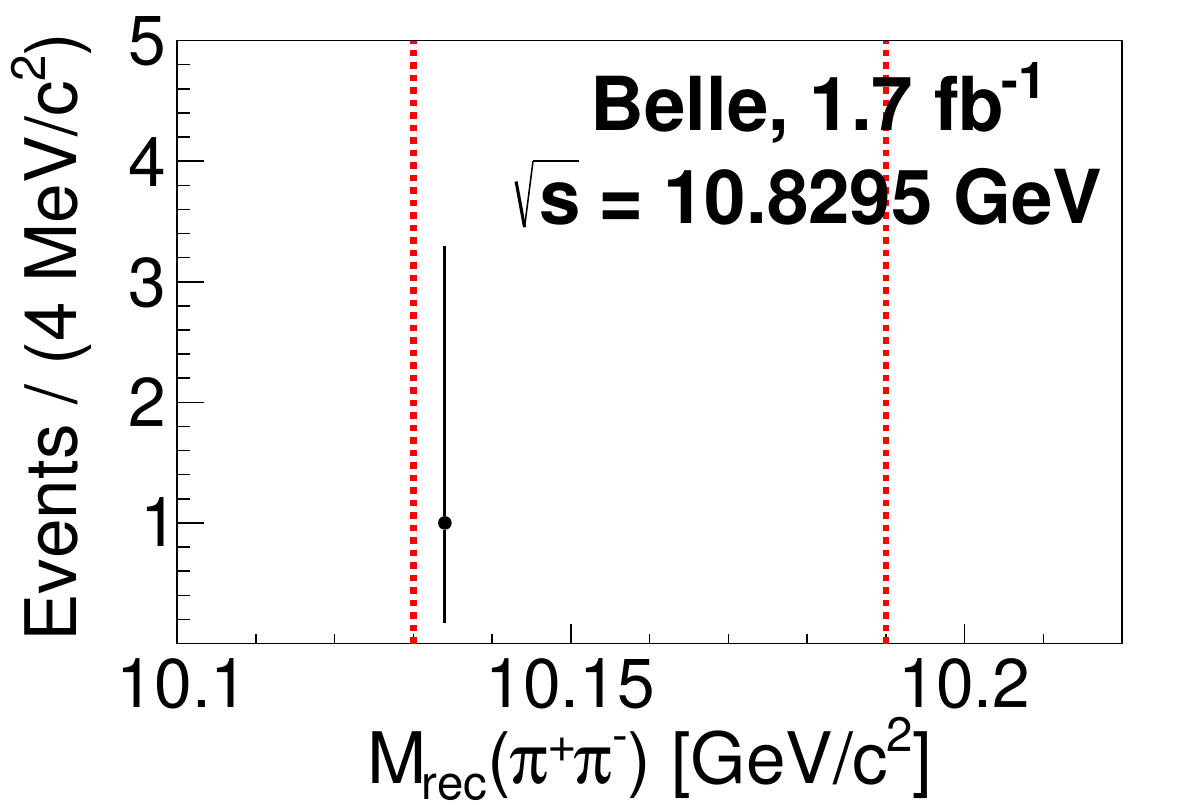}
\includegraphics[width=2.9cm]{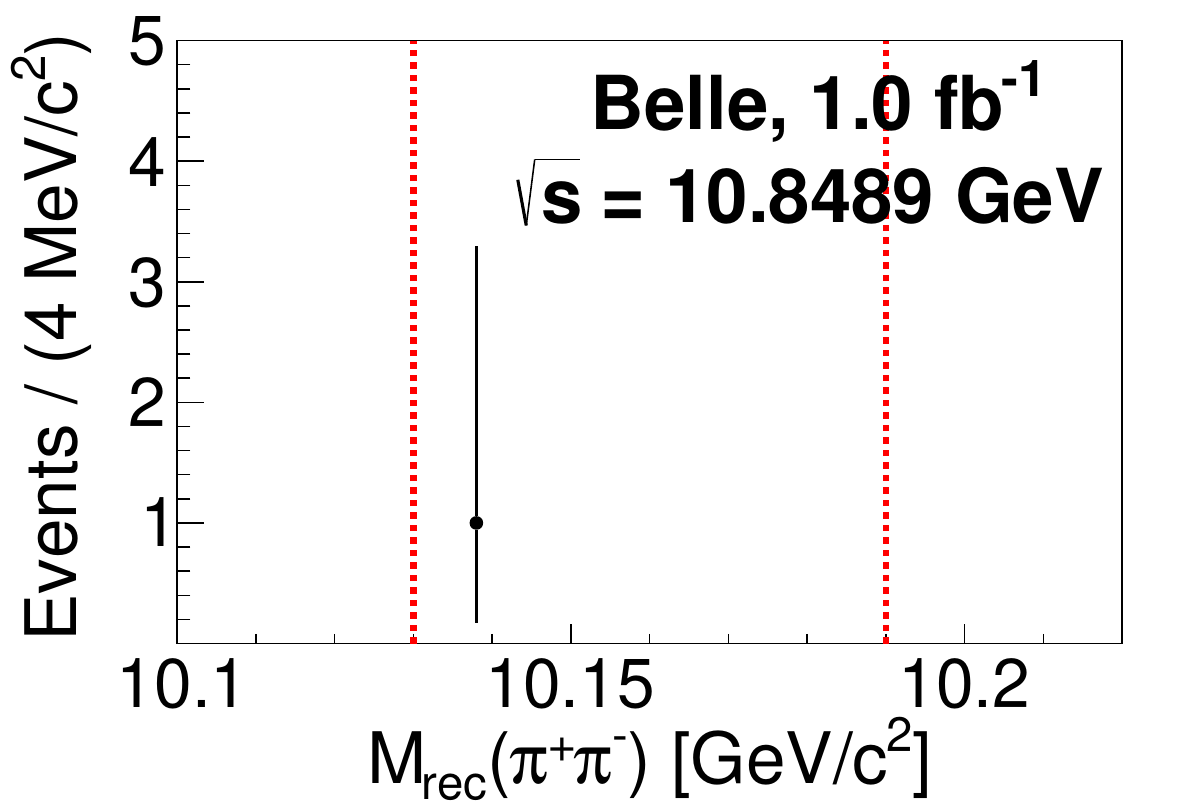}
\includegraphics[width=2.9cm]{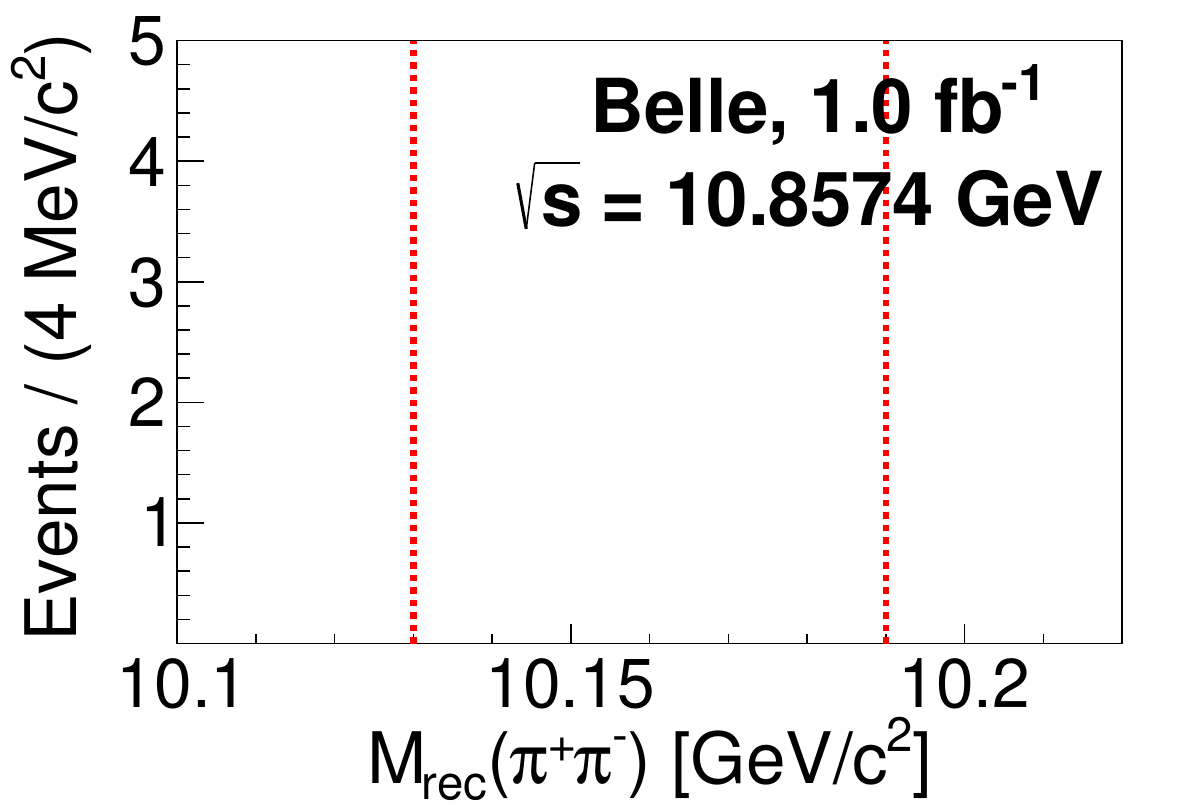}
\includegraphics[width=2.9cm]{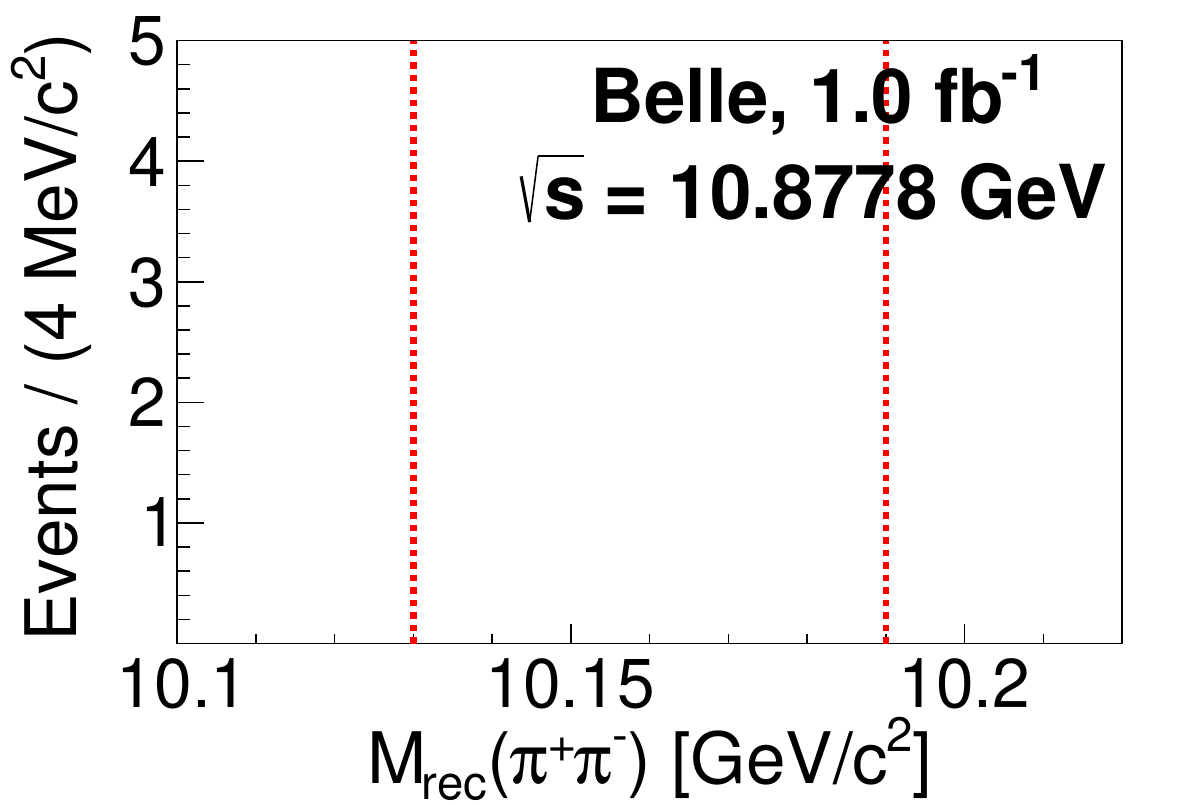}

\includegraphics[width=2.9cm]{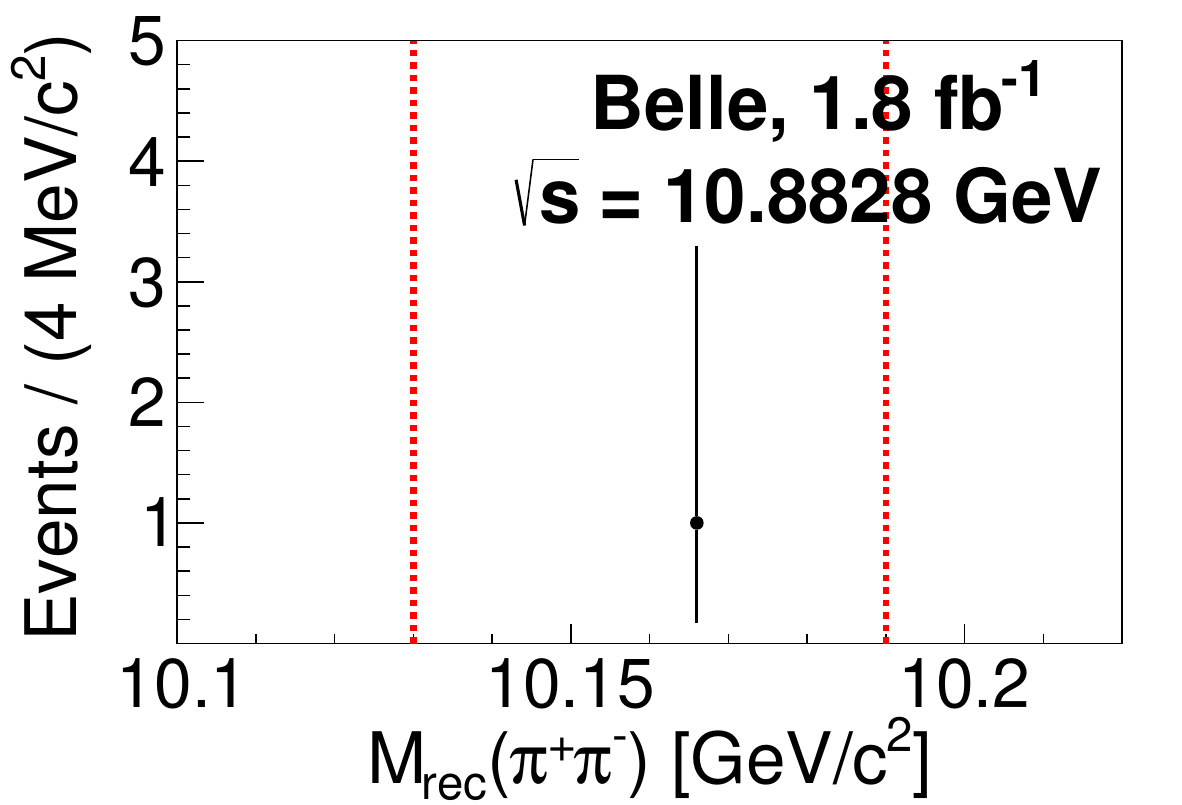}
\includegraphics[width=2.9cm]{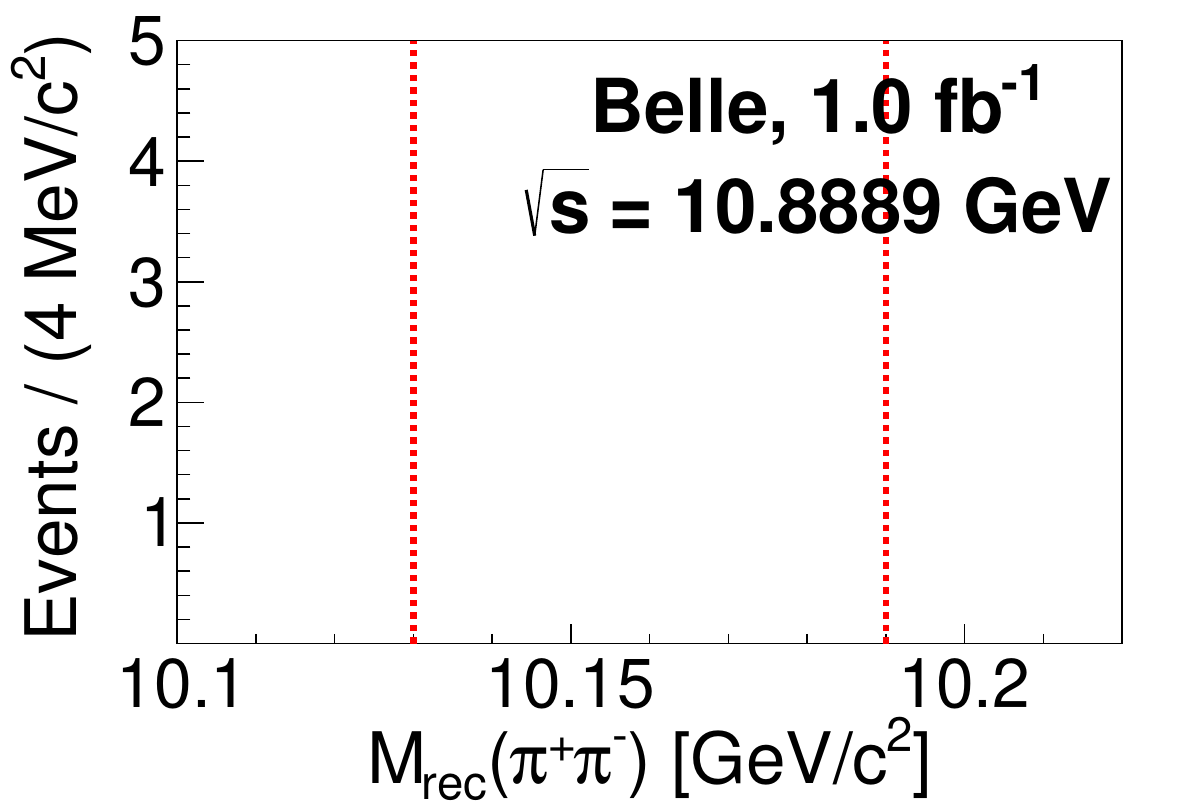}
\includegraphics[width=2.9cm]{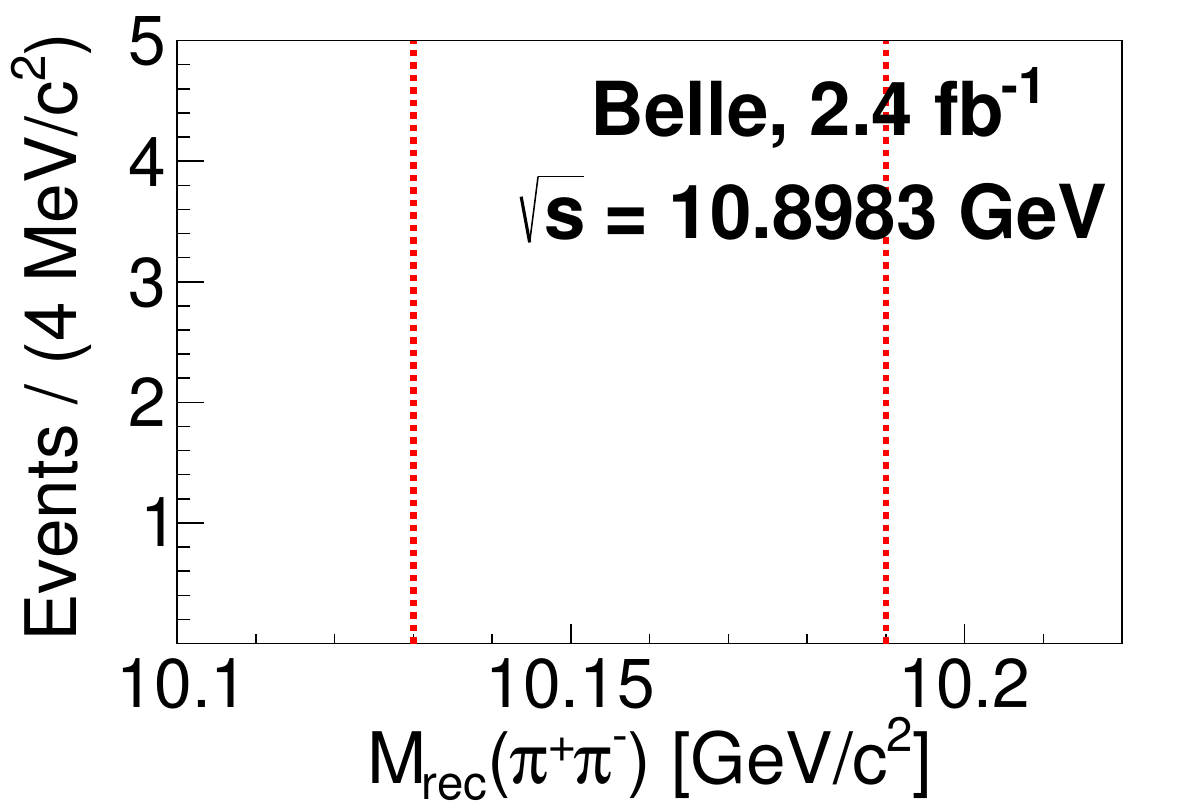}
\includegraphics[width=2.9cm]{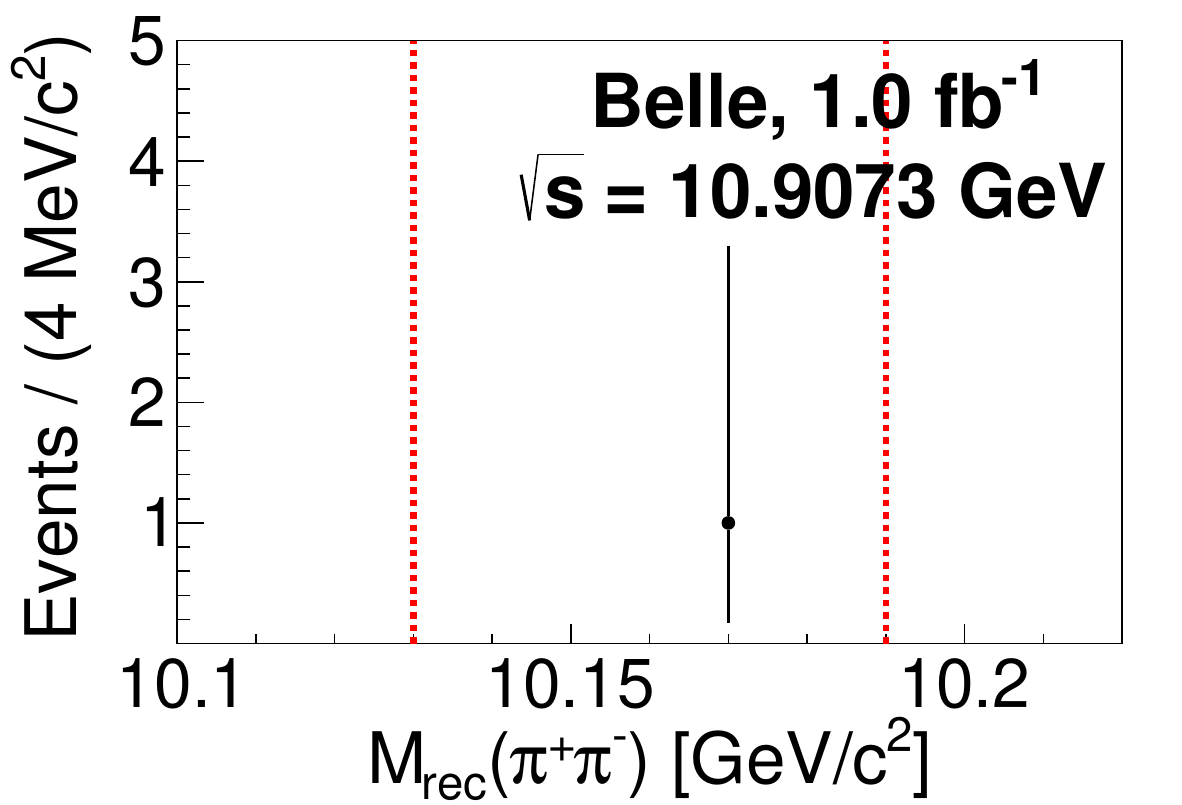}
\includegraphics[width=2.9cm]{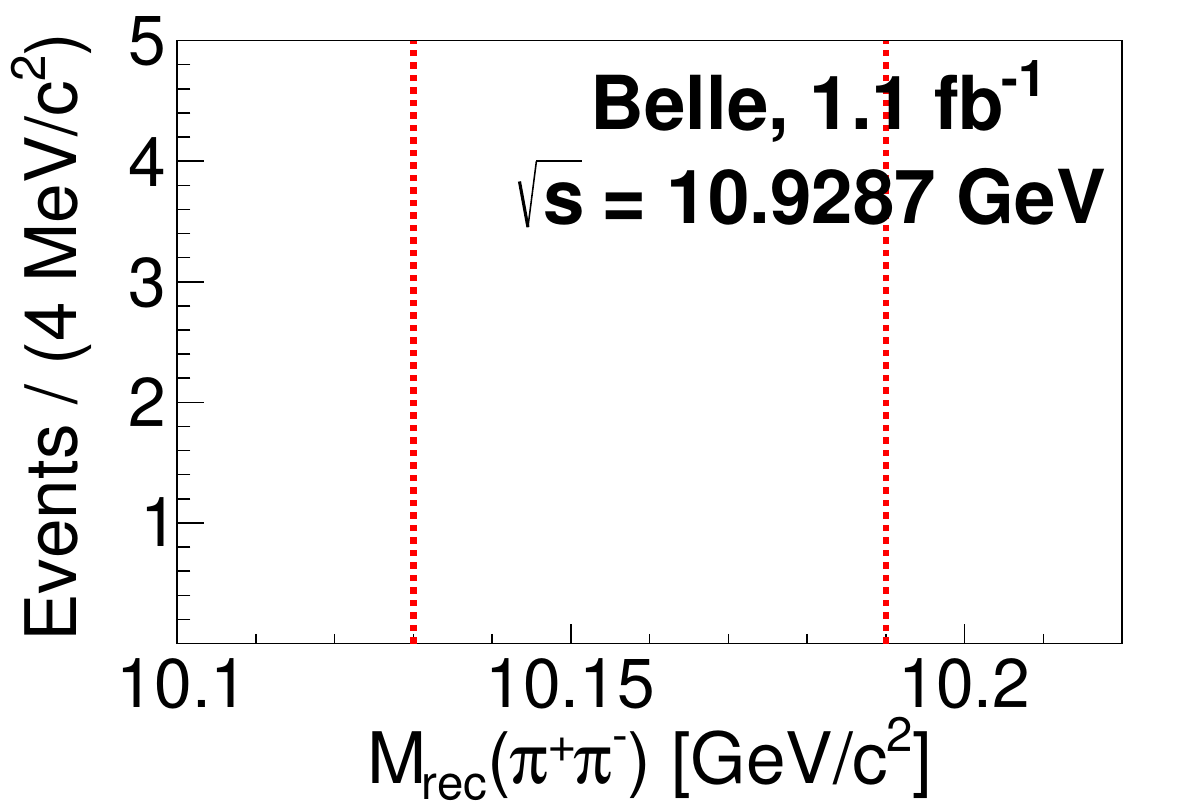}
\includegraphics[width=2.9cm]{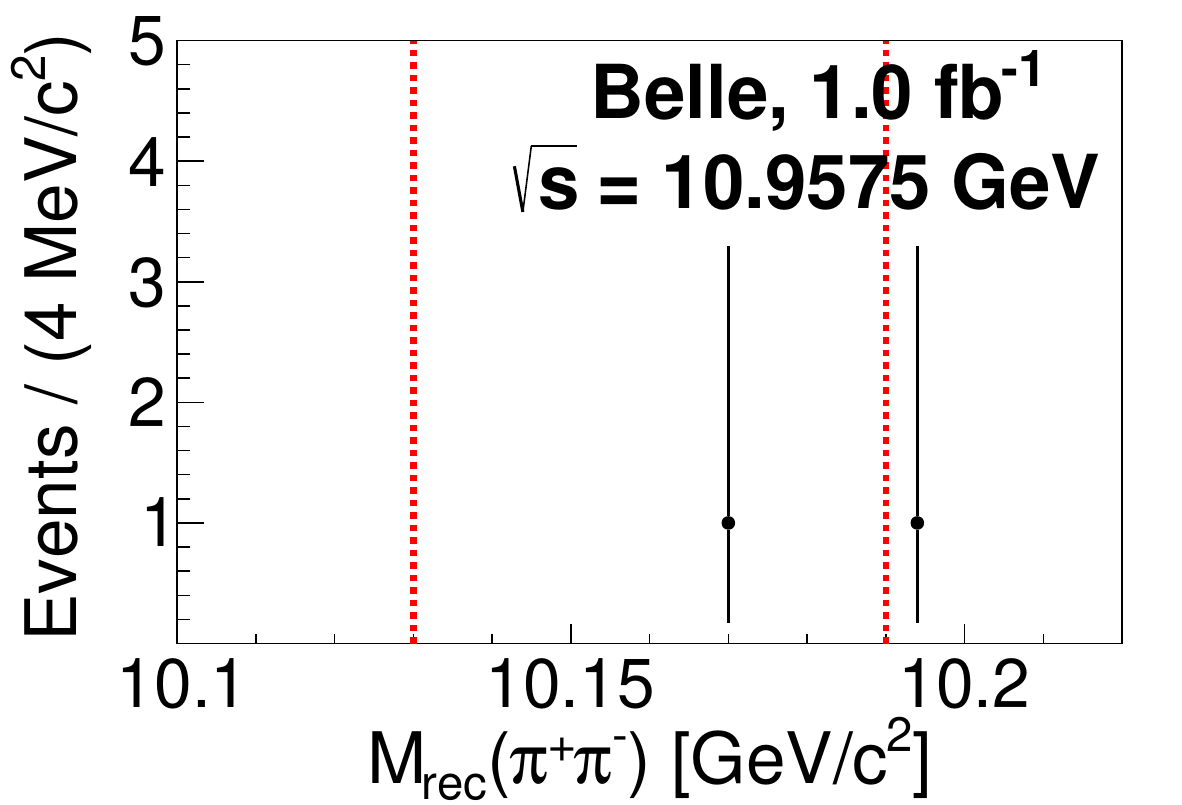}

\includegraphics[width=2.9cm]{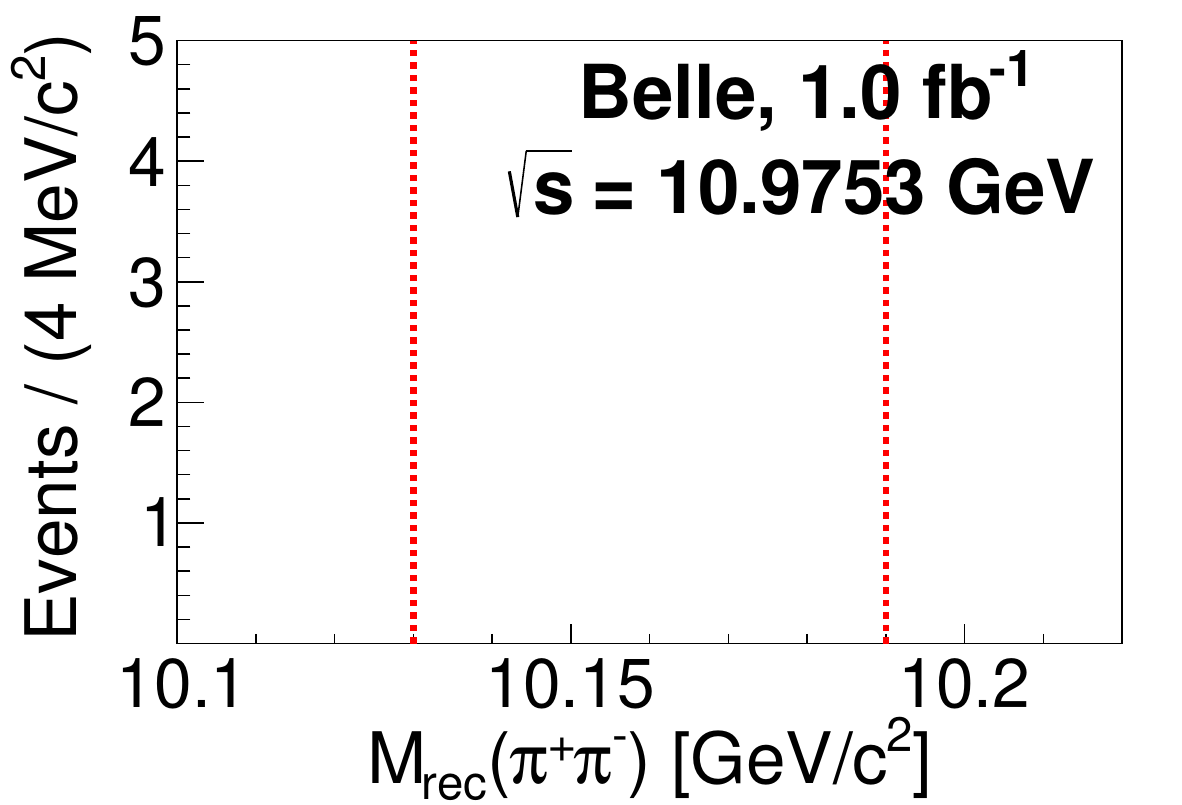}
\includegraphics[width=2.9cm]{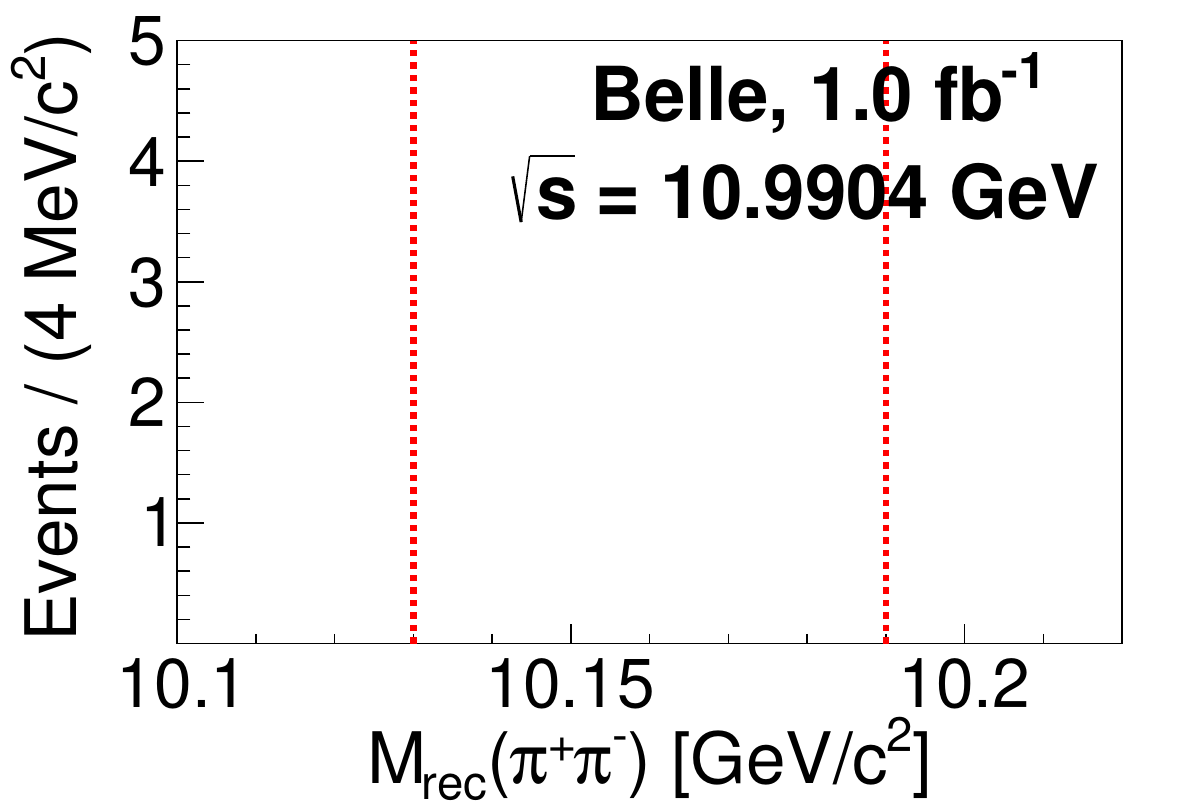}
\includegraphics[width=2.9cm]{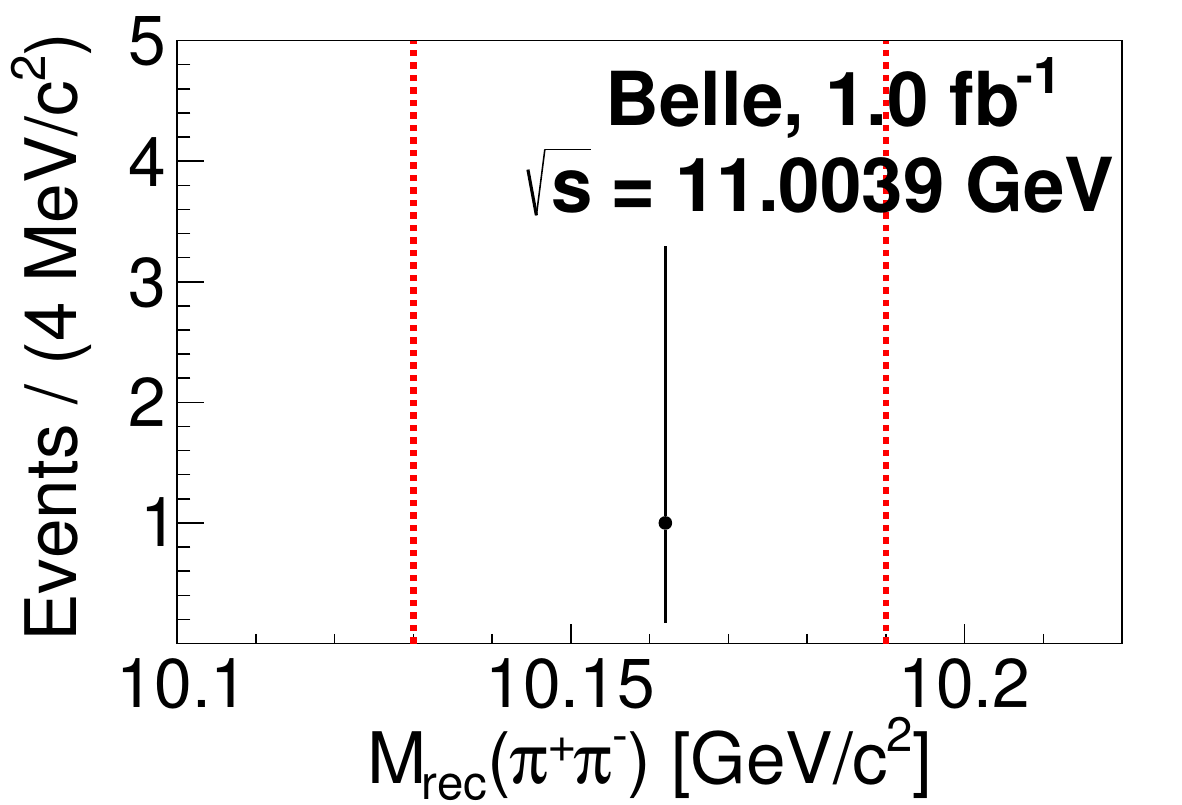}
\includegraphics[width=2.9cm]{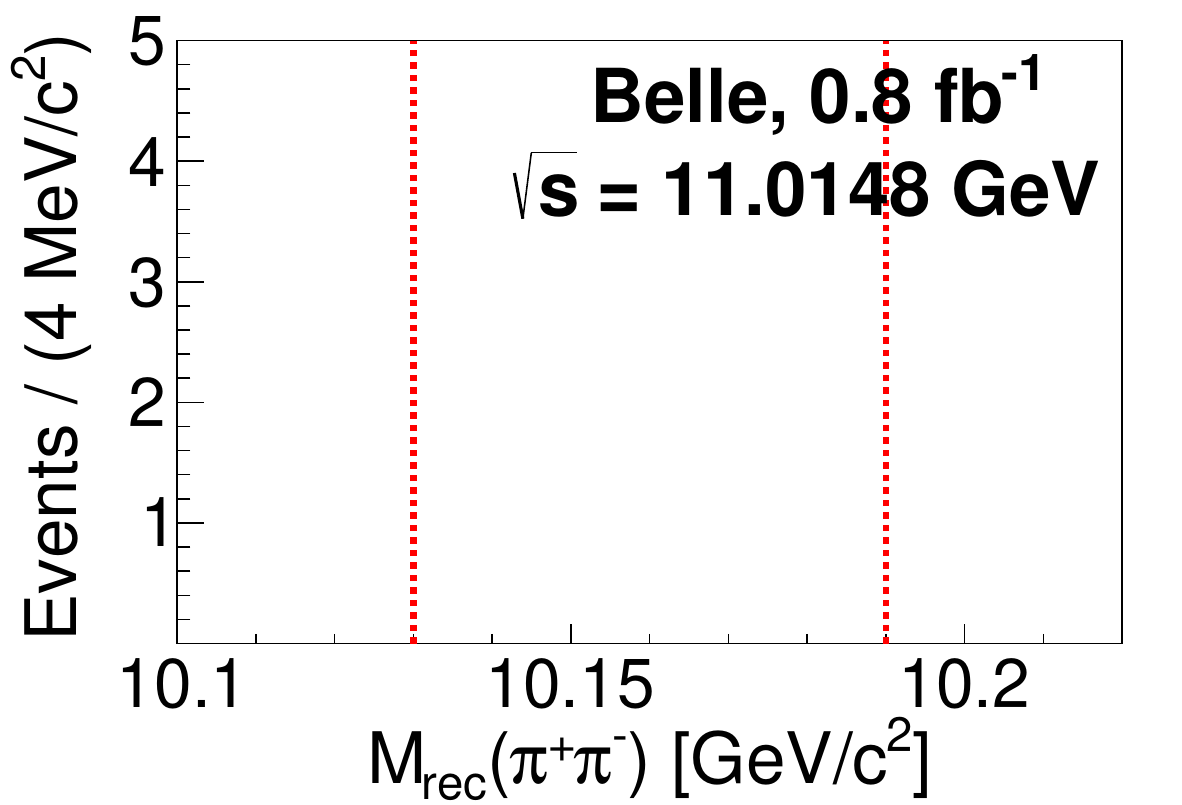}
\includegraphics[width=2.9cm]{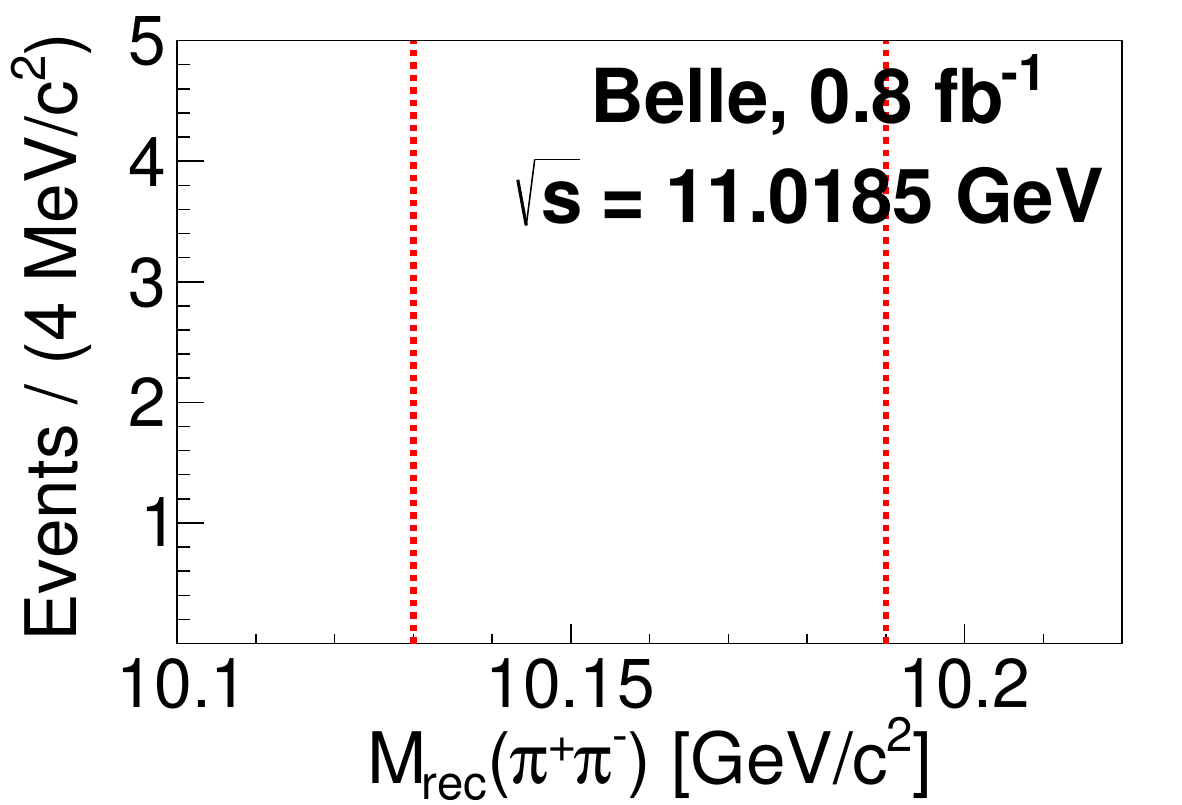}
\includegraphics[width=2.9cm]{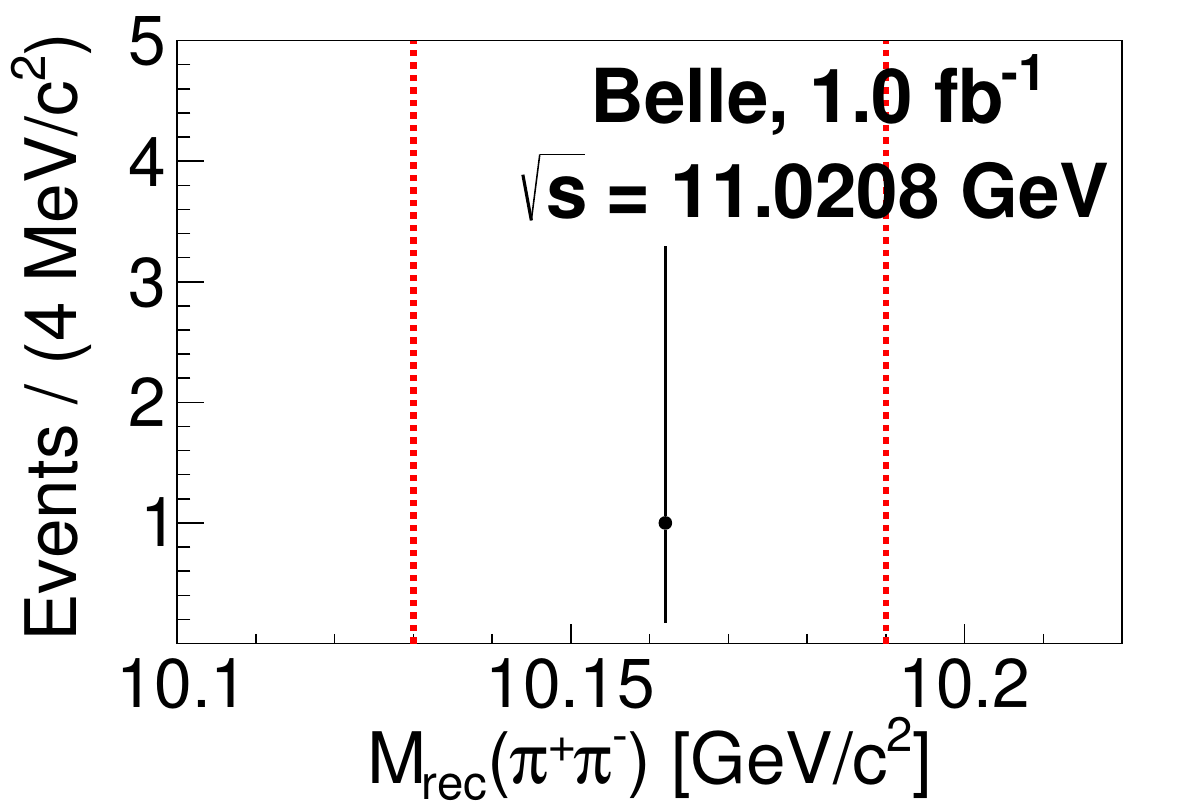}
\caption{The $M_{\rm rec}(\pi^+\pi^-)$ distributions in data at each energy point at Belle. The vertical dashed lines show the $\Upsilon_J(1D)$ signal region.
}\label{scanpipi}
\end{figure}

\newpage

\item The signal yields, Born cross sections, and upper limits on the Born cross sections for $e^+e^-\to\Upsilon_J(1D)\eta$ at Belle are listed in Table~\ref{table1}.

The signal yields, Born cross sections, and upper limits on the Born cross sections for $e^+e^-\to\Upsilon_J(1D)\pi^+\pi^-$ at Belle and Belle II are listed in Table~\ref{table2}.

\begin{table*}[htbp]
\renewcommand\arraystretch{1.2}
\setlength{\tabcolsep}{2pt}
\centering
\caption{Results for $e^+e^-\to\Upsilon_J(1D)\eta$ as a function of C.M.\ energy at Belle. We do not resolve the $\Upsilon_1(1D)$, $\Upsilon_2(1D)$, and $\Upsilon_3(1D)$ states and list results for combined $\Upsilon_1(1D)$, $\Upsilon_2(1D)$, and $\Upsilon_3(1D)$ states.
The ${\cal L}$ is the integrated luminosity of the data sample; $\varepsilon_i$ is the reconstruction efficiency; $(1+\delta_{\rm ISR})$ is the radiative correction factor; 
$N^{\rm obs}$ is the number of observed events;
$N^{\rm bg}$ is the number of background events;
$N^{\rm sig}$ is the signal yield;
$\sigma^{(3)}_{\rm Born}=\sigma_{\rm Born}(e^+e^-\to\Upsilon_J(1D)\eta)\,\BR(\Upsilon_J(1D)\to\chibJ\gamma)\BR(\chibJ\to\Upsilon(1S)\gamma)$;
$\sigma^{\rm UL}_{\rm Born}$ is the upper limit at 90\% C.L.\ on the $\sigma^{(3)}_{\rm Born}$.
}\label{table1}
\vspace{0.2cm}
\begin{tabular}{ccccccccc}
\hline\hline
$\sqrt{s}$ (MeV) & $\cal L$ (fb$^{-1}$) & $\varepsilon_{\eta\to\gamma\gamma,\,\eta\to\pi^+\pi^-\pi^0}$ & $1+\delta_{\rm ISR}$ & $N^{\rm obs}$ & $N^{\rm bg}$ & $N^{\rm sig}$ & $\sigma^{(3)}_{\rm Born}$ (pb) & $\sigma^{\rm UL}_{\rm Born}$ (pb) \\\hline
10.7313	&	0.946	&	0.100,	0.039	&	0.65	&	1.0	&	0.1	&	$	0.9	^{+	1.3	}_{-	0.7	}$	&	$	0.59	^{+	0.83	}_{-	0.45	}$	&	2.63	\\
10.7712	&	0.955	&	0.100,	0.040	&	0.71	&	0.0	&	0.1	&	$	0.0	^{+	0.5	}_{-	0.1	}$	&	$	0.00	^{+	0.29	}_{-	0.06	}$	&	1.23	\\
10.8295	&	1.697	&	0.101,	0.041	&	0.70	&	0.0	&	0.1	&	$	0.0	^{+	0.5	}_{-	0.1	}$	&	$	0.00	^{+	0.16	}_{-	0.03	}$	&	0.68	\\
10.8489	&	0.989	&	0.102,	0.042	&	0.68	&	0.0	&	0.1	&	$	0.0	^{+	0.5	}_{-	0.1	}$	&	$	0.00	^{+	0.29	}_{-	0.06	}$	&	1.20	\\
10.8574	&	0.988	&	0.102,	0.042	&	0.67	&	0.0	&	0.1	&	$	0.0	^{+	0.5	}_{-	0.1	}$	&	$	0.00	^{+	0.29	}_{-	0.06	}$	&	1.22	\\
10.8658	&	122.0	&	0.103,	0.042	&	0.65	&	57.0	&	10.0	&	$	47.0	^{+	7.9	}_{-	7.2	}$	&	$	0.22	^{+	0.04	}_{-	0.03	}$	&	-	\\
10.8778	&	0.978	&	0.103,	0.043	&	0.64	&	1.0	&	0.1	&	$	0.9	^{+	1.3	}_{-	0.7	}$	&	$	0.56	^{+	0.78	}_{-	0.42	}$	&	2.47	\\
10.8828	&	1.848	&	0.103,	0.044	&	0.65	&	0.0	&	0.2	&	$	0.0	^{+	0.5	}_{-	0.1	}$	&	$	0.00	^{+	0.16	}_{-	0.03	}$	&	0.66	\\
10.8889	&	0.990	&	0.103,	0.046	&	0.68	&	0.0	&	0.1	&	$	0.0	^{+	0.5	}_{-	0.1	}$	&	$	0.00	^{+	0.28	}_{-	0.06	}$	&	1.17	\\
10.8983	&	2.408	&	0.104,	0.046	&	0.75	&	1.0	&	0.2	&	$	0.8	^{+	1.3	}_{-	0.7	}$	&	$	0.16	^{+	0.27	}_{-	0.14	}$	&	0.84	\\
10.9073	&	0.980	&	0.104,	0.047	&	0.83	&	0.0	&	0.1	&	$	0.0	^{+	0.5	}_{-	0.1	}$	&	$	0.00	^{+	0.23	}_{-	0.05	}$	&	0.96	\\
10.9287	&	1.149	&	0.105,	0.049	&	0.96	&	0.0	&	0.1	&	$	0.0	^{+	0.5	}_{-	0.1	}$	&	$	0.00	^{+	0.17	}_{-	0.03	}$	&	0.70	\\
10.9575	&	0.969	&	0.105,	0.049	&	1.00	&	0.0	&	0.1	&	$	0.0	^{+	0.5	}_{-	0.1	}$	&	$	0.00	^{+	0.19	}_{-	0.04	}$	&	0.79	\\
10.9753	&	0.999	&	0.105,	0.051	&	1.00	&	0.0	&	0.1	&	$	0.0	^{+	0.5	}_{-	0.1	}$	&	$	0.00	^{+	0.18	}_{-	0.04	}$	&	0.75	\\
10.9904	&	0.985	&	0.106,	0.052	&	0.99	&	0.0	&	0.1	&	$	0.0	^{+	0.5	}_{-	0.1	}$	&	$	0.00	^{+	0.18	}_{-	0.04	}$	&	0.76	\\
11.0039	&	0.976	&	0.106,	0.053	&	0.98	&	0.0	&	0.1	&	$	0.0	^{+	0.5	}_{-	0.1	}$	&	$	0.00	^{+	0.18	}_{-	0.04	}$	&	0.78	\\
11.0148	&	0.771	&	0.106,	0.053	&	0.98	&	0.0	&	0.1	&	$	0.0	^{+	0.5	}_{-	0.1	}$	&	$	0.00	^{+	0.24	}_{-	0.05	}$	&	0.99	\\
11.0185	&	0.859	&	0.106,	0.055	&	0.97	&	0.0	&	0.1	&	$	0.0	^{+	0.5	}_{-	0.1	}$	&	$	0.00	^{+	0.21	}_{-	0.04	}$	&	0.88	\\
11.0208	&	0.982	&	0.106,	0.056	&	0.97	&	0.0	&	0.1	&	$	0.0	^{+	0.5	}_{-	0.1	}$	&	$	0.00	^{+	0.18	}_{-	0.04	}$	&	0.77	\\
\hline\hline
\end{tabular}
\end{table*}

\begin{table*}[htbp]
\renewcommand\arraystretch{1.2}
\setlength{\tabcolsep}{2pt}
\centering
\caption{Results for $e^+e^-\to\Upsilon_J(1D)\pi^+\pi^-$ as a function of C.M.\ energy at Belle II (points marked with an
asterisk) and Belle (all other points). 
We do not resolve the $\Upsilon_1(1D)$, $\Upsilon_2(1D)$, and $\Upsilon_3(1D)$ states and list results for combined $\Upsilon_1(1D)$, $\Upsilon_2(1D)$, and $\Upsilon_3(1D)$ states.
The ${\cal L}$ is the integrated luminosity of the data sample; $\varepsilon$ is the reconstruction efficiency; $(1+\delta_{\rm ISR})$ is the radiative correction factor; 
$N^{\rm obs}$ is the number of observed events;
$N^{\rm bg}$ is the number of background events;
$N^{\rm sig}$ is the signal yield;
$\sigma^{(4)}_{\rm Born}=\sigma_{\rm Born}(e^+e^-\to\Upsilon_J(1D)\pi^+\pi^-)\,\BR(\Upsilon_J(1D)\to\chibJ\gamma)\BR(\chibJ\to\Upsilon(1S)\gamma)$;
$\sigma^{\rm UL}_{\rm Born}$ is the upper limit at 90\% C.L.\ on the $\sigma^{(4)}_{\rm Born}$.}
\label{table2}
\vspace{0.2cm}
\begin{tabular}{cccccccccc}
\hline\hline
&$\sqrt{s}$ (MeV) & $\cal L$ (fb$^{-1}$) & $\varepsilon$ & $1+\delta_{\rm ISR}$ & $N^{\rm obs}$ & $N^{\rm bg}$ & $N^{\rm sig}$ & $\sigma^{(4)}_{\rm Born}$ (pb) & $\sigma^{\rm UL}_{\rm Born}$ (pb) \\\hline
$\ast$	&	10.6530	&	3.500	&	0.179			&	0.82	&	2.0	&	2.9	&	$	0.0	^{+	1.0	}_{-	0.2	}$	&	$	0.00	^{+	0.04	}_{-	0.01	}$	&	0.12	\\
$\ast$	&	10.7010	&	1.600	&	0.184			&	0.90	&	0.0	&	1.3	&	$	0.0	^{+	0.5	}_{-	0.1	}$	&	$	0.00	^{+	0.04	}_{-	0.01	}$	&	0.09	\\
	&	10.7313	&	0.946	&	0.169			&	0.67	&	0.0	&	0.1	&	$	0.0	^{+	0.5	}_{-	0.1	}$	&	$	0.00	^{+	0.09	}_{-	0.02	}$	&	0.38	\\
$\ast$	&	10.7450	&	9.800	&	0.199			&	0.88	&	16.0	&	8.0	&	$	8.0	^{+	5.0	}_{-	4.1	}$	&	$	0.09	^{+	0.06	}_{-	0.05	}$	&	0.18	\\
	&	10.7712	&	0.955	&	0.171			&	0.72	&	1.0	&	0.1	&	$	0.9	^{+	1.3	}_{-	0.7	}$	&	$	0.15	^{+	0.21	}_{-	0.11	}$	&	0.67	\\
$\ast$	&	10.8050	&	4.700	&	0.225			&	0.73	&	2.0	&	3.8	&	$	0.0	^{+	0.8	}_{-	0.2	}$	&	$	0.00	^{+	0.02	}_{-	0.00	}$	&	0.06	\\
	&	10.8295	&	1.697	&	0.179			&	0.71	&	1.0	&	0.1	&	$	0.9	^{+	1.3	}_{-	0.7	}$	&	$	0.08	^{+	0.11	}_{-	0.06	}$	&	0.36	\\
	&	10.8489	&	0.989	&	0.180			&	0.69	&	1.0	&	0.1	&	$	0.9	^{+	1.3	}_{-	0.7	}$	&	$	0.14	^{+	0.20	}_{-	0.11	}$	&	0.64	\\
	&	10.8574	&	0.988	&	0.180			&	0.67	&	0.0	&	0.1	&	$	0.0	^{+	0.5	}_{-	0.1	}$	&	$	0.00	^{+	0.08	}_{-	0.02	}$	&	0.34	\\
	&	10.8658	&	122.0	&	0.180			&	0.66	&	79.0	&	8.6	&	$	70.4	^{+	9.8	}_{-	8.0	}$	&	$	0.09	^{+	0.01	}_{-	0.01	}$	&	-	\\
	&	10.8778	&	0.978	&	0.180			&	0.65	&	0.0	&	0.1	&	$	0.0	^{+	0.5	}_{-	0.1	}$	&	$	0.00	^{+	0.08	}_{-	0.02	}$	&	0.35	\\
	&	10.8828	&	1.848	&	0.181			&	0.65	&	1.0	&	0.1	&	$	0.9	^{+	1.3	}_{-	0.7	}$	&	$	0.08	^{+	0.11	}_{-	0.06	}$	&	0.36	\\
	&	10.8889	&	0.990	&	0.182			&	0.68	&	0.0	&	0.1	&	$	0.0	^{+	0.5	}_{-	0.1	}$	&	$	0.00	^{+	0.08	}_{-	0.02	}$	&	0.33	\\
	&	10.8983	&	2.408	&	0.183			&	0.75	&	0.0	&	0.2	&	$	0.0	^{+	0.5	}_{-	0.1	}$	&	$	0.00	^{+	0.03	}_{-	0.01	}$	&	0.12	\\
	&	10.9073	&	0.980	&	0.184			&	0.82	&	1.0	&	0.1	&	$	0.9	^{+	1.3	}_{-	0.7	}$	&	$	0.12	^{+	0.17	}_{-	0.09	}$	&	0.53	\\
	&	10.9287	&	1.149	&	0.184			&	0.93	&	0.0	&	0.1	&	$	0.0	^{+	0.5	}_{-	0.1	}$	&	$	0.00	^{+	0.05	}_{-	0.01	}$	&	0.20	\\
	&	10.9575	&	0.969	&	0.186			&	0.89	&	1.0	&	0.1	&	$	0.9	^{+	1.3	}_{-	0.7	}$	&	$	0.11	^{+	0.15	}_{-	0.08	}$	&	0.49	\\
	&	10.9753	&	0.999	&	0.189			&	0.78	&	0.0	&	0.1	&	$	0.0	^{+	0.5	}_{-	0.1	}$	&	$	0.00	^{+	0.06	}_{-	0.01	}$	&	0.27	\\
	&	10.9904	&	0.985	&	0.191			&	0.67	&	0.0	&	0.1	&	$	0.0	^{+	0.5	}_{-	0.1	}$	&	$	0.00	^{+	0.08	}_{-	0.02	}$	&	0.32	\\
	&	11.0039	&	0.976	&	0.191			&	0.70	&	1.0	&	0.1	&	$	0.9	^{+	1.3	}_{-	0.7	}$	&	$	0.13	^{+	0.19	}_{-	0.10	}$	&	0.60	\\
	&	11.0148	&	0.771	&	0.193			&	0.85	&	0.0	&	0.1	&	$	0.0	^{+	0.5	}_{-	0.1	}$	&	$	0.00	^{+	0.08	}_{-	0.02	}$	&	0.32	\\
	&	11.0185	&	0.859	&	0.193			&	0.91	&	0.0	&	0.1	&	$	0.0	^{+	0.5	}_{-	0.1	}$	&	$	0.00	^{+	0.06	}_{-	0.01	}$	&	0.27	\\
	&	11.0208	&	0.982	&	0.193			&	0.94	&	1.0	&	0.1	&	$	0.9	^{+	1.3	}_{-	0.7	}$	&	$	0.10	^{+	0.14	}_{-	0.08	}$	&	0.44	\\
\hline\hline
\end{tabular}
\end{table*}

\newpage

\item Figure~\ref{Belle2v1} show the $M(\pi^+\pi^-)^{\rm recoil}$ distributions for data events in the $\chi_{b1}$ and $\chi_{b2}$ signal regions of 9.869--9.915 GeV/$c^2$ at each energy point at Belle II~\cite{2602.19807}.
We use the unified counting method to calculate the $\sigma^{(4)}_{\text{dressed}}$ for Belle II energy points.

\begin{figure*}[htbp]
\centering
\includegraphics[width=4.4cm]{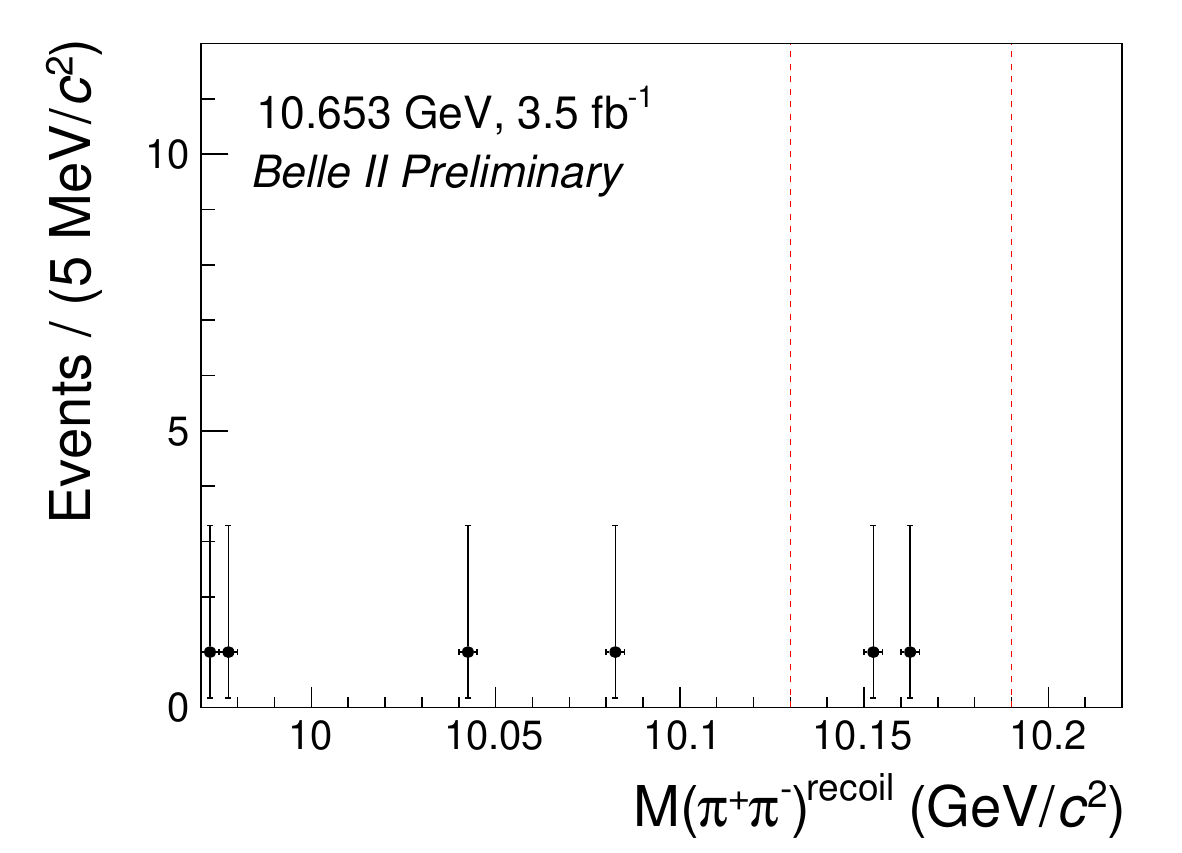}
\includegraphics[width=4.4cm]{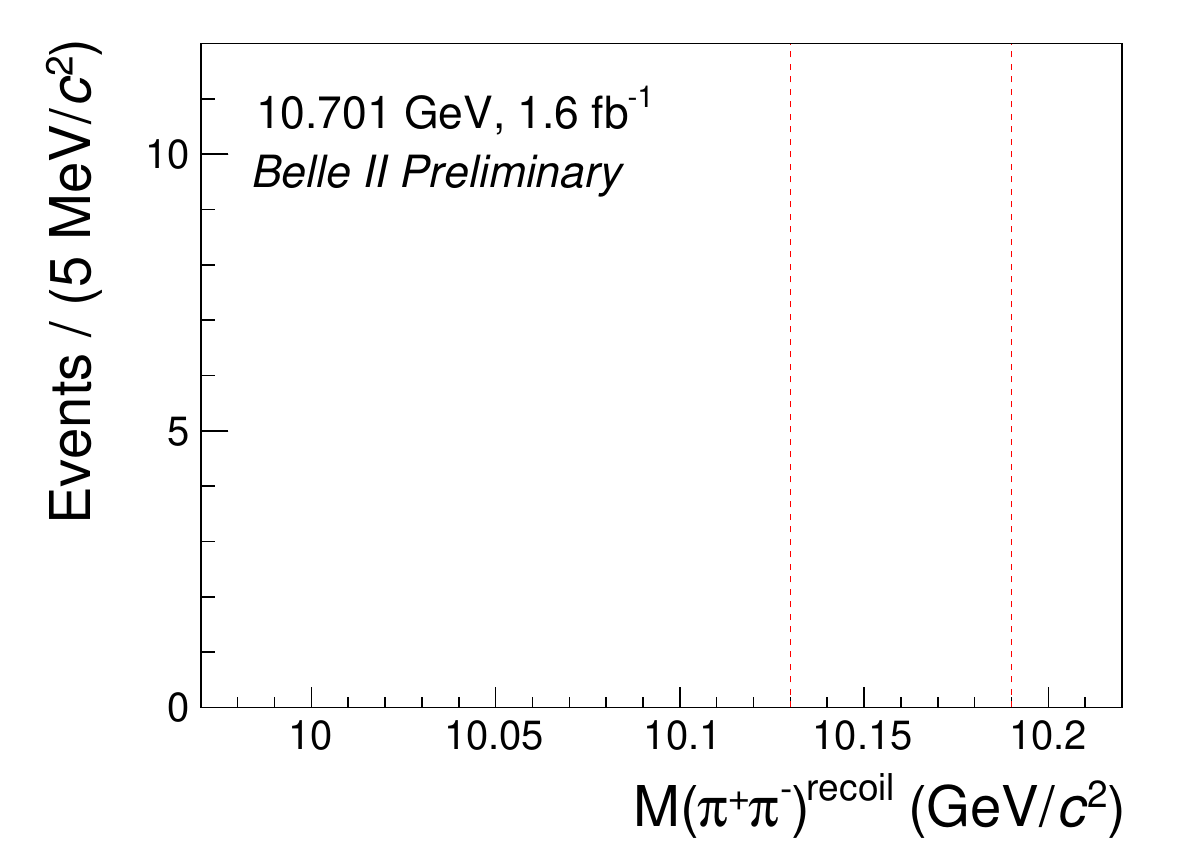}
\includegraphics[width=4.4cm]{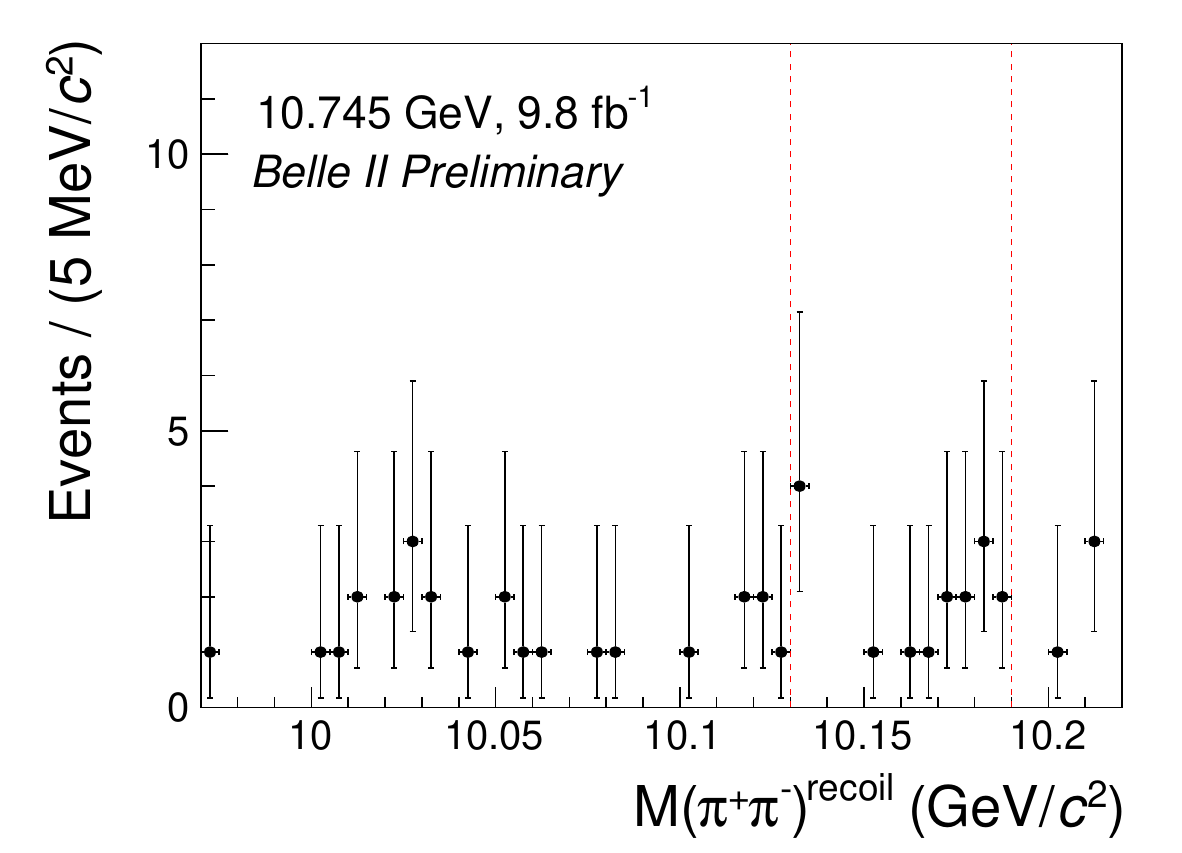}
\includegraphics[width=4.4cm]{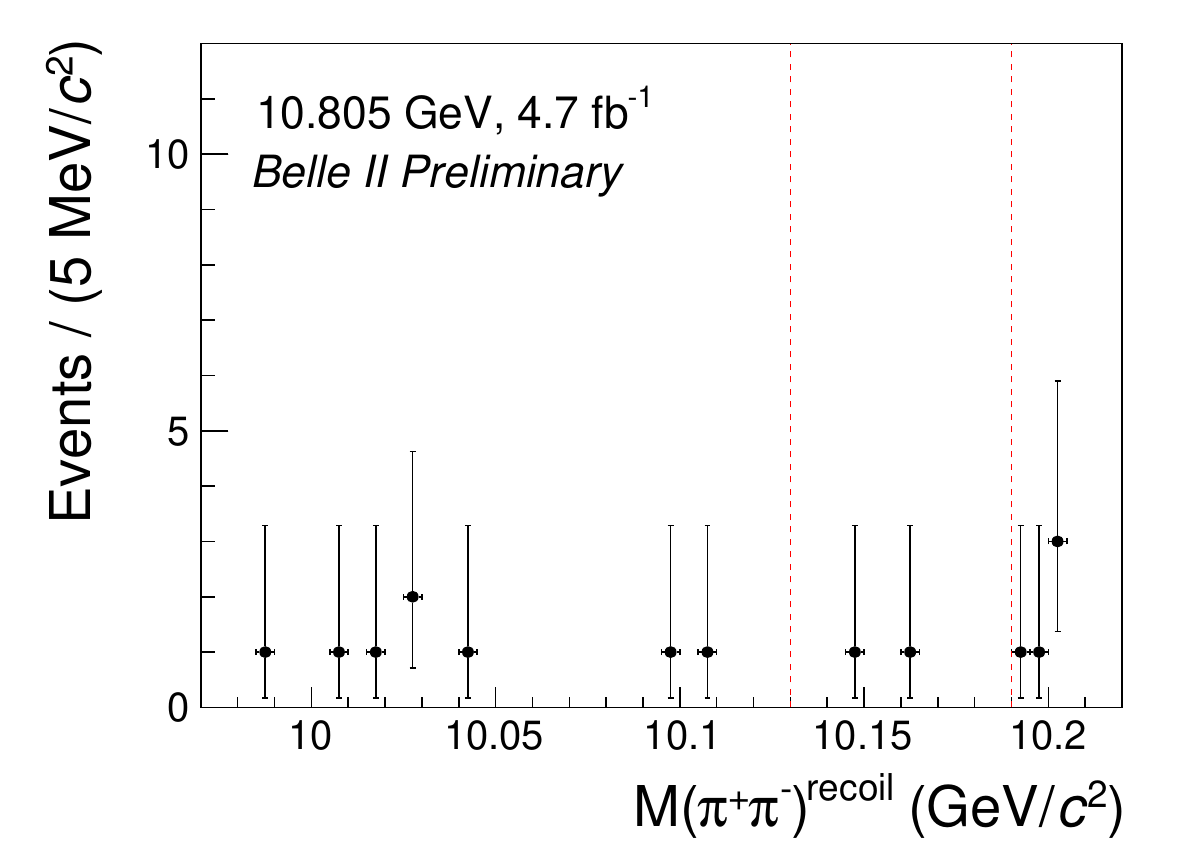}
\caption{The $M(\pi^+\pi^-)^{\rm recoil}$ distributions for data events in the $\chi_{b1}$ and $\chi_{b2}$ signal regions of 9.869--9.915 GeV/$c^2$ at $\sqrt{s}$ = 10.653, 10.701, 10.745, and 10.805 GeV at Belle II~\cite{2602.19807}.}
\label{Belle2v1}
\end{figure*}

\newpage

\item At C.M.\ energies with small numbers of candidates, Gaussian uncertainties give a poor description of the contribution to the likelihood. 
Therefore, we follow the procedure in Refs.~\cite{a220,a2510.25461} to use the profile likelihoods as inputs to determine the energy dependences of the cross sections for $e^+e^-\to\Upsilon_J(1D)\eta$ and $e^+e^-\to\Upsilon_J(1D)\pi^+\pi^-$.

For $N^{\rm obs}=0$, the dependence is parameterized using the proportionality function
\begin{equation}\label{eq:likelihood1}
f(x)=ax.
\end{equation}
When $N^{\rm obs}=1$ or higher, the dependence is parameterized by the function~\cite{a220}
\begin{equation}\label{eq:likelihood2}
f(x)=2(p_2x + p_3 - p_1 + p_1 {\rm ln} \frac{p_1}{p_2x + p_3}) P_4(x),
\end{equation}
where $P_4=1+q_1 x+q_2 x^2+q_3 x^3$. 
The parameters $a$, $p_i$, and $q_i$ of Eqs.~(\ref{eq:likelihood1}) and (\ref{eq:likelihood2}) are determined from fits. 
When fitting the energy dependence of the cross sections, instead of Gaussian uncertainties, we use Eqs.~(\ref{eq:likelihood1}) and (\ref{eq:likelihood2}) to account for the contributions of all samples to the function being minimized. The fitted values of $a$, $p_i$, and $q_i$ could be found in Ref.~\cite{likelihood}.

\item Figure~\ref{pipi} shows the distribution of $M(\pi^+\pi^-)$ in $e^+e^-\to\Upsilon_J(1D)\pi^+\pi^-$ in data at $\sqrt{s}$ = 10.8658 GeV. No clear structure is observed.

\begin{figure*}[htbp]
\centering
\includegraphics[width=8cm]{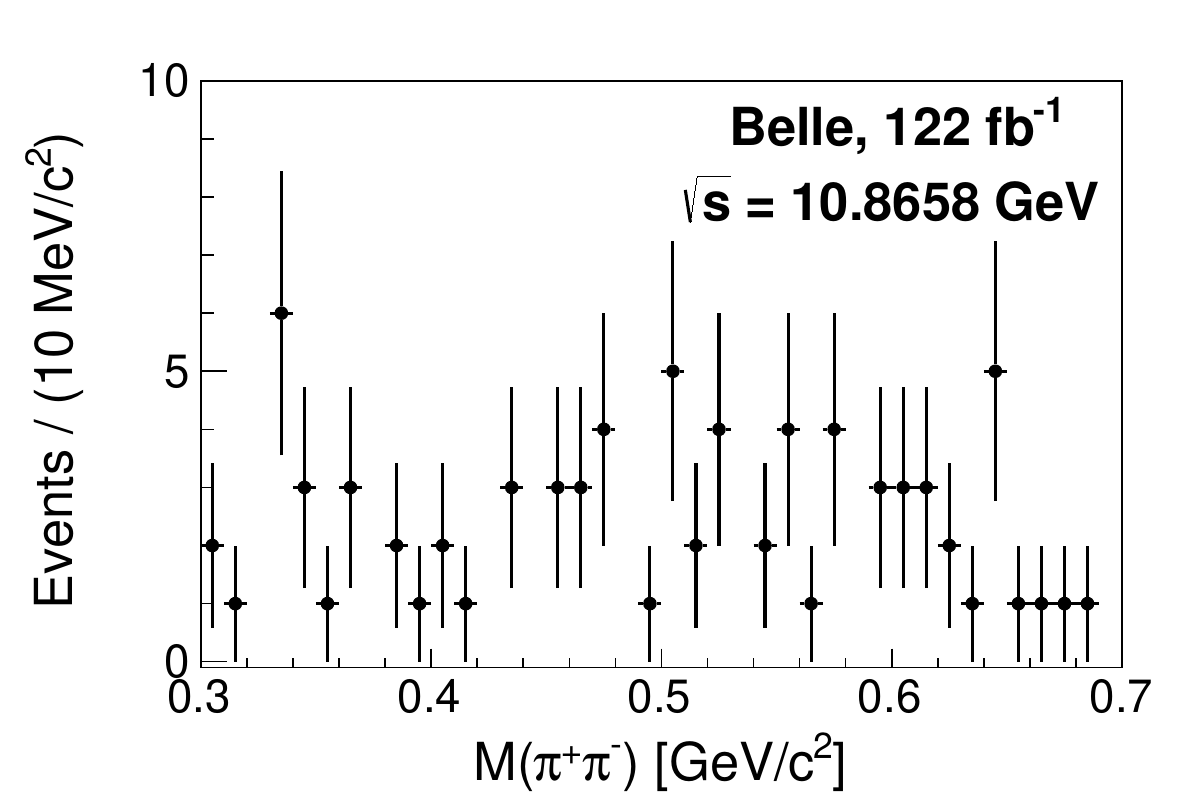}
\put(-205, 146){\large \bf Preliminary}
\caption{The distribution of $M(\pi^+\pi^-)$ in $e^+e^-\to\Upsilon_J(1D)\pi^+\pi^-$ in data at $\sqrt{s}$ = 10.8658 GeV.}
\label{pipi}
\end{figure*}

\end{itemize}

%%%%%%%%%%%%%%%%%%%%%%%%%%%%%%%%%%%%%%%%%%%%%%%%%%%%%%%%%%%%%%%%%%%%%%%%%%%%
\end{document}